\documentclass[
twocolumn,
aps,
prd,
superscriptaddress,
floatfix,
longbibliography
]{revtex4-1}
\usepackage{graphicx}
\usepackage{dcolumn}
\usepackage{bm}
\usepackage{amsmath}
\usepackage{amssymb}
\usepackage{color} 
\usepackage{xcolor}
\usepackage{mathrsfs} % mathscr
\usepackage{algorithm}
\usepackage{algorithmicx}
\usepackage{algpseudocode}

\usepackage{hyperref}
\hypersetup{
  colorlinks=true,
  linkcolor=[rgb]{0.00,0.80,0.0},
  citecolor=[rgb]{0.0,0.0,1.0},
  urlcolor=[rgb]{1.0,0.0,0.4},
  setpagesize=false
}

\DeclareRobustCommand
  \Compactcdots{\mathinner{\cdotp\mkern-2mu\cdotp\mkern-2mu\cdotp}}

\begin{document}

\title{Automatic quantum circuit encoding of a given arbitrary quantum state}

\author{Tomonori Shirakawa} 
\affiliation{Computational Materials Science Research Team, RIKEN Center for Computational Science (R-CCS), Kobe, Hyogo 650-0047, Japan}
\affiliation{Quantum Computational Science Research Team, RIKEN Center for Quantum Computing (RQC), Wako, Saitama 351-0198, Japan}

\author{Hiroshi Ueda}
\affiliation{Center for Quantum Information and Quantum Biology, Osaka University, Toyonaka, Osaka 560-8531, Japan}
\affiliation{Computational Materials Science Research Team, RIKEN Center for Computational Science (R-CCS), Kobe, Hyogo 650-0047, Japan}
\affiliation{JST, PRESTO, Kawaguchi, 332-0012, Japan}

\author{Seiji Yunoki}
\affiliation{Computational Materials Science Research Team, RIKEN Center for Computational Science (R-CCS), Kobe, Hyogo 650-0047, Japan}
\affiliation{Quantum Computational Science Research Team, RIKEN Center for Quantum Computing (RQC), Wako, Saitama 351-0198, Japan}
\affiliation{Computational Quantum Matter Research Team, RIKEN Center for Emergent Matter Science (CEMS), Wako, Saitama 351-0198, Japan}
\affiliation{Computational Condensed Matter Physics Laboratory, RIKEN Cluster for Pioneering Research (CPR), Saitama 351-0198, Japan}

\date{\today}

\begin{abstract}
  We propose a quantum-classical hybrid algorithm, named automatic quantum circuit encoding (AQCE), 
  to encode a given arbitrarily quantum state $\vert \Psi \rangle$ 
  onto an optimal quantum circuit $\hat{\mathcal{C}}$ with
  a finite number of single- and two-qubit quantum gates. 
  The proposed algorithm employs as an objective function the absolute value of fidelity 
  $F = \langle 0 \vert \hat{\mathcal{C}}^{\dagger} \vert \Psi \rangle$, which is maximized iteratively 
  to construct an optimal quantum circuit $\hat{\mathcal{C}}$ with controlled accuracy. 
  Here, $\vert 0 \rangle$ is a trivial product state in the computational basis of a quantum computer.
  The key ingredient of the algorithm is the sequential determination of a set of optimal two-qubit unitary operators 
  one by one via the singular value decomposition of the fidelity tensor.  
  Once the optimal unitary operators are determined, including the location of qubits on which each unitary operator acts, 
  elementary quantum gates are assigned algebraically. 
  These procedures are deterministic without assuming a quantum circuit ansatz 
  and thus do not introduce any parameter optimization of parametrized quantum gates.   
  With noiseless numerical simulations, we demonstrate the AQCE algorithm to encode a ground state of 
  quantum many-body systems, including the spin-1/2 antiferromagnetic Heisenberg model and the spin-1/2 XY model. 
  The results are also compared with the quantum circuit encoding of the same quantum state onto a quantum circuit in a given 
  circuit structure, i.e., a circuit ansatz, such as Trotter-like and MERA-like circuit structures.  
  Moreover, we demonstrate that the AQCE algorithm can also be applied to construct an optimal quantum circuit 
  for classical data such as a classical image that is represented as a quantum state by the amplitude encoding. 
  This scheme allows us to flexibly vary the required quantum resource. i.e., the number of qubits, 
  by dividing classical data into multiple pieces.
  Therefore, this is potentially useful for a near-term application in quantum machine learning, 
  e.g, as a state preparation of classical data for an input quantum state to be processed. 
  Finally, we also experimentally demonstrate that a quantum circuit generated by the AQCE algorithm
  can indeed represent the original quantum state reasonably on a noisy real quantum device.
\end{abstract}

\maketitle

\section{Introduction} \label{sec:intro}

It has been of crucial importance to find useful applications of quantum computers 
specially since the realization of real programable quantum devices.
Considering that currently available quantum devices are prone to noise and decoherence, 
it is highly desirable to find applications that can work effectively with a less number of quantum gates and qubits. 
Under these conditions, 
one of the promising and appealing approaches is based on variational quantum algorithms~\cite{Cerezo2021b} 
because it can be applied to a wide range of applications including quantum chemistry~\cite{Peruzzo2014,Yung2014,OMalley2016,Kandala2017,Yaangchao2017,Romero2018,Evangelista2019,McArdle2020}
and quantum machine learning~\cite{Schuld2015,WittekBook,Adcock2015,Biamonte2017,SchuldBook,Perdomo2018,Mitarai2018}.

In variational quantum algorithms, a quantum circuit is composed of parametrized quantum gates 
and these parameters are optimized classically on a classical computer so as to minimize or maximize a cost function 
by using standard optimization techniques such as a natural gradient descent method~\cite{Amari1998,Seki2020vqe,Stokes2020} 
and a sequential optimization technique~\cite{Nakanishi2020}.
In this context, there is an issue, known as barren plateau (BP) phenomena, where
the partial derivatives of the cost function vanish exponentially with increasing
the number of qubits and quantum gates~\cite{McClean2018,Cerezo2021,Wang2021,Marrero2021}.
The basic tool used to discuss the BP phenomena is the unitary $t$-design~\cite{Dankert2009}
related to the representability of a quantum circuit via the Haar measure~\cite{SimSukin2019,Hubregtsen2021}.
The theory of the BP phenomenon~\cite{McClean2018} claims that a quantum circuit
which shows unitary 2-design exhibits the BP phenomena. 
Generally, one tends to increase the number of quantum gates to represent a quantum state that one intends to prepare 
in the first place. 
However, such a quantum circuit with enhanced capability of representation 
can easily fall into the class of unitary operators belonging to the unitary 2-design~\cite{Harrow2009,Diniz2011},
thus suggesting the emergence of the BP phenomena~\cite{Holmes2021}.

Several routes have been discussed to alleviate and even avoid the BP phenomena. 
A simplest way is to select a cost function appropriately because the BP phenomena is 
cost function dependent~\cite{Cerezo2021}.
The importance of properly setting the initial variational parameters has also been pointed out~\cite{Mitarai2020qc}.
Another route to address the BP issue is to construct an appropriate quantum circuit.
For instance, it has been reported that a quantum circuit with a structure like 
the multi-scale entanglement renormalization group ansatz (MERA)~\cite{Vidal2007,Vidal2008,Evenbly2009}, 
known also as quantum convolutional neural network~\cite{Cong2019}, can avoid the BP phenomena~\cite{Pesah2021}. 
However, the tensor network structure of MERA is originally constructed to capture the quantum entanglement of a 
particular quantum state, e.g., in a one-dimensional critical system, 
and therefore it is not obvious at all and is probably not appropriate that the MERA-type quantum circuit can be 
applied to general problems. 
An alternative approach in this regard is a method such as adapted variational quantum eigensolver (ADAPT-VQE)~\cite{Grimsley2019,Tang2021,Liu2021,Yao2021,FengZhang2021},
where an appropriate quantum circuit is automatically generated by selecting quantum gates sequentially among a predetermined 
set of quantum gates accordingly to a problem to be sloved.

Following a similar strategy of the ADAPT-VQE, 
here in this paper, we propose a method that constructs an appropriate quantum circuit automatically, 
named automatic quantum circuit encoding (AQCE). The AQCE algorithm proposed here is
to construct a quantum circuit $\hat{\mathcal{C}}$ that approximates a given quantum state $\vert \Psi \rangle$ such that
$\vert \Psi \rangle \approx \hat{\mathcal{C}} \vert 0 \rangle$ with controlled accuracy. 
Here, $\hat{\mathcal{C}}$ is composed of a standard set of
quantum gates acting on up to two qubits and $|0\rangle$ is a trivial product state in the computational basis.
The algorithm is not based on a parametrized circuit ansatz but determines sequentially optimal two-qubit unitary operators, 
including an optimal location of qubits on which each unitary operator acts, by maximizing the fidelity with 
a technique inspired by the optimization algorithm in the tensor-network method~\cite{Evenbly2009}.
A standard set of quantum gates is assigned algebraically to these optimally determined unitary operators.
Therefore, the AQCE algorithm does not requires any derivatives of a cost function.

With noiseless numerical simulations,
we demonstrate the AQCE algorithm to encode a ground state of quantum many-body systems 
including the spin-1/2 isotropic antiferromagnetic Heisenberg model and the spin-1/2 XY model. 
We also compare the results with the quantum circuit encoding of the same quantum state onto a quantum circuit 
in a given circuit structure (i.e, a quantum circuit ansatz) such as the Trotter-like~\cite{Trotter1959,Suzuki1976,Lloyd1996} and MERA-like circuit structures.  
Furthermore, we apply this algorithm to encode classical data that is represented as a quantum state 
via the amplitude encoding~\cite{SchuldBook}, demonstrating a potential near-term application for 
a quantum state preparation of input data in quantum machine learning.
In addition, we employ a real quantum device provided by IBM Quantum~\cite{IBM} 
to demonstrate experimentally that a quantum circuit generated
by the AQCE algorithm can indeed represent the original quantum state reasonably.

The rest of this paper is organized as follows.
The AQCE algorithm is first introduced in Sec.~\ref{sec:encode}.
The performance of this algorithm 
is then demonstrated by numerical simulations in Sec.~\ref{sec:benchmark}. 
The method is first applied to encode the ground states of the spin-1/2 Heisenberg models in Sec.~\ref{sec:res:state},
and the results are compared with those for the quantum circuit encoding of the same quantum states 
onto quantum circuits with fixed circuit structures in Sec.~\ref{sec:res:comp}. 
The application of the AQCE algorithm to classical data such as a classical image 
represented by a quantum state via the amplitude encoding is also discussed in Sec.~\ref{sec:res:data}. 
Moreover, the AQCE algorithm is partially demonstrated experimentally with a real quantum device in Sec.~\ref{sec:exp}. 
Finally, the paper is concluded with a brief summary in Sec.~\ref{sec:summary}. 
The details of the gate assignment of unitary operators acting on a single qubit and on two qubits are 
described in Appendix~\ref{app:assign} and Appendix~\ref{app:phase}.

\section{\label{sec:encode}Quantum circuit encoding algorithm}

We first introduce the fidelity as an objective function for quantum circuit encoding in Sec.~\ref{sec:encode:obj}.
We then describe how to determine a unitary matrix of a quantum gate operation
by maximizing the objective function in Sec.~\ref{sec:encode:unitary}
and briefly explain how to assign an arbitrary unitary matrix to a standard set of single- and  two-qubit 
quantum gates in Sec.~\ref{sec:encode:assign}.
Based on these techniques, we introduce a prototype of the algorithm for the quantum circuit encoding in Sec.~\ref{sec:encode:algorithm}.
We also explain how to evaluate the fidelity tensor elements on a quantum computer in Sec.~\ref{sec:encode:imp}. 
Although the encoding algorithm can be applied in any cases, 
it might meet some difficulty when the fidelity tensor is essentially zero due to a particular symmetry reason. 
We discuss this issue and introduce an alternative approach to overcome this problem in Sec.~\ref{sec:encode:init}. 
This approach can be used for the initialization of the quantum circuitencoding. 
Combining with these methods in Secs.~\ref{sec:encode:algorithm} and \ref{sec:encode:init},
we finally introduce an algorithm, i.e., the AQCE algorithm, to construct a quantum circuit automatically 
in Sec.~\ref{sec:encode:aqce}.

\subsection{\label{sec:encode:obj}Objective for quantum circuit encoding}

We consider a quantum state defined on $L$ qubits that are enumerated as $\mathbb{L} = \{ 1,2, \Compactcdots, i, \Compactcdots, L \}$.
Let $\hat{X}_i$, $\hat{Y}_i$, and $\hat{Z}_i$
denote the $x$, $y$, and $z$ components of the Pauli operators, respectively, acting on qubit $i$.
We also introduce the notation $\hat{I}_i$ for representing the identity operator on qubit $i$. 
Let $\vert \sigma_i \rangle_i=\vert 0 \rangle_i$ and $\vert 1 \rangle_i$ denote the eigenstates of the Pauli operator $\hat{Z}_i$ at 
qubit $i$, i.e.,  
$\hat{Z}_i \vert 0 \rangle_i = \vert 0 \rangle_i$ and $\hat{Z}_i \vert 1 \rangle = - \vert 1 \rangle_i$.
The Hilbert space $\mathbb{H}_L$ on the $L$-qubit system $\mathbb{L}$
is spanned by the basis $\{ \vert \sigma_1 \sigma_2 \Compactcdots \sigma_L \rangle \}$, where
$\vert \sigma_1 \sigma_2 \Compactcdots \sigma_L \rangle = \underset{i=1}{\overset{L}{\otimes}} \vert \sigma_i \rangle_i$.
We can label the state $\vert \sigma_1 \sigma_2 \Compactcdots \sigma_L \rangle$ 
by introducing the integer number 
\begin{equation}
  n = \sum_{i=1}^L 2^{i-1} \sigma_i \label{eq:computational:label}
\end{equation}
as $\{ \vert n \rangle = \vert \sigma_1 \sigma_2 \Compactcdots \sigma_L \rangle \}_{n=0}^{2^L-1}$
%computational basis, and 
in the Hilbert space $\mathbb{H}_L = \text{span}\{ \vert \sigma_1 \sigma_2 \Compactcdots \sigma_L \rangle \}$.

Let $\vert \Psi \rangle$ be an arbitrary quantum state defined on $\mathbb{L}$ and let us assume that $\vert \Psi \rangle$ is normalized.
In addition, we can assume that $\vert \Psi \rangle$ is given generally 
by a linear combination of many different quantum circuit states, i.e., 
\begin{equation}
  \vert \Psi \rangle = \sum_{\gamma=1}^{\Gamma} \chi_{\gamma} \vert \psi^{(\gamma)} \rangle
  \label{eq:psi:lc}
\end{equation}
with
\begin{equation}
  \vert \psi^{(\gamma)} \rangle = \hat{\psi}^{(\gamma)} \vert 0 \rangle, 
  \label{eq:psi_lc}
\end{equation}
where $\chi_{\gamma}$ and $\hat{\psi}^{(\gamma)}$ with $\gamma=1,2,\Compactcdots,\Gamma$ are 
complex-valued coefficients and quantum circuits, respectively.
Although $\langle  \psi^{(\gamma)}   | \psi^{(\gamma)}  \rangle=1$, i.e., $\hat{\psi}^{(\gamma)}$ being unitary, 
here we do not assume that the states $ \vert \psi^{(\gamma)} \rangle$ with different values of $\gamma$ are mutually orthogonal. 
Note that Eq.~(\ref{eq:psi:lc}) may include the simplest and most extreme case 
where the circuits $\hat{\psi}_{\gamma}$ are composed simply of 
products of Pauli-X operators 
\begin{equation}
  \hat{\mathcal{P}}_{ \{ \sigma_i \}_{i=1}^L  } \equiv \prod_{i=1}^L \hat{X}_{i}^{\sigma_i}, 
\end{equation}
i.e., $\hat{\mathcal{P}}_{ \{ \sigma_i \}_{i=1}^L} \vert 0 \rangle = \vert \sigma_1 \sigma_2 \Compactcdots \sigma_L \rangle$, 
as in the case of a quantum state representing a classical image via the amplitude encoding discussed in Sec.~\ref{sec:res:data}. 
The objective here is to represent $\vert \Psi \rangle$ by a quantum circuit $\hat{\mathcal{C}}|0\rangle$ 
that is a priori unknown.
The algorithm proposed here constructs a quantum circuit $\hat{\mathcal{C}}$ that approximately 
represents $\vert \Psi \rangle\approx\hat{\mathcal{C}} \vert 0\rangle$ with controlled accuracy. 
We should note that this can be considered 
as a special case in the variational quantum state eigensolver for a density operator recently reported in Ref.~\cite{Cerezo2020a}, 
although the optimization method introduced here is different, as it will be described below.

For this purpose, we consider as an objective function to be maximized 
the absolute value of the overlap $F$ between $\vert \Psi \rangle$ and $\hat{\mathcal{C}} \vert 0 \rangle$, i.e., 
\begin{equation}
  F = \langle 0 \vert \hat{\mathcal{C}}^{\dagger} \vert \Psi \rangle.
\end{equation}
Assuming that the quantum circuit $\hat{\mathcal{C}}$ is composed of a product of unitary operators 
$\hat{\mathcal{U}}_{m}$, i.e.,
\begin{equation}
  \hat{\mathcal{C}}^{\dagger} = \prod_{m=1}^{M} \hat{\mathcal{U}}_{m}^{\dagger}
  = \hat{\mathcal{U}}_1^{\dagger} \hat{\mathcal{U}}_{2}^{\dagger} \Compactcdots \hat{\mathcal{U}}_M^{\dagger}, 
\end{equation}
$F$ can be expressed as
\begin{equation}
  F_m = \langle \Phi_{m-1} \vert \hat{\mathcal{U}}_m^{\dagger} \vert \Psi_{m+1} \rangle, 
\end{equation}
where we have introduced the subscript $m$ explicitly for the reason clarified below 
and the quantum states $\vert \Psi_m \rangle$ and 
$\langle \Phi_m \vert$ defined respectively as 
\begin{equation}
  \vert \Psi_m \rangle = \prod_{k=m}^M \hat{\mathcal{U}}_k^{\dagger} \vert \Psi \rangle 
  =  \hat{\mathcal{U}}_m^\dag \hat{\mathcal{U}}_{m+1}^\dag \Compactcdots \hat{\mathcal{U}}_M^\dag \vert \Psi \rangle
  \label{eq:ket:m}
\end{equation}
and 
\begin{equation}
  \langle \Phi_m \vert = \langle 0 \vert \prod_{k=1}^{m} \hat{\mathcal{U}}_k^{\dagger}
  = \langle 0 \vert 
  \hat{\mathcal{U}}_1^{\dagger} \hat{\mathcal{U}}_{2}^{\dagger} \Compactcdots \hat{\mathcal{U}}_m^{\dagger}.
  \label{eq:bra:m}
\end{equation}

\subsection{Determination of unitary operators} \label{sec:encode:unitary}

In order to appropriately determine each unitary operator $\hat{\mathcal{U}}_m$ composing the quantum circuit $\hat{\mathcal{C}}$, 
here we propose a method inspired by a tensor-network algorithm~\cite{Evenbly2009} by 
introducing a fidelity tensor operator. 

Let $\mathbb{I}_m = \{ i_1, i_2, \Compactcdots, i_K \}$ be a subsystem in the total qubit system $\mathbb{L}$ 
and assume that an unitary operator $\hat{\mathcal{U}}_m$ is defined on the subsystem $\mathbb{I}_m$.
By labeling the basis states $\{ \vert n \rangle = \vert \sigma_{i_1} \sigma_{i_2} \Compactcdots \sigma_{i_K} \rangle \}_{n=0}^{2^K-1}$
on the subsystem $\mathbb{I}_m$,
the unitary operator $\hat{\mathcal{U}}_m$ can be represented generally as
\begin{equation}
  \hat{\mathcal{U}}_m 
  = \sum_{n=0}^{2^K-1} \sum_{n^{\prime}=0}^{2^K-1} \vert n \rangle [ {\bm U}_m ]_{nn^{\prime}} \langle n^{\prime} \vert,
\end{equation}
where ${\bm U}_m$ is a $2^K \times 2^K$ unitary matrix and $[ {\bm A} ]_{nn^{\prime}}$ denotes a matrix element in 
the $n$th row and the $n^{\prime}$th column of matrix ${\bm A}$.

We shall now introduce the following fidelity tensor operator $\hat{\mathcal{F}}_m$: 
\begin{equation}
  \hat{\mathcal{F}}_m = {\rm Tr}_{ \bar{\mathbb{I}}_m } \left[ \vert \Psi_{m+1} \rangle \langle \Phi_{m-1} \vert \right],
  \label{eq:op:fm}
\end{equation}
where $\bar{\mathbb{I}}_m$ is the complement of the subsystem $\mathbb{I}_m$ in $\mathbb{L}$
and ${\rm Tr}_{ \mathbb{A} }\hat{\mathcal{O}}$ indicates the trace of operator $\hat{\mathcal{O}}$
over the Hilbert space spanned by the basis states for subsystem $\mathbb{A} = \{ i_1 ,i_2, \Compactcdots, i_A \} \subset \mathbb{L}$, i.e., 
\begin{equation}
  {\rm Tr}_{\mathbb{A}} [ \hat{\mathcal{O}} ] = 
  \sum_{\sigma_{i_1}=0}^1 \sum_{\sigma_{i_2}=0}^1 \Compactcdots
  \sum_{\sigma_{i_A}=0}^1 \langle \sigma_{i_1} \sigma_{i_2} \Compactcdots \sigma_{i_A} \vert 
  \hat{\mathcal{O}}
  \vert \sigma_{i_1} \sigma_{i_2} \Compactcdots \sigma_{i_A} \rangle, 
\end{equation}
with $\vert \sigma_{i_1} \Compactcdots \sigma_{i_A} \rangle = \vert \sigma_{i_1} \rangle_{i_1} \vert \sigma_{i_2} \rangle_{i_2} \Compactcdots\vert \sigma_{i_A} \rangle_{i_A}$.
Since $\hat{\mathcal{F}}_m$ is an operator defined on the Hilbert space spanned by the basis states for the subsystem 
$\mathbb{I}_m$, one can represent the operator $\hat{\mathcal{F}}_m$ in the matrix form as 
\begin{equation}
  \hat{\mathcal{F}}_m = \sum_{n=0}^{2^K-1} \sum_{n^{\prime}=0}^{2^K-1} \vert n \rangle [ {\bm F}_m ]_{nn^{\prime}} \langle n^{\prime} \vert.
  \label{eq:mat:fm}
\end{equation}
We can now readily find that
\begin{equation}
  {\rm Tr}_{ \mathbb{I}_m } [ \hat{\mathcal{F}}_m \hat{\mathcal{U}}_m^{\dagger} ] = \langle \Phi_{m-1} \vert \hat{\mathcal{U}}_m^{\dagger} \vert \Psi_{m+1} \rangle = F_m.
\end{equation}
We also find that
\begin{equation}
  {\rm Tr}_{ \mathbb{I}_m } [ \hat{\mathcal{F}}_m \hat{\mathcal{U}}_m^{\dagger} ] = {\rm tr} [ {\bm F}_m {\bm U}_m^{\dagger} ],
\end{equation}
where ${\rm tr} {\bm A}$ indicates the trace of matrix ${\bm A}$. 
Note that ${\bm F}_m$ is a $2^K\times 2^K$ matrix and is neither Hermitian nor unitary in general.

Let us now perform the singular-value decomposition (SVD) for ${\bm F}_m$ as 
${\bm F}_m = {\bm X} {\bm D} {\bm Y}$, where ${\bm X}$ and ${\bm Y}$ are $2^{K} \times 2^{K}$ unitary matrices,
and ${\bm D}$ is a non-negative real diagonal matrix with its diagonal elements being the singular values $d_n$ ($n=0,1,2,\Compactcdots,2^K-1$)
of matrix ${\bm F}_m$. Note that the $m$ dependence of these matrices 
${\bm X}$, ${\bm Y}$, and ${\bm D}$ is implicitly assumed. 
We then find that
\begin{equation}
  F_m = {\rm tr} [ {\bm X} {\bm D} {\bm Y} {\bm U}_m^{\dagger} ] = {\rm tr} [ {\bm D} {\bm Z} ]
  = \sum_{n=0}^{2^K-1} d_n [ {\bm Z} ]_{nn},
  \label{eq:fm_svd}
\end{equation}
where ${\bm Z} = {\bm Y} {\bm U}_m^{\dagger} {\bm X}$ is a unitary matrix. 
The absolute value of $F_m$ thus satisfies that
\begin{equation}
\vert F_m \vert = \left|  \sum_{n=0}^{2^K-1} d_n [ {\bm Z} ]_{nn} \right| \le \sum_{n=0}^{2^K-1} d_n \left| [ {\bm Z} ]_{nn} \right| .
\label{eq:abF}
\end{equation}
The equality in Eq.~(\ref{eq:abF}) holds if and only if $\text{arg} \left[[ {\bm Z} ]_{nn} \right]$ is the same for all $n$, 
where $\text{arg}\left[ {\bm c} \right]$ denotes the phase of complex number ${\bm c}$. 
Noticing also that $\sum_{n'}\left|  [ {\bm Z} ]_{nn'} \right|^2 = \sum_{n'}\left|  [ {\bm Z} ]_{n'n} \right|^2 =1$
because ${\bm Z}$ is unitary,
the absolute value of $F_m$ is thus maximized by choosing $[ {\bm Z} ]_{nn'} = \delta_{nn'}$~\cite{phase}, i.e., 
${\bm Z} = {\bm I}$, where ${\bm I}$ is the identity matrix. 
Hence, the unitary matrix ${\bm U}_m$ that maximizes the absolute value of $F_m$ is obtained as
\begin{equation}
  {\bm U}_m = {\bm X} {\bm Y} 
  \label{eq:optU}
\end{equation}
and therefore we can determine the optimal unitary operator $\hat{\mathcal{U}}_m$.

Three remarks are in order. First, as already stated above, a similar idea is used in the optimization of tensor 
network states~\cite{Evenbly2009}.
Second, although the fidelity tensor $\hat{\mathcal{F}}_m$ can be defined for a subsystem containing many qubits,
we focus mostly on the two-qubit case in this paper. 
This is simply because the assignment of elementary quantum gates for an optimal unitary operator determined in 
Eq.~(\ref{eq:optU}) can be made rather simply, as described in the next section.
Third, the fidelity $F_m$ and the fidelity tensor operator $\hat{\mathcal{F}}_m$ can be more explicitly expressed for 
the case when the state $|\Psi\rangle$ is given by a linear combination of several quantum states as in Eq.~(\ref{eq:psi:lc}), i.e., 
\begin{equation}
F_m = \sum_{\gamma=1}^\Gamma \chi_\gamma f_m^{(\gamma)} 
\label{eq:fm_lc}
\end{equation} 
and
\begin{equation}
\hat{\mathcal{F}}_m = \sum_{\gamma=1}^\Gamma \chi_\gamma \hat{f}_m^{(\gamma)}, 
\end{equation}
where 
\begin{equation}
f_m^{(\gamma)} = {\rm Tr}_{ \mathbb{I}_m } [ \hat{f}_m^{(\gamma)} \hat{\mathcal{U}}_m^{\dagger} ] 
\end{equation} 
and
\begin{equation}
  \hat{f}_m^{(\gamma)} = {\rm Tr}_{\bar{\mathbb{I}}_m} [ \vert \psi_{m+1}^{(\gamma)} \rangle \langle \Phi_{m-1} \vert ]
  \label{eq:f:lc}
\end{equation}
with
\begin{equation}
\vert \psi_{m+1}^{(\gamma)}\rangle = \hat{\mathcal{U}}_m^\dag \hat{\mathcal{U}}_{m+1}^\dag \Compactcdots \hat{\mathcal{U}}_M^\dag 
\vert \psi^{(\gamma)} \rangle.
\label{eq:phi_gamma}
\end{equation} 
The optimal unitary operator $ \hat{\mathcal{U}}_m$ that maximizes the absolute value of $F_m$ 
is still determined by Eqs.~(\ref{eq:fm_svd})-(\ref{eq:optU}).

\subsection{\label{sec:encode:assign}Assignment of quantum gates for a general unitary operator}

Once we obtain the matrix representation ${\bm U}_m$ for the unitary operator $\hat{\mathcal{U}}_m$ in Eq.~(\ref{eq:optU}), 
we have to assign a standard set of elementary quantum gates to this operator $\hat{\mathcal{U}}_m$.
It is well known that 
any unitary operator can be compiled as a product of two-qubit quantum gates 
by using the method that proves the universality of the quantum computation~\cite{NielsenChuangBook}. 
However, this method of decomposing a unitary operator acting on $K$ qubits generates an exponentially large number of 
elementary single- and two-qubit quantum gates with $K$. 
Therefore, it is not practical for our purpose.

% --- NOTE ---
% This method is composed of the two-level decomposition by using the two-component unitary matrix.
% The number of the two-level decomposition necessary is the same as half of the off-diagonal elements of $\hat{\mathcal{U}}_m$,
% suggesting the exponential increase as a function of the size of the subset $\mathbb{I}$ for the unitary gate $\hat{\mathcal{U}}_m$.
% In addition, the two-component matrix for the two-level decomposition is represented by using the multi-controlled unitary gate (general Toffoli gate).
% To represent the multi-controlled unitary gate, we have to decompose it into the product of the two-qubit gates.
% A simple method to generate the multi-controlled unitary gate is the use of the Gray code,
% and thus it also increases exponentially with the increase of the target qubit size of the unitary operator $\hat{\mathcal{U}}_m$.
% ------------

\begin{figure}
  \includegraphics[width=\hsize]{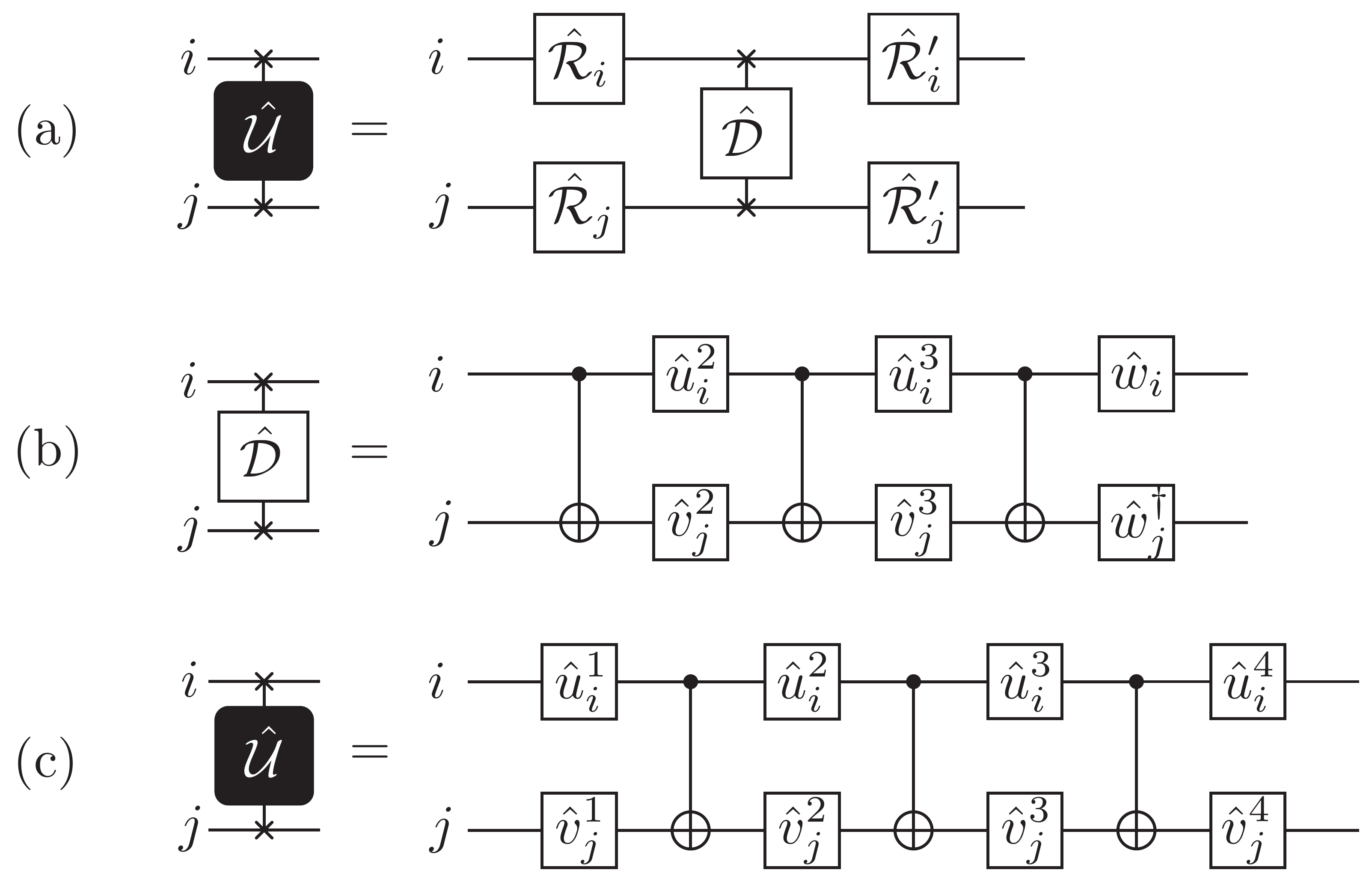}
  \caption{General form of a two-qubit unitary gate acting on qubits $i$ and $j$. 
  (a) Any unitary operator $\hat{\mathcal U}$ is decomposed into four single-qubit Euler rotations and two-qubit operator 
  $\hat{\mathcal D}$. 
  (b) $\hat{\mathcal D}$ is further decomposed into a product of the most standard quantum gates, including 
  three controlled NOT gates. 
  (c) Decomposition of a unitary operator $\hat{\mathcal U}$ into a standard set of the most elementary quantum gates. 
  Each single-qubit operation is 
  defined in the text. 
  }
  \label{fig:twogate}
\end{figure}

In contrast, focusing on an unitary operator acting on two qubits, there exists an optimal form decomposing it into elementary quantum gates~\cite{Kraus2001}, 
which can be determined from the matrix representation ${\bm U}_m$. Here we briefly outline this procedure.
As proved in Ref.~\onlinecite{Kraus2001}, any two-qubit unitary operator $\hat{\mathcal{U}}$ acting on qubits $i$ and $j$
can be decomposed into a product of elementary gate operations in the following canonical form [also see Fig.~\ref{fig:twogate}(a)]:
\begin{equation}
  \hat{\mathcal{U}} = {\rm e}^{-{\rm i}\alpha_0} \hat{\mathcal{R}}_i^{\prime} \hat{\mathcal{R}}_j^{\prime}
  \hat{\mathcal{D}} \hat{\mathcal{R}}_i \hat{\mathcal{R}}_j,
  \label{eq:twogate:physical}
\end{equation}
where $\alpha_0$ is an overall phase factor, not relevant for the assignment, 
$\hat{\mathcal{R}}_q^{\prime}$ and $\hat{\mathcal{R}}_q$ are single-qubit Euler rotations acting on qubit $q\,(=i,j)$ 
given by 
\begin{equation}
  \hat{\mathcal{R}}_q =
      {\rm e}^{-{\rm i}\xi_1^q \hat{Z}_q/2}
      {\rm e}^{-{\rm i}\xi_2^q \hat{Y}_q/2}
      {\rm e}^{-{\rm i}\xi_3^q \hat{Z}_q/2}
\end{equation}
and
\begin{equation}
  \hat{\mathcal{R}}_q^{\prime} =
      {\rm e}^{-{\rm i}\zeta_1^q \hat{Z}_q/2}
      {\rm e}^{-{\rm i}\zeta_2^q \hat{Y}_q/2}
      {\rm e}^{-{\rm i}\zeta_3^q \hat{Z}_q/2},
\end{equation}
and $\hat{\mathcal{D}}$ is a two-qubit entangled operator
\begin{equation}
  \hat{\mathcal{D}} = {\rm e}^ {-{\rm i} ( \alpha_1 \hat{X}_i \hat{X}_j
    + \alpha_2 \hat{Y}_i \hat{Y}_j 
    +  \alpha_3 \hat{Z}_i \hat{Z}_j )}.
  \label{eq:exgate}
\end{equation}
By following the proof of Eq.~(\ref{eq:twogate:physical}) in Ref.~\onlinecite{Kraus2001},
the parameters $\xi_q^k$ ($k=1,2,3$, $q=i,j$) and $\zeta_q^k$ ($k=1,2,3$, $q=i,j$) for the Euler rotaions 
and $\alpha_k$ ($k=1,2,3$) for $\hat{\mathcal{D}}$ as well as $\alpha_0$ are determined algebraically. 
The details are provided in Appendix~\ref{app:assign} and Appendix~\ref{app:phase}. 
The total number of parameters in the right hand side of Eq.~(\ref{eq:twogate:physical}) is 16
and is identical to the number of free real parameters in a general U(4) matrix.

Next, as shown in Fig.~\ref{fig:twogate}(b), 
$\hat{\mathcal{D}}$ can be represented by a product of the most standard quantum gates~\cite{Vidal2004,Mark2008} 
\begin{equation}
  \hat{\mathcal{D}} = \hat{w}_i \hat{w}_j^{\dagger}
  \hat{\rm C}_i (\hat{X}_j) \hat{u}_i^3 \hat{v}_j^3
  \hat{\rm C}_i (\hat{X}_j) \hat{u}_i^2 \hat{v}_j^2
  \hat{\rm C}_i (\hat{X}_j).
  \label{eq:exgate:decomposed}
\end{equation}
Here, $\hat{\rm C}_i (\hat{X}_j)$ denotes the controlled NOT gate
where the NOT operation acting on the $j$th qubit is controlled by the $i$th qubit, 
and other gates are single qubit gates given by 
\begin{align}
  & \hat{w}_i = {\rm e}^{{\rm i}\pi \hat{X}_i/4},\ 
  \hat{w}_j^{\dagger} = {\rm e}^{-{\rm i}\pi \hat{X}_j/4},\\
  & \hat{u}_i^3 = \hat{H}_i \hat{S}_i,\
  \hat{v}_j^3 = {\rm e}^{-{\rm i}\alpha_2 \hat{Z}_j}, \\
  & \hat{u}_i^2 = \hat{H}_i {\rm e}^{{\rm i}\alpha_1 \hat{X}_i},\
  \hat{v}_j^2 = {\rm e}^{{\rm i}\alpha_3 \hat{Z}_j} 
\end{align}
with $\hat{H}_i$ and $\hat{S}_i$ being the Hadamard and shift gates, respectively.
The matrix representations ${\bm H}$ and ${\bm S}$ for these quantum gates
$\hat{H}_i$ and $\hat{S}_i$ in the computational basis are given respectively by
\begin{equation}
  {\bm H} = \frac{1}{\sqrt{2}} \left(
  \begin{array}{cc}
    1 & 1 \\
    1 & -1 \\
  \end{array}
  \right),\ 
  {\bm S} = \left(
  \begin{array}{cc}
    1 & 0 \\
    0 & {\rm i} \\
  \end{array}
  \right).
\end{equation}

Inserting the expression of Eq.~(\ref{eq:exgate:decomposed}) into Eq.~(\ref{eq:twogate:physical}),
we obtain that 
\begin{equation}
  \hat{\mathcal{U}} =
  \hat{u}_i^4 \hat{v}_j^4
  \hat{C}_i(\hat{X}_j) 
  \hat{u}_i^3 \hat{v}_j^3
  \hat{C}_i(\hat{X}_j) 
  \hat{u}_i^2 \hat{v}_j^2
  \hat{C}_i(\hat{X}_j) 
  \hat{u}_i^1 \hat{v}_j^1,
\end{equation}
where
\begin{align}
  & \hat{u}_i^1 = \hat{\mathcal{R}}_i,\
  \hat{v}_j^1 = \hat{\mathcal{R}}_j,\\
  & \hat{u}_i^4 = \hat{\mathcal{R}}_i^{\prime} \hat{w}_i,\ 
  \hat{v}_j^4 = \hat{\mathcal{R}}_j^{\prime} \hat{w}_j^{\dag}.
\end{align}
This is also schematically shown in Fig.~\ref{fig:twogate}(c). 
Note that once the matrix representation for a single-qubit unitary operator is obtained, 
we can reparametrize any sequential product of single-qubit operators by using an overall phase factor and
a single Euler rotation algebraically (see Appendix~\ref{app:phase}),
suggesting that all $\hat{u}_i^{k}$ ($k=1,2,3,4$) and $\hat{v}_j^{k}$ ($k=1,2,3,4$)
can be represented as single Euler rotations (apart from an overall phase factor). 
We should also note that when a matrix representation ${\bm U}$ of an unitary operator $\hat{\mathcal{U}}$ happens to 
be O(4), instead of U(4), the corresponding two-qubit operator for $\hat{\mathcal{U}}$ in Fig.~\ref{fig:twogate}(c) can be constructed 
with two controlled NOT gates~\cite{Vatan2004}.

\subsection{Quantum circuit encoding algorithm} \label{sec:encode:algorithm}

Using the procedures described above, 
we can now introduce an algorithm to construct a quantum circuit $\hat{\mathcal{C}}$ that approximately represents a given 
quantum state $\vert \Psi \rangle$. 
Without loss of generality, let us assume that 
$\hat{\mathcal{C}}^{\dagger} = \prod_{m=1}^M \hat{\mathcal{U}}_m^{\dagger}$ is given.
For example, we set $\hat{\mathcal{U}}_m = \hat{I}$ for all $m$ as the initial condition,
where $\hat{I}$ is the identity operator of the subspace defining $\hat{\mathcal{U}}_m$.
In the algorithm, we sequentially replaces $\hat{\mathcal{U}}_m$ to a new $\hat{\mathcal{U}}_m^{\prime}$
that maximizes the absolute value of the fidelity
\begin{equation}
  F_m = \langle \Phi_{m-1} \vert ( \hat{\mathcal{U}}_m^{\prime} ) ^{\dagger} \vert \Psi_{m+1} \rangle,
\end{equation}
where $\langle \Phi_{m-1} \vert$ and $\vert \Psi_{m+1} \rangle$ are given respectively by
\begin{equation}
  \begin{split}
    \langle \Phi_{m-1} \vert = & \langle 0 \vert \prod_{m^{\prime}=1}^{m-1} (\hat{\mathcal{U}}_{m^{\prime}}^\prime)^{\dagger} \\
    = & \langle 0 \vert (\hat{\mathcal{U}}_{1}^\prime)^{\dagger} (\hat{\mathcal{U}}_{2}^\prime)^{\dagger} 
    \Compactcdots (\hat{\mathcal{U}}_{m-1}^\prime)^{\dagger} 
  \end{split}
\end{equation}
and 
\begin{equation}
  \begin{split}
  \vert \Psi_{m+1} \rangle = & \prod_{m^{\prime}=m+1}^M ( \hat{\mathcal{U}}_{m^{\prime}} )^{\dagger} \vert \Psi \rangle \\
   = & ( \hat{\mathcal{U}}_{m+1} )^{\dagger} ( \hat{\mathcal{U}}_{m+2} )^{\dagger} \Compactcdots ( \hat{\mathcal{U}}_M )^{\dagger} \vert \Psi \rangle
  \end{split}
\end{equation}
with $\langle \Phi_{0} \vert = \langle 0 \vert$ and $\vert \Psi_{M+1} \rangle = \vert \Psi \rangle$. 
Furthermore, we assume that the $m$th two-qubit unitary operator $\hat{\mathcal{U}}_m$ 
acting at $\mathbb{I}_m = \{ i_m, j_m \}$ is replaced with an unitary operator $\hat{\mathcal{U}}_m^\prime$ 
acting at $\mathbb{I}_k = \{ i_k, j_k \}$ that is properly selected among a set of bonds 
$\mathbb{B} = \{ \mathbb{I}_1, \mathbb{I}_2, \Compactcdots, \mathbb{I}_B \}$. 
However, the generalization to $K$-qubit unitary operators with $K>2$ is straightforward.

\begin{figure}
  \includegraphics[width=\hsize]{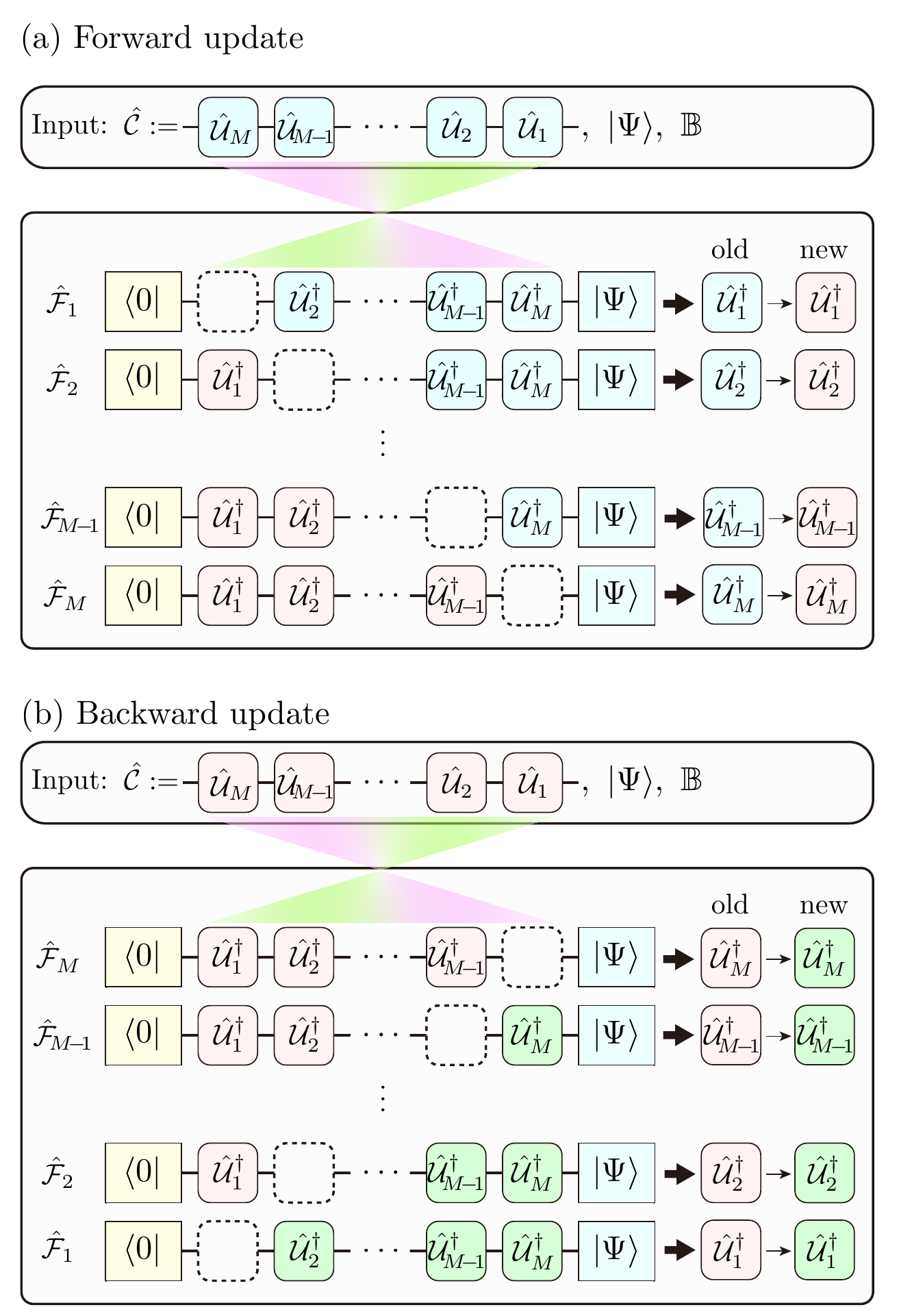}
  \caption{(a) Forward and (b) backward updates that optimize sequentially unitary operators $\hat{\mathcal U}_m$ 
  for $m=1,2,\Compactcdots,M$. 
  }
  \label{fig:qce}
\end{figure}

A prototype of the algorithm is then given as follows:
\begin{itemize}
\item[(1)] Set $m := 1$, $\hat{\mathcal{C}}_{0} := 1$, $\vert \Psi_{2} \rangle := \prod_{m'=2}^M  ( \hat{\mathcal{U}}_{m^{\prime}})^{\dagger} \vert \Psi \rangle$, and $\langle \Phi_{0} \vert := \langle 0 \vert$. 
\item[(2)] Evaluate matrix ${\bm F}_m^{(k)}$ of the fidelity tensor  $\hat{\mathcal{F}}_m^{(k)} = {\rm Tr}_{\bar{\mathbb{I}}_k} [ \vert \Psi_{m+1} \rangle \langle \Phi_{m-1} \vert ]$ for all $\mathbb{I}_k \in \mathbb{B}$.
\item[(3)] Perform SVD ${\bm F}_m^{(k)} = {\bm X}_k {\bm D}_k {\bm Y}_k$ for all ${\bm F}_m^{(k)}$, and calculate $S_k = \sum_{n=0}^3 [ {\bm D}_k ]_{nn}$. 
\item[(4)] Find $k=k^*$ that mximizes $S_k$.
\item[(5)] Set ${\bm U} := {\bm X}_{k^*} {\bm Y}_{k^*}$ and assign the quantum gates for the new $m$th unitary operator 
$\hat{\mathcal{U}}_m^\prime$, represented by the matrix ${\bm U}$, which acts on $\mathbb{I}_m := \{ i_{k^*}, j_{k^*} \}$.
\item[(6)] Set $\hat{\mathcal{C}}_{m} := \hat{\mathcal{U}}_m^\prime \hat{\mathcal{C}}_{m-1} $, 
$\vert \Psi_{m+2} \rangle := \hat{\mathcal{U}}_{m+1} \vert \Psi_{m+1} \rangle$,
and
$\langle \Phi_{m} \vert :=  \langle \Phi_{m-1} \vert (\hat{\mathcal{U}}_{m}^{\prime})^{\dagger}$.
\item[(7)] Set $m := m+1$ and go to (2) if $m \leq M$. Otherwise, quit and return $\hat{\mathcal{C}} := \hat{\mathcal{C}}_M$.
\end{itemize}
A schematic representation of this procedure is given in Fig.~\ref{fig:qce}(a).
Note that the procedure (2) is most time consuming and should be done on a quantum computer (see Sec.~\ref{sec:encode:imp}),
while the other procedures are performed on a classical computer. 
Since the evaluation of ${\bm F}_m^{(k)}$ for different bonds $\mathbb{I}_k$ is independent, 
one can trivially parallelize this part.

Although we have described the case for updating the unitary operators from $m=1$ to $m=M$,
it is apparent that we can reverse the order of updating the unitary operators from $m=M$ to $m=1$, 
as schematically shown in Fig.~\ref{fig:qce}(b). 
We shall call the algorithm for updating the unitary operators from $m=1$ to $m=M$
forward update, and the algorithm for updating the unitary operators from $m=M$ to $m=1$
backward update. Furthermore, let us refer to a single set of consecutive updates, 
forward update followed by backward update, as a sweep.

\subsection{Implementation on a quantum computer} \label{sec:encode:imp}

The most demanding part computationally in the quantum circuit encoding algorithm is to evaluate 
the fidelity tensor operator $\hat{\mathcal{F}}_m$ in Eq.~(\ref{eq:op:fm}). 
As explained here, this part can be evaluated directly using a quantum computer
when the quantum state $\vert \Psi \rangle$ is given by a linear combination of several quantum circuits 
as in Eq.~(\ref{eq:psi:lc}), which includes the extreme case where $\vert \Psi \rangle$ is given by a linear combination 
of direct product states in the computational basis.

Although this procedure can be extended to the case for any number of qubits in principle,
here we consider a subsystem composed of two qubits, i.e., $\mathbb{I}_m = \{ i, j \}$, on which an unitary operator 
$\hat{\mathcal U}_m$ acts. 
Let us introduce the following notation 
\begin{equation}
  \hat{\mathcal{P}}_i^{\alpha} = \left\{
  \begin{array}{cc}
    \hat{I}_i & (\alpha = 0 ) \\
    \hat{X}_i & (\alpha = 1 ) \\
    \hat{Y}_i & (\alpha = 2 ) \\
    \hat{Z}_i & (\alpha = 3 ) \\
  \end{array}
  \right.
\end{equation}
for the identity and Pauli operators acting on qubit $i$. 
Then the fidelity tensor operator $\hat{\mathcal{F}}_m$ in the two-qubit subsystem $\mathbb{I}_m = \{ i,j \}$
can be expressed generally as 
\begin{equation}
  \hat{\mathcal{F}}_m = \sum_{\alpha=0}^3 \sum_{\alpha^{\prime}=0}^3 \tilde{f}_{\alpha,\alpha^{\prime}} \hat{\mathcal{P}}_i^{\alpha} \hat{\mathcal{P}}_j^{\alpha^{\prime}},
  \label{eq:pauli:fm}
\end{equation}
where $\tilde{f}_{\alpha,\alpha^{\prime}}$ are complex numbers.
This is simply because the operator $\vert n \rangle \langle n^{\prime} \vert$ in Eq.~(\ref{eq:mat:fm}) for all $n,n^{\prime}=0,1,2,3$
can be expanded with a polynomial of the Pauli and identity operators.
We thus find that
\begin{equation}
  {\rm Tr}_{\mathbb{I}_m} [ \hat{\mathcal{F}}_{m} \hat{\mathcal{P}}_i^{\alpha} \hat{\mathcal{P}}_j^{\alpha^{\prime}} ]
  = 2^2 \tilde{f}_{\alpha,\alpha^{\prime}}
\end{equation}
because
\begin{equation}
  {\rm Tr}_{\mathbb{I}_m} [ \hat{\mathcal{P}}_{i}^{\beta} \hat{\mathcal{P}}_{j}^{\beta^{\prime}}
  \hat{\mathcal{P}}_i^{\alpha} \hat{\mathcal{P}}_j^{\alpha^{\prime}} ] = 2^2 \delta_{\alpha,\beta} \delta_{\alpha^{\prime},\beta^{\prime}}. 
\end{equation}

\begin{figure}
  \includegraphics[width=\hsize]{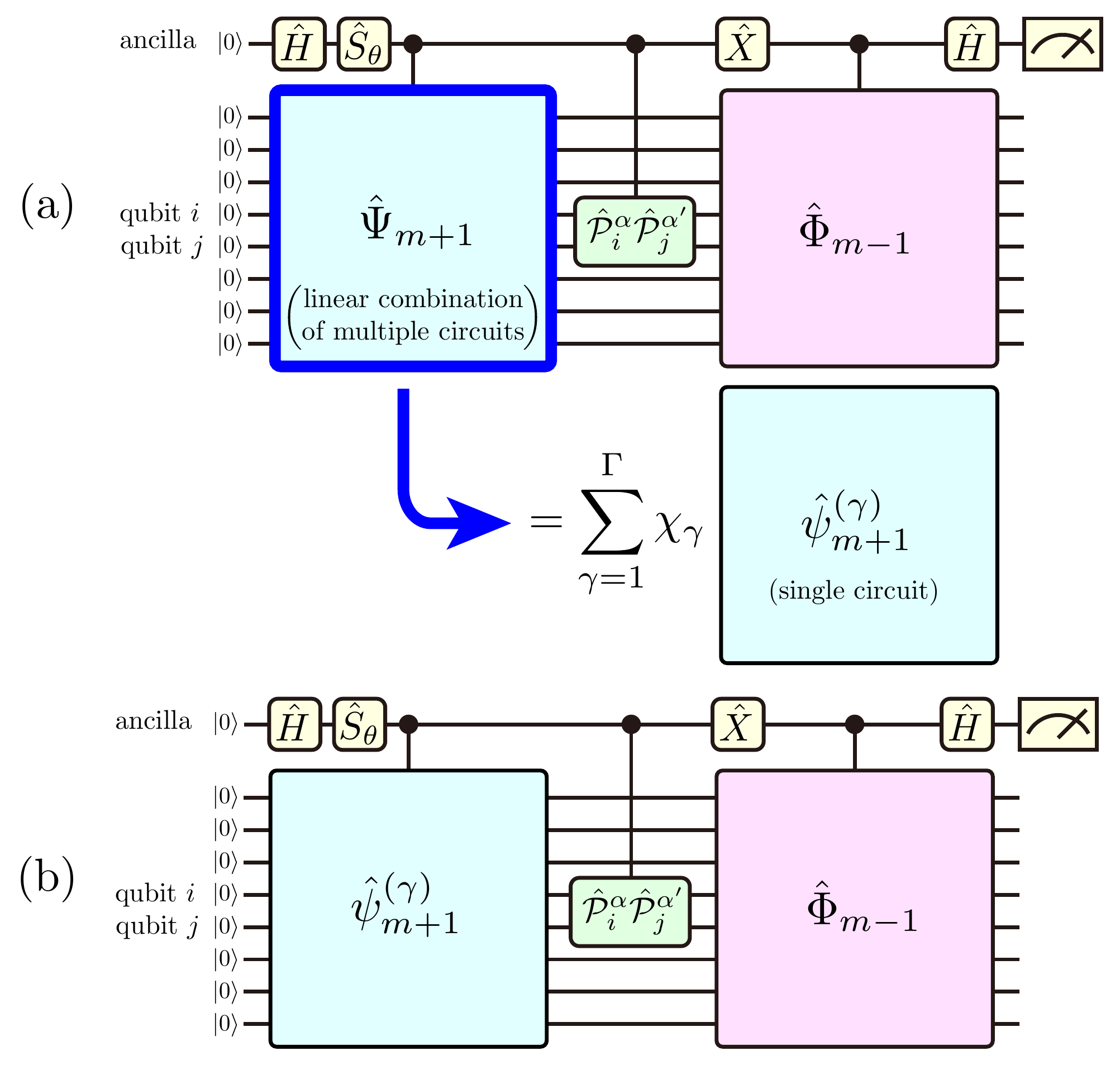}
  \caption{(a) A quantum circuit to evaluate Eq.~(\ref{eq:tr_fm}).  
    $\hat{H}$ and $\hat{X}$ are Hadamard and Pauli-X gates, respectively.
    $\hat{S}_{\theta}$ is a phase shift gate given by $\hat{S}_{\theta} \vert 0 \rangle = \vert 0 \rangle$
    and $\hat{S}_{\theta} \vert 1 \rangle = {\rm e}^{-{\rm i}\theta} \vert 1 \rangle$. 
    The quantum circuits $\hat{\Psi}_{m+1}$ and $\hat{\Phi}_{m-1}$ are defined as $|\Psi_{m+1}\rangle = 
     \prod_{k=m+1}^{M} \hat{\mathcal{U}}_{k}^{\dagger} |\Psi\rangle = \hat{\Psi}_{m+1} |0\rangle $ and 
     $\vert \Phi_{m-1} \rangle = \prod_{k=m-1}^{1} \hat{\mathcal{U}}_{k} \vert 0 \rangle = \hat{\Phi}_{m-1} \vert 0 \rangle$, respectively. 
    In our protocol,  we assume that $|\Psi\rangle$ is given by a linear combination of quantum circuit states $|\psi^{(\gamma)}\rangle$, 
    including the extreme case where it is given by a linear combination of direct product states in the computational basis.  
    Therefore, as indicated in the figure, $\hat{\Psi}_{m+1}$ is given by a linear combination of different quantum circuits, i.e., 
    $\hat{\Psi}_{m+1} = \sum_{\gamma=1}^{\Gamma} \chi_{\gamma} \hat{\psi}_{m+1}^{(\gamma)}$, where 
    $\hat{\psi}^{(\gamma)}_{m+1} = \prod_{k=m+1}^{M} \hat{\mathcal{U}}_{k}^{\dagger} \hat{\psi}^{(\gamma)}$ and 
    $|\psi^{(\gamma)}\rangle=\hat{\psi}^{(\gamma)}|0\rangle$. 
    Hence, $f_{\alpha,\alpha^{\prime}}^{(\gamma)} = \langle {\Phi}_{m-1} \vert \hat{\mathcal{P}}_i^{\alpha} \hat{\mathcal{P}}_j^{\alpha^{\prime}} \vert \psi_{m+1}^{(\gamma)} \rangle$ can be evaluated separately for each $\gamma$, as shown in (b), by a 
    Hadamard test like circuit. 
    By measuring $\hat{Z}$ at the ancilla qubit, we can evaluate 
    ${\rm Re}[ \langle \Phi_{m-1} \vert \hat{P}_i^{\alpha} \hat{P}_j^{\alpha^{\prime}} \vert \psi_{m+1}^{(\gamma)} \rangle ]$ for $\theta = 0 $
     and 
    ${\rm Im}[ \langle \Phi_{m-1} \vert \hat{P}_i^{\alpha} \hat{P}_j^{\alpha^{\prime}} \vert \psi_{m+1}^{(\gamma)} \rangle ]$ for 
    $\theta = \pi/2$.
    A black circle in the circuits indicates a control qubit for a control gate.
    }
  \label{fig:hadamard}
\end{figure}

On the other hand, by using the definition of the fidelity tensor operator $\hat{\mathcal{F}}_{m}$ in Eq.~(\ref{eq:op:fm}),
we find that
\begin{equation}
  {\rm Tr}_{\mathbb{I}_m} [ \hat{\mathcal{F}}_m \hat{\mathcal{P}}_i^{\alpha} \hat{\mathcal{P}}_j^{\alpha^{\prime}} ]
  = \langle {\Phi}_{m-1} \vert \hat{\mathcal{P}}_i^{\alpha} \hat{\mathcal{P}}_j^{\alpha^{\prime}} \vert \Psi_{m+1} \rangle.
\label{eq:tr_fm}
\end{equation}
Therefore, $\tilde{f}_{\alpha,\alpha^{\prime}}$ can be determined by
estimating the overlap between $\hat{\mathcal{P}}_i^{\alpha} \hat{\mathcal{P}}_j^{\alpha^{\prime}} \vert \Psi_{m+1} \rangle$
and $\vert \Phi_{m-1} \rangle$ for all $\alpha,\alpha^{\prime}=0,1,2,3$. 
This overlap can be evaluated using a Hadamard test like circuit shown in Fig.~\ref{fig:hadamard}(a), 
provided that a quantum circuit $\hat{\Psi}$ generating the quantum state $|\Psi\rangle=\hat{\Psi}|0\rangle$ is already known.  
However, this is generally not the case but rather the main task of the quantum circuit encoding algorithm is to finding a quantum 
circuit $\hat{\mathcal{C}}$ that approximately represents $\hat\Psi$.
Instead, here we assume that $\vert \Psi \rangle$ is given by a linear combination of quantum circuit states 
as in Eqs.~(\ref{eq:psi:lc}) and (\ref{eq:psi_lc}). In this case, Eq.~(\ref{eq:tr_fm}) can be more explicitly written as 
\begin{eqnarray}
{\rm Tr}_{\mathbb{I}_m} [ \hat{\mathcal{F}}_m \hat{\mathcal{P}}_i^{\alpha} \hat{\mathcal{P}}_j^{\alpha^{\prime}} ] 
&=& \sum_{\gamma=1}^\Gamma \chi_\gamma 
\langle {\Phi}_{m-1} \vert \hat{\mathcal{P}}_i^{\alpha} \hat{\mathcal{P}}_j^{\alpha^{\prime}} \vert \psi_{m+1}^{(\gamma)} \rangle \\
&=& \sum_{\gamma=1}^\Gamma \chi_\gamma
{\rm Tr}_{\mathbb{I}_m} [ \hat{f}_m^{(\gamma)} \hat{\mathcal{P}}_i^{\alpha} \hat{\mathcal{P}}_j^{\alpha^{\prime}} ], 
\end{eqnarray}
where $\hat{f}_m^{(\gamma)} $ and $\vert \psi_{m+1}^{(\gamma)} \rangle$ are defined in Eqs.~(\ref{eq:f:lc}) 
and (\ref{eq:phi_gamma}), respectively. 
As shown in Fig.~\ref{fig:hadamard}(b), 
$f_{\alpha,\alpha^{\prime}}^{(\gamma)} = {\rm Tr}_{\mathbb{I}_m} [ \hat{f}_m^{(\gamma)} \hat{\mathcal{P}}_i^{\alpha} \hat{\mathcal{P}}_j^{\alpha^{\prime}} ]$ can now be evaluated separately for each $\gamma$ by using a Hadamard test like circuit on a quantum computer.

\subsection{Initialization algorithm} \label{sec:encode:init}

Although the quantum circuit encoding algorithm described above in Sec.~\ref{sec:encode:algorithm} 
can be applied to general cases,
there are some exceptions for which care must be taken. 
For example, when we consider a ground state $\vert \Psi \rangle$ of a quantum spin system, 
the state $\vert \Psi \rangle$ is often in the spin singlet sector. 
In this case, there is no overlap between $\vert \Psi \rangle$ and $\vert 0 \rangle$ 
because the product state $\vert 0 \rangle$ represents the fully polarized state with the maximum spin value. 
Therefore, an alternative algorithm is required to construct an initial circuit $\hat{\mathcal{C}}$, 
for which $\hat{\mathcal{C}}^{\dagger} \vert \Psi \rangle$ has a finite overlap with $\vert 0 \rangle$.

Let us consider the reduced density matrix $\hat{\rho}$ of a quantum state $\vert \Psi \rangle$
on the subsystem $\mathbb{I} = \{ i_1, i_2, \Compactcdots, i_K \}$ that is given by
\begin{equation}
  \hat{\rho} = {\rm Tr}_{\bar{\mathbb{I}}} [ \vert \Psi \rangle \langle \Psi \vert ] 
\end{equation}
with the associated eigenstates and eigenvalues being denoted as 
$\vert \lambda_n \rangle$ and $\lambda_n$, respectively. Here we assume that $\lambda_n$ is in the descending order, i.e., 
$\lambda_0 \geq \lambda_1 \geq \lambda_2 \geq \Compactcdots$.
The reduced density matrix $\hat{\rho}$ is then represented as
\begin{equation}
  \hat{\rho} = \sum_{n} \vert \lambda_n \rangle \lambda_n \langle \lambda_n \vert. 
\end{equation}
We shall now find the unitary operator $\hat{\mathcal{V}}_1$ in the subsystem $\mathbb{I}$ such that 
\begin{equation}
  \underset{\hat{\mathcal{V}}_1}{\text{max}} \langle 0 \vert \hat{\mathcal{V}}^\dag_1 \hat{\rho} \hat{\mathcal{V}}_1 \vert 0 \rangle, 
  \label{eq:max:init}
\end{equation}
where $\vert 0 \rangle = \vert 0 \rangle_{i_1} \vert 0 \rangle_{i_2} \Compactcdots \vert 0 \rangle_{i_K}$ 
in the computational basis defined in the subsystem $\mathbb{I}$. 

For this end, let us first expand $\hat{\mathcal{V}}_1$ in the following general form: 
\begin{equation}
  \hat{\mathcal{V}}_1 = \sum_{l=0}^{2^K-1} \sum_{n=0}^{2^K-1}  v_{ln} \vert  \lambda_l \rangle \langle n \vert,
\end{equation}
where $\{ \vert n \rangle = \vert \sigma_{i_1} \sigma_{i_2} \Compactcdots \sigma_{i_K} \rangle \}_{n=0}^{2^K-1}$ are the basis states 
in the subsystem $\mathbb{I}$. 
We then find that
\begin{equation}
  \langle 0 \vert \hat{\mathcal{V}}^\dag_1 \hat{\rho} \hat{\mathcal{V}}_1 \vert 0 \rangle
  = \sum_{l=0}^{2^K-1} \lambda_l \vert v_{l0} \vert^2.
\end{equation}
It is now easy to find that
\begin{equation}
  \hat{\mathcal{V}}_1 = \sum_{n=0}^{2^K-1} \vert \lambda_n \rangle \langle n \vert 
  \label{eq:u:init}
\end{equation}
yields one of the solutions for Eq.~(\ref{eq:max:init}).
Once we determine the unitary operator $\hat{\mathcal{V}}_1$ in Eq.~(\ref{eq:u:init}), we can assign quantum gates 
for this operator by following the prescription described in Sec.~\ref{sec:encode:assign} for the two-qubit case, if it is required.

In numerical simulations, we can determine $\hat{\mathcal{V}}_1$ in Eq.~(\ref{eq:u:init}) by directly evaluating the eigenstates
$\vert \lambda_n\rangle$ of the reduced density operator $\hat{\rho}$. $\hat{\mathcal{V}}_1$ can also be determined via a quantum computer. 
To show this, let us consider the subsystem $\mathbb{I}$ composed of two qubits $\mathbb{I} = \{ i,j \}$, for simplicity. 
Expanding the reduced density matrix $\hat{\rho}$ with a polynomials of the Pauli and identity operators 
\begin{equation}
  \hat{\rho} = \sum_{\alpha,\alpha^{\prime}} \tilde{r}_{\alpha,\alpha^{\prime}} \hat{\mathcal{P}}_i^{\alpha} \hat{\mathcal{P}}_j^{\alpha^{\prime}},
  \label{eq:dopt}
\end{equation}
we find that
\begin{equation}
  \langle \Psi \vert \hat{\mathcal{P}}_i^{\alpha} \hat{\mathcal{P}}_j^{\alpha^{\prime}} \vert \Psi \rangle
  = 
  {\rm Tr}_{\mathbb{I}} \left[ \hat{\rho} \hat{\mathcal{P}}_i^{\alpha} \hat{\mathcal{P}}_j^{\alpha^{\prime}} \right]
  = 2^2 \tilde{r}_{\alpha,\alpha^{\prime}}.
  \label{eq:dopt2}
\end{equation}
This implies that the matrix elements $ \tilde{r}_{\alpha,\alpha^{\prime}}$ of $\hat{\rho}$ in Eq.~(\ref{eq:dopt})
can be determined
by measuring all possible pairs of products of the Pauli and identity operators
$\hat{\mathcal{P}}_i^{\alpha} \hat{\mathcal{P}}_j^{\alpha^{\prime}}$ ($\alpha,\alpha^{\prime} = 0,1,2,3$). 
This can be performed on a quantum computer directly
if the state $\vert \Psi\rangle$ is given in a quantum circuit, 
or by using the procedure described in Sec.~\ref{sec:encode:imp} (also see Fig.~\ref{fig:hadamard}), otherwise.  
Having estimated the reduced density matrix for $\hat{\rho}$, one can determine the unitary operator $\hat{\mathcal{V}}_1$ 
in Eq.~(\ref{eq:u:init}) classically. The extension to the subsystem $\mathbb{I}$ composed of more than two qubits is straightforward. 
We should also note that constructing a reduced density matrix by measuring a set of Pauli and identity operators 
on a quantum computer is known as quantum state tomography~\cite{NielsenChuangBook}, 
and the technique described in Sec.~\ref{sec:encode:imp} is also along this line.

This procedure can be easily extended for further adding unitary operators $\hat{\mathcal{V}}_2, \hat{\mathcal{V}}_3, \dots$. 
Let us assume that we have already determined the first unitary operator $\hat{\mathcal{V}}_1$ acting on 
$\mathbb{I} = \{ i_1, i_2, \Compactcdots, i_K \}$. The location of these qubits is selected among a set of clusters of $K$ qubits 
$\mathbb{C} = \{ \mathbb{I}_1, \mathbb{I}_2, \Compactcdots, \mathbb{I}_C \}$ so as to maximize Eq.~(\ref{eq:max:init}), i.e., 
\begin{equation}
\underset{\mathbb{I} \in \mathbb{C}}{\text{max}} \left[ \underset{\hat{\mathcal{V}}_1}{\text{max}} \langle 0 \vert \hat{\mathcal{V}}^\dag_1 \hat{\rho} \hat{\mathcal{V}}_1 \vert 0 \rangle \right]. 
\label{eq:max_v1}
\end{equation}
Let us now define a quantum state $|\tilde{\Psi}_1\rangle$ incorporating $\hat{\mathcal{V}}_1^\dag$ into the quantum state 
$|\Psi\rangle$, i.e., 
\begin{equation}
|\tilde{\Psi}_1\rangle = \hat{\mathcal{V}}_1^\dag |\Psi\rangle,  
\end{equation}
and consider the reduced density matrix $\hat{\rho}_1$ of $|\tilde{\Psi}_1\rangle$ on 
the subsystem $\mathbb{I}^\prime = \{ i'_1, i'_2, \Compactcdots, i'_K \}$ given by 
\begin{equation}
  \hat{\rho}_1 = {\rm Tr}_{\bar{\mathbb{I}}^\prime} [ \vert \tilde{\Psi}_1 \rangle \langle \tilde{\Psi}_1 \vert ].
\end{equation}
$\hat{\mathcal{V}}_2$ is then determined by maximizing $ \langle 0 \vert \hat{\mathcal{V}}^\dag_2 \hat{\rho}_1 \hat{\mathcal{V}}_2 \vert 0 \rangle$, i.e., 
\begin{equation}
\underset{\mathbb{I}^\prime \in \mathbb{C}}{\text{max}} \left[ \underset{\hat{\mathcal{V}}_2}{\text{max}} \langle 0 \vert \hat{\mathcal{V}}^\dag_2 \hat{\rho}_1 \hat{\mathcal{V}}_2 \vert 0 \rangle \right]. 
\end{equation}
This procedure can be continued until the desired number $\delta M$ of unitary operators 
$\hat{\mathcal{V}}_1, \hat{\mathcal{V}}_2, \dots, \hat{\mathcal{V}}_{\delta M}$ are added, i.e., 
$\hat{\mathcal{V}}_{\delta M}^\dag\Compactcdots\hat{\mathcal{V}}_2^\dag\hat{\mathcal{V}}_1^\dag|\Psi\rangle$. 
Note that the reduced density operators $\hat{\rho}_1, \hat{\rho}_2, \dots$ can be evaluated on a quantum computer 
by the quantum state tomography.

\subsection{Automatic quantum circuit encoding algorithm} \label{sec:encode:aqce}

\begin{figure}
  \includegraphics[width=0.95\hsize]{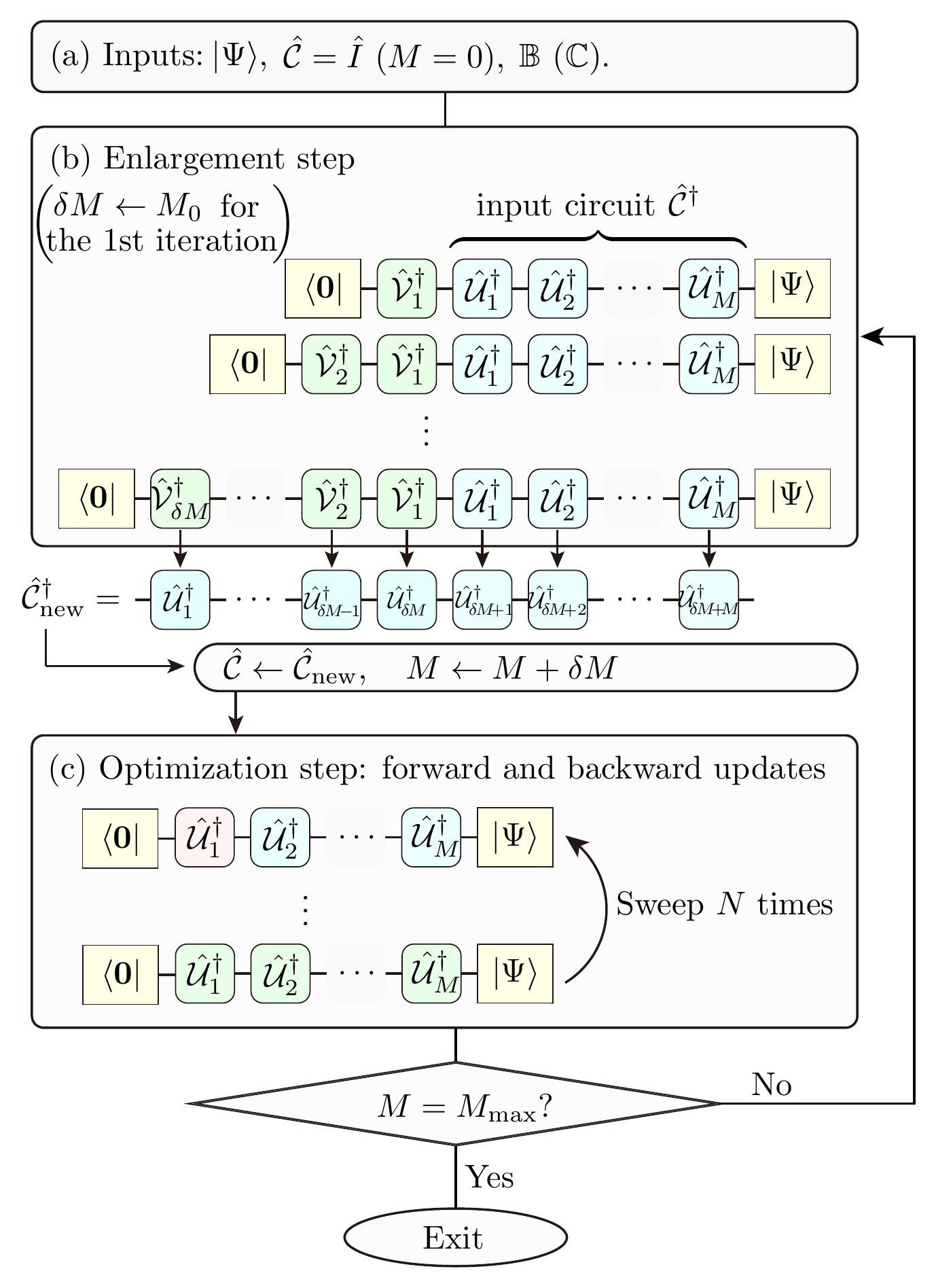}
  \caption{Automatic quantum circuit encoding (AQCE) algorithm.
    (a) Inputs: a quantum states $\vert \Psi \rangle$, a quantum circuit $\hat{\mathcal{C}}$, and a set of bonds $\mathbb{B}$ 
    (or clusters $\mathbb{C}$) of quabits 
    on which unitary operators act.  We set $\hat{\mathcal{C}}=\hat{I}$ and thus the number $M$ of unitary operators in the 
    circuit is zero.
    (b) Enlargement step. $\delta M$ unitary operators acting on two qubits (or $K$ qubits) are newly inserted in the circuit, 
    according to the optimization algorithm (precisely, the backward update) described in Sec.~\ref{sec:encode:algorithm} (also see the text).  
    In the initial step, $\delta M$ is set to $M_0$ and the initialization algorithm in Sec.~\ref{sec:encode:init} is employed to properly 
    construct the number $M_0$ of unitary operators. 
    (c) Optimization step. The circuit consisting of $M$ unitary operators is optimized by the forward and backward updates $N$ times, 
    in which each unitary operator is updated one by one, including the location of qubits that is acted on, 
    according to the quantum circuit encoding algorithm described in Sec.~\ref{sec:encode:algorithm}. 
    The overall iteration is terminated when $M$ reaches to the desired number $M_{\rm max}$ of unitary operators. Otherwise,  
    the algorithm goes back to the enlargement step (b) to additionally insert $\delta M$ new unitary operators. 
    The enlargement step (b) outputs the new quantum circuit $\hat{\mathcal{C}}_{\rm new}$ with $M+\delta M$ unitary operators 
    and this circuit is used as the input of the optimization step (c).  
    Light-green and light-red squares indicate updated unitary operators, while light-blue squares indicate input unitary operators.
    Control parameters in the AQCE algorithm are $M_0$, $M_{\rm max}$, $\delta M$, and $N$.
  }
  \label{fig:aqce}
\end{figure}

Finally, we combine a prototype algorithm of the quantum circuit encoding described in Sec.~\ref{sec:encode:algorithm} 
(also see Fig.~\ref{fig:qce}) and the initialization algorithm explained in Sec.~\ref{sec:encode:init}, 
in order to automatically construct an optimal quantum circuit for encoding a give quantum state.
Figure~\ref{fig:aqce} summarizes the resulting algorithm that is referred to as automatic quantum circuit encoding (AQCE) algorithm. 

The AQCE algorithm is composed of two steps, i.e., the enlargement step in Fig.~\ref{fig:aqce}(b) and 
the optimization step in Fig.~\ref{fig:aqce}(c).
The inputs of the AQCE algorithm are a target quantum state $\vert \Psi \rangle$,
a quantum circuit $\hat{\mathcal{C}}$ set to be the identity operator $\hat{I}$,
and a set of bonds $\mathbb{B}$ of two qubits 
(or a set of clusters $\mathbb{C}$ of $K$ qubits), as shown in Fig.~\ref{fig:aqce}(a). 
In the first enlargement step, we employ the initialization algorithm as in Fig.~\ref{fig:aqce}(b)
to construct a quantum circuit $\hat{\mathcal{C}}$ having $M_0$ number of unitary operators 
and finite overlap between $\hat{\mathcal{C}} \vert 0 \rangle$ and $\vert \Psi \rangle$.
Then, in the following optimization step, we perform the forward and backward updates of the quantum circuit encoding algorithm 
to optimize these unitary operators as in Fig.~\ref{fig:aqce}(c).
The total number of sweeps for the optimization is set to $N$. 
  
Next, we enlarge the quantum circuit by increasing the number $M$ of unitary operators by $\delta M$, i.e., 
$M = M_0+\delta M$, as in Fig.~\ref{fig:aqce}(b). This is done one by one by inserting single identity operator next to $|0\rangle$ 
and perform the backward update of the quantum circuit encoding algorithm to optimize the whole unitary operators. 
After adding $\delta M$ new unitary operators, 
we move to the optimization step and perform the forward and backward updates again to optimize each unitary operator 
as in Fig.~\ref{fig:aqce}(c). 
Notice that expect for the initial enlargement step, 
we perform in the enlargement step the backward update of the quantum circuit encoding algorithm 
described in Sec.~\ref{sec:encode:algorithm}, 
which is more efficient than the initialization algorithm described in Sec.~\ref{sec:encode:init}.
We repeat this whole iteration of the enlargement and optimization steps
until the quantum circuit contains the desired number $M_{\rm max}$ of unitary operators. 
The control parameters in the AQCE algorithm are thus $M_0$, $M_{\rm max}$, $\delta M$, and $N$.

\section{Numerical simulation} \label{sec:benchmark}

In this section, we demonstrate by numerical simulations the AQCE algorithm for quantum many-body states and 
for classical data. In particular, the latter application is potentially useful for quantum machine learning in preparing 
an input quantum state that represents classical data~\cite{Schuld2019}. 
For the purpose of demonstration, we consider the unitary operators $\hat{\mathcal{U}}_m$ and $\hat{\mathcal{V}}_m$ 
acting on two qubits. However, the AQCE algorithm can also be applied to general cases for $K$ qubits with $K>2$.

\subsection{Quantum circuit encoding of quantum many-body states} \label{sec:res:state}

Here, we show the numerical demonstration of the quantum circuit encoding for the ground states of the
one-dimensional $S=1/2$ isotropic antiferromagnetic Heisenberg model and XY model.
The Hamiltonian of these models is given as 
\begin{equation}
  \hat{\mathcal{H}} = \sum_{i=1}^{L} (\hat{X}_i \hat{X}_{i+1} + \hat{Y}_i \hat{Y}_{i+1} + \Delta \hat{Z}_i \hat{Z}_{i+1}), 
\end{equation}
where $\hat{X}_i$, $\hat{Y}_i$, and $\hat{Z}_i$ are the $x$-, $y$-, and $z$-components of Pauli operators, respectively, 
at site $i$ on a one-dimensional chain with $L$ sites under periodic boundary conditions, i.e., 
$\hat{X}_{L+1} = \hat{X}_{1}$, $\hat{Y}_{L+1} = \hat{Y}_1$, 
and $\hat{Z}_{L+1} = \hat{Z}_1$.
The Hamiltonian $\hat{\mathcal{H}}$ with $\Delta =1$ and $0$ corresponds to the isotropic Heisenberg and XY models, respectively, 
and the ground states of these two models are at criticality with algebraically decaying correlation functions.

The ground states $\vert \Psi \rangle$ of these models are calculated numerically 
by the standard Lancozs method within the accuracy of the ground state energy $10^{-12}$.
Although the AQCE algorithm is formulated deterministically, 
it turns out that the resulting structure of the quantum circuit depends on
the numerical tiny error of the quantum state $\vert \Psi \rangle$ obtained by finite precision arithmetic. 
This is simply because even when the fidelity tensor $\hat{\mathcal{F}}_m$ for equivalent pairs of qubits 
is exactly the same theoretically, a particular pair of qubits $\mathbb{I} = \{ i,j \}$ may be selected because of 
the numerical error due to finite precision calculations. 
Therefore, here we perform 100 AQCE calculations for each system size $L$, in which 
the ground state $\vert \Psi \rangle$ is prepared by the Lanczos method with 100 different initial Lanczos vectors, 
thus implying that the ground state $\vert \Psi \rangle$ to be encoded is slightly different among these 
100 different calculations, 
and select the best circuit $\hat{\mathcal{C}}$ in terms of the fidelity $|\langle 0|\hat{\mathcal{C}}^\dag|\Psi\rangle|$. 
In addition, we perform 1000 sweeps to further optimize the unitary operators in the best circuit using the quantum 
circuit encoding algorithm.  
The parameters for the AQCE algorithm are 
$(M_0,N,\delta M, M_{\rm max}) = (L, 20, L/2, L(L-5)/2)$ for the XY model
and $(M_0, N, \delta M, M_{\rm max})=(L, 20, L/2, L^2/2)$ for the isotropic Heisenberg model. 
We set that $\mathbb{B}$ is composed of all pairs of two sites (i.e., qubits) $\{ i, j \}$ with $i, j\in \mathbb{L}$, 
thus including pairs of distant sites.

\begin{figure}
  \includegraphics[width=\hsize]{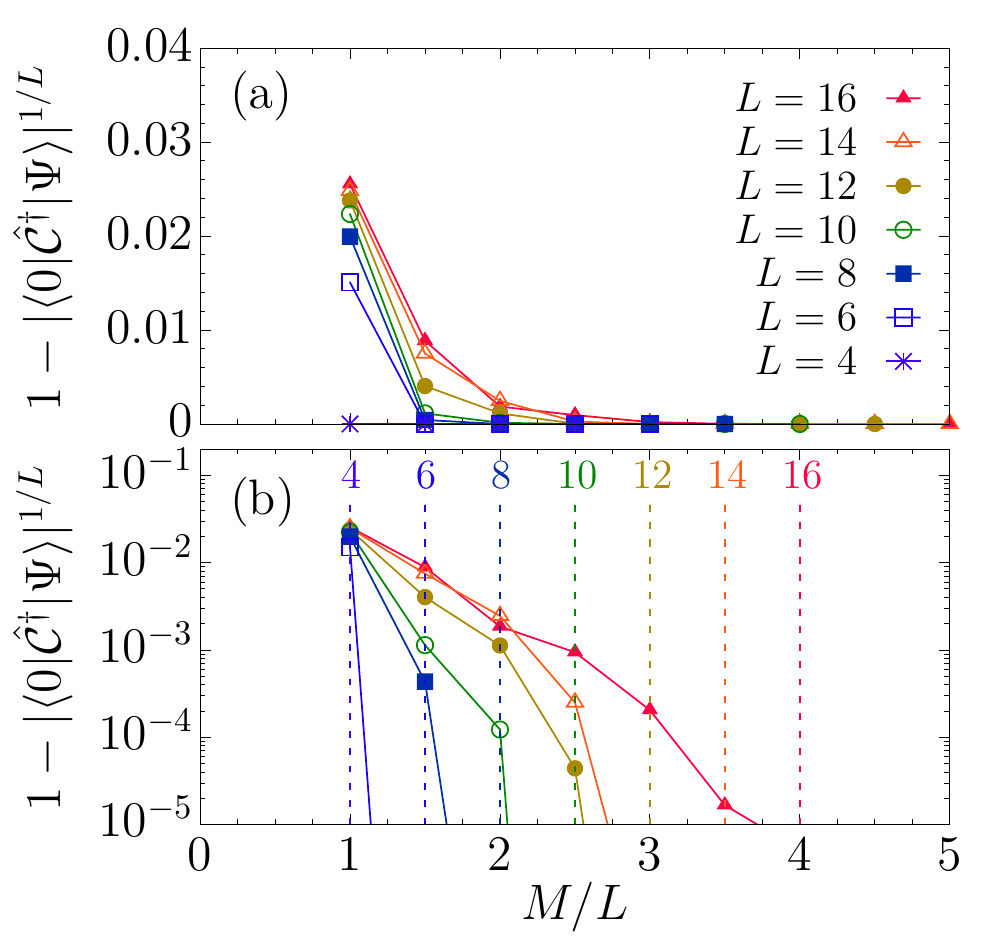}
  \caption{
  (a) Fidelity per site between the ground state $\vert \Psi \rangle$ of the one-dimensional $S=1/2$ XY model and
    the quantum circuit state $\hat{\mathcal{C}} \vert 0 \rangle$ 
    optimized by the AQCE algorithms for different system sizes $L$. (b) Semi-log plot of (a).
    Vertical dotted lines with numbers in (b) indicate the number of local two-qubit unitary operators $M_{\rm e}=L^2/4$
    required to represent the exact ground state by the DQAP ansatz~\cite{Shirakawa2021}.
    }
  \label{fig:olap:xy}
\end{figure}

Figure~\ref{fig:olap:xy} shows the fidelity between the ground state $\vert \Psi \rangle$ of the XY model and
the optimized quantum circuit state $\hat{\mathcal{C}} \vert 0 \rangle$ obtained by the AQCE algorithm. 
We should first recall the previous results by the discretized quantum adiabatic process (DQAP) ansatz~\cite{Shirakawa2021}, 
a similar approach to digitized adiabatic quantum computing reported in Refs.~\onlinecite{Barends2015,Barends2016}, 
where a parametrized quantum circuit is constructed on the basis of digitized quantum adiabatic 
process expressed by a product of local time-evolution unitary operators and the variational parameters 
are optimized so as to minimize the expectation value of energy, as in the 
variational quantum eigensolver~\cite{Peruzzo2014}. 
It is found in Ref.~\onlinecite{Shirakawa2021} that the optimized DQAP ansatz gives 
the exact ground state of the XY model with the minimum number $M_{\rm e} = L^2/4$ 
of local two-qubit unitary operators set by the Lieb-Robinson bound. 
For comparison, this number $M_{\rm e}$ is also indicated for each system size $L$ in Fig.~\ref{fig:olap:xy}(b). 
We find that the AQCE algorithm can generate the quantum circuit state $\hat{\mathcal{C}} \vert 0 \rangle$ 
that represents essentially the exact ground state $\vert \Psi \rangle$ with $M=M_{\rm e}$ for all system sizes studied 
except for $L=16$, for which the convergence of the quantum circuit state $\hat{\mathcal{C}} \vert 0 \rangle$ towards 
the ground state $\vert \Psi \rangle$ appears slower with increasing $M$. 
However, we should note that the quantum circuit state $\hat{\mathcal{C}} \vert 0 \rangle$ with $M < M_{\rm e}$ is better 
in terms of the fidelity than the DQAP ansatz composed of the same number $M$ of local time-evolution
unitary operators even for $L=16$ [see Fig.~\ref{fig:olap:compare} (a)].

\begin{figure}
  \includegraphics[width=\hsize]{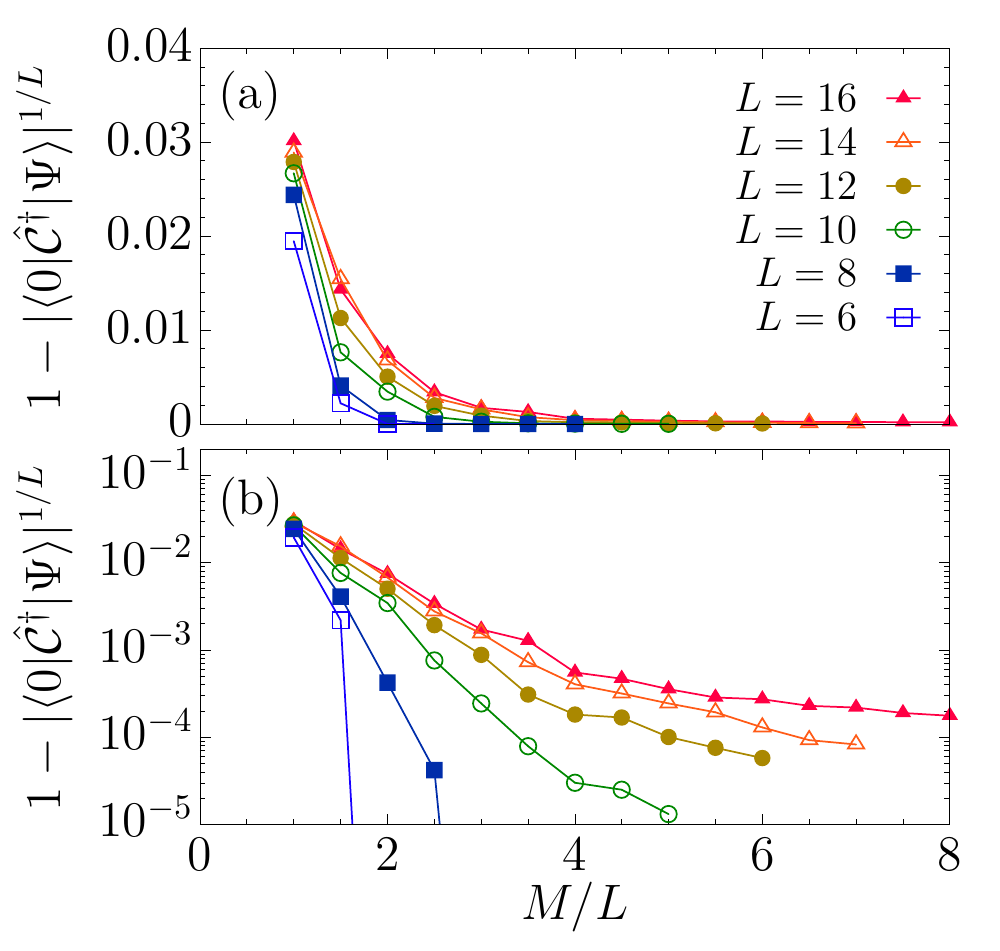}
  \caption{
  Same as Fig.~\ref{fig:olap:xy} but for the ground states $\vert \Psi \rangle$ of the one-dimensional $S=1/2$ 
  isotropic antiferromagnetic Heisenberg model. 
  }
  \label{fig:olap:heisen}
\end{figure}

\begin{figure}
  \includegraphics[width=\hsize]{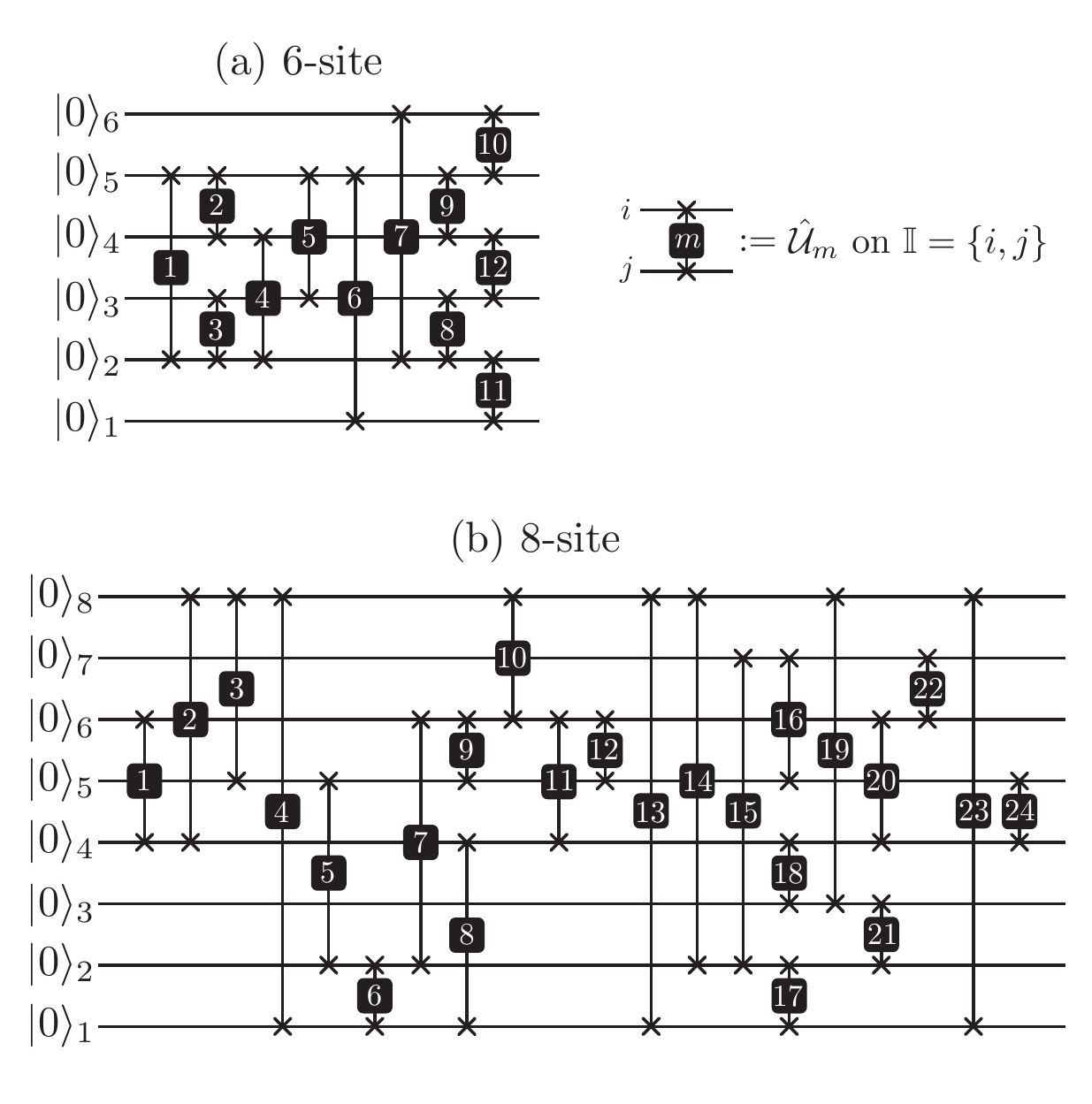}
  \caption{
  Optimized quantum circuit states $\hat{\mathcal{C}} \vert 0 \rangle$ obtained by the AQCE algorithm, 
  which represent essentially exactly the ground states of the one-dimensional $S=1/2$ isotropic antiferromagnetic 
  Heisenberg model for (a) $L=6$ and (b) $L=8$, containing 12 and 24 two-qubit unitary operators $\hat{\mathcal{U}}_m$, respectively, 
  denoted by black squares with number $m$ in them. The location of two qubits 
  on which each unitary operator acts is indicated by crosses. 
  Each two-qubit unitary operator can be decomposed into a standard set of quantum 
  gates (see Fig.~\ref{fig:twogate}) with having 15 independent real parameters (apart from a single global phase). Since adjacent two 
  single-qubit Euler rotations are combined into a single single-qubit Euler rotation, the total number of independent real parameters 
  is 126 for $L=6$ in (a) and 240 for $L=8$ in (b). 
    }
  \label{fig:qce:circuit}
\end{figure}

Figure~\ref{fig:olap:heisen} shows the fidelity between the ground state $\vert \Psi \rangle$ of the isotropic antiferromagnetic 
Heisenberg model and
the optimized quantum circuit state $\hat{\mathcal{C}} \vert 0 \rangle$ obtained by the AQCE algorithm.
For smaller systems with $L \leq 8$, the AQCE algorithm can construct a quantum circuit state $\hat{\mathcal{C}} \vert 0 \rangle$ 
that represents numerically exactly the ground state $\vert \Psi \rangle$ with a less number of $M$. 
For example, one of the resulting quantum circuits describing the ground state for $L=6$ is shown in  Fig.~\ref{fig:qce:circuit}(a). 
The number $M$ of two-qubit unitary operators contained in this particular circuit is $M=12$ 
and the number of independent parameters, once these unitary operators are represented by a standard set of quantum gates 
(see Fig.~\ref{fig:twogate}), is 
$9\times 12 + 6 \times 3 = 126$ if we combine adjacent two single-qubit Euler rotations into a single-qubit Euler rotation.  
On the other hand, the dimension of the Hilbert space for the $L=6$ system is $2^L = 64$, 
suggesting that there are $128-2 = 126$ independent real parameters, where two is subtracted because of the overall phase factor 
and the normalization factor.
It is hence interesting to find that the number of the independent real parameters in this quantum circuit $\hat{\mathcal{C}} $ 
with $M=12$ coincides with that for the Hilbert space on which the quantum state $\vert \Psi \rangle$ is defined. 
However, it is highly nontrivial whether the quantum circuit $\hat{\mathcal{C}}$ composed of the limited number of two-qubit 
unitary operators can always represent any quantum state whenever the number of parameters 
in a quantum circuit is the same as that for the Hilbert space.

We should also note that the two-qubit unitary operators in the optimized quantum circuit $\hat{\mathcal{C}}$, 
representing the ground state $\vert \Psi \rangle$ essentially exactly for $L=6$ and $8$ (see Figure~\ref{fig:qce:circuit}), are not  
uniformly distributed, even though the ground state $\vert \Psi \rangle$ represented by the quantum circuit is translational 
symmetric (apart from the finite precision numerical error). 
Figure~\ref{fig:qce:circuit}(b) shows one of the resulting quantum circuits describing the ground state for $L=8$.
The circuit structure is much more complicated than that for $L=6$ shown in Fig.~\ref{fig:qce:circuit}(a).
Nonetheless, we have confirmed numerically that the resulting quantum circuit states $\hat{\mathcal{C}}|0\rangle$ for $L=6$ 
and $8$ are essentially translational symmetric and also spin rotation symmetric.

In contrast, for the systems with $L > 8$, we find 
that the convergence of the optimized quantum circuit state $\hat{\mathcal{C}}|0\rangle$ 
towards the ground state $\vert \Psi \rangle$ is slower with the number $M$ of unitary operators, 
although the convergence is still approximately exponential, as shown in Fig.~\ref{fig:olap:heisen}(b). 
For example, the error in fidelity of the optimized quantum circuit state $\hat{\mathcal{C}}|0\rangle$ for $L=16$ 
is still relatively large even when $M/L = 8$. 
Moreover, as observed in Fig.~\ref{fig:olap:heisen}(b), the slope of the fidelity in the semi-log plot becomes more flattered 
with increasing the system size $L$. 
We speculate that this is due to a difficulty of sequentially optimizing each unitary operator $\hat{\mathcal{U}}_m$, including 
the location of qubits on which the unitary operator $\hat{\mathcal{U}}_m$ acts, when the system size is large. 
Since much more computational resources are required for further systematic analysis with larger system sizes, 
we leave this issue for a future study.

\subsection{Quantum circuit encoding with fixed Trotter- and MERA-like circuit structures} \label{sec:res:comp}

\begin{figure}
  \includegraphics[width=\hsize]{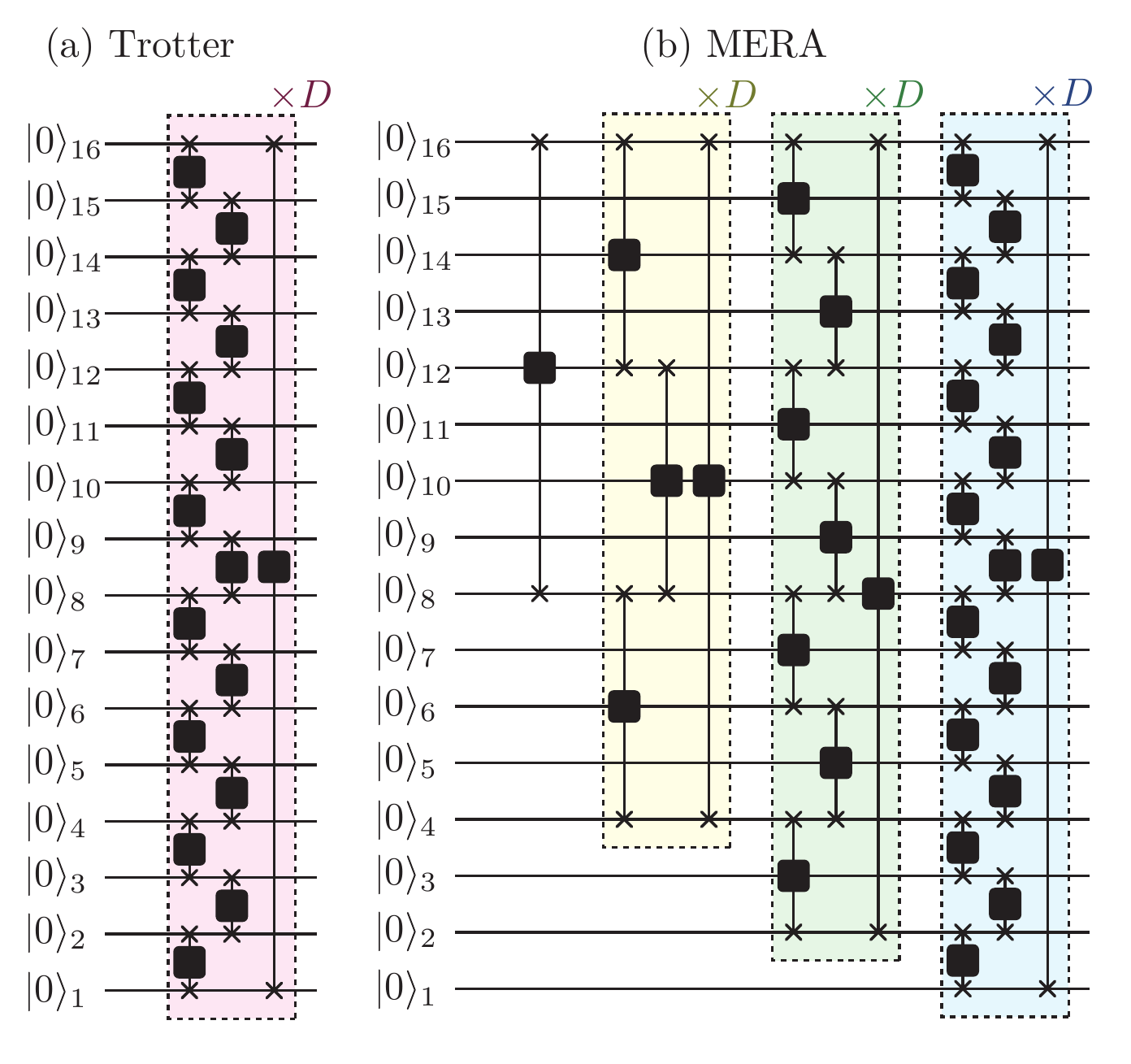}
  \caption{(a) Trotter-like circuit and (b) MERA-like circuit for a 16-qubit system. 
  Two-qubit unitary operators $\hat{\mathcal{U}}_m$ are indicated by back squares and the location of two qubits 
  on which each unitary operator acts is indicated by crosses. 
  Shaded color layers are repeated $D$ times. 
  }
  \label{fig:fix:circuit}
\end{figure}

In this section, using numerical simulations, 
we shall compare the results obtained by the AQCE algorithm, which can automatically construct a quantum circuit with 
a self-assembled optimal structure, and those obtained for a quantum circuit with a fixed circuit structure. 
For this purpose, here we consider two particular fixed circuit structures. 
One is a Trotter-like circuit structure schematically shown in Fig.~\ref{fig:fix:circuit}(a). 
In this Trotter-like circuit, two-qubit unitary operators $\hat{\mathcal{U}}_m$ acting on adjoining qubits are 
distributed as if a time evolution operator of the whole system is Trotter decomposed into two parts in a one dimensionally aligned 
qubit ring. 
The quantum circuit is composed of $D$ layers and each layer corresponds to one Trotter step, containing $L$ two-qubit unitary 
operators $\hat{\mathcal{U}}_m$. Therefore, the total number $M$ of unitary operators $\hat{\mathcal{U}}_m$ 
in the Trotter-like circuit is $D L$.

The other circuit structure considered here is inspired by the MERA
and is shown schematically in Fig.~\ref{fig:fix:circuit}(b). 
In this MERA-like circuit, each basic layer indicated by different shaded color in Fig.~\ref{fig:fix:circuit}(b) represents a different length 
scale and thus two-qubit unitary operators $\hat{\mathcal{U}}_m$ in different basic layers act on two qubits that are 
located in different (adjoining as well as distinct) distances.  
To improve the accuracy, we also increase the number of layers in each basic layer $D$ times [see Fig.~\ref{fig:fix:circuit}(b)], 
and therefore the total number $M$ of unitary operators $\hat{\mathcal{U}}_m$ in the MERA-like circuit is 
$D(L+L/2+L/2^2+L/2^3+\Compactcdots+4)+(2-1)=2D(L-2)+1$, assuming that the system size $L$ is factorial of 2. 
In order to optimize two-qubit unitary operators $\hat{\mathcal{U}}_m$ in the Trotter- and MERA-like circuits 
for encoding a quantum state $\vert \Psi\rangle$, we perform 1000 sweeps of the forward and backward updates using the algorithm 
described in Sec.~\ref{sec:encode:algorithm} (also see Fig.~\ref{fig:qce}), 
i.e., the quantum circuit encoding algorithm, but with the fixed circuit structures.

\begin{figure}
  \includegraphics[width=\hsize]{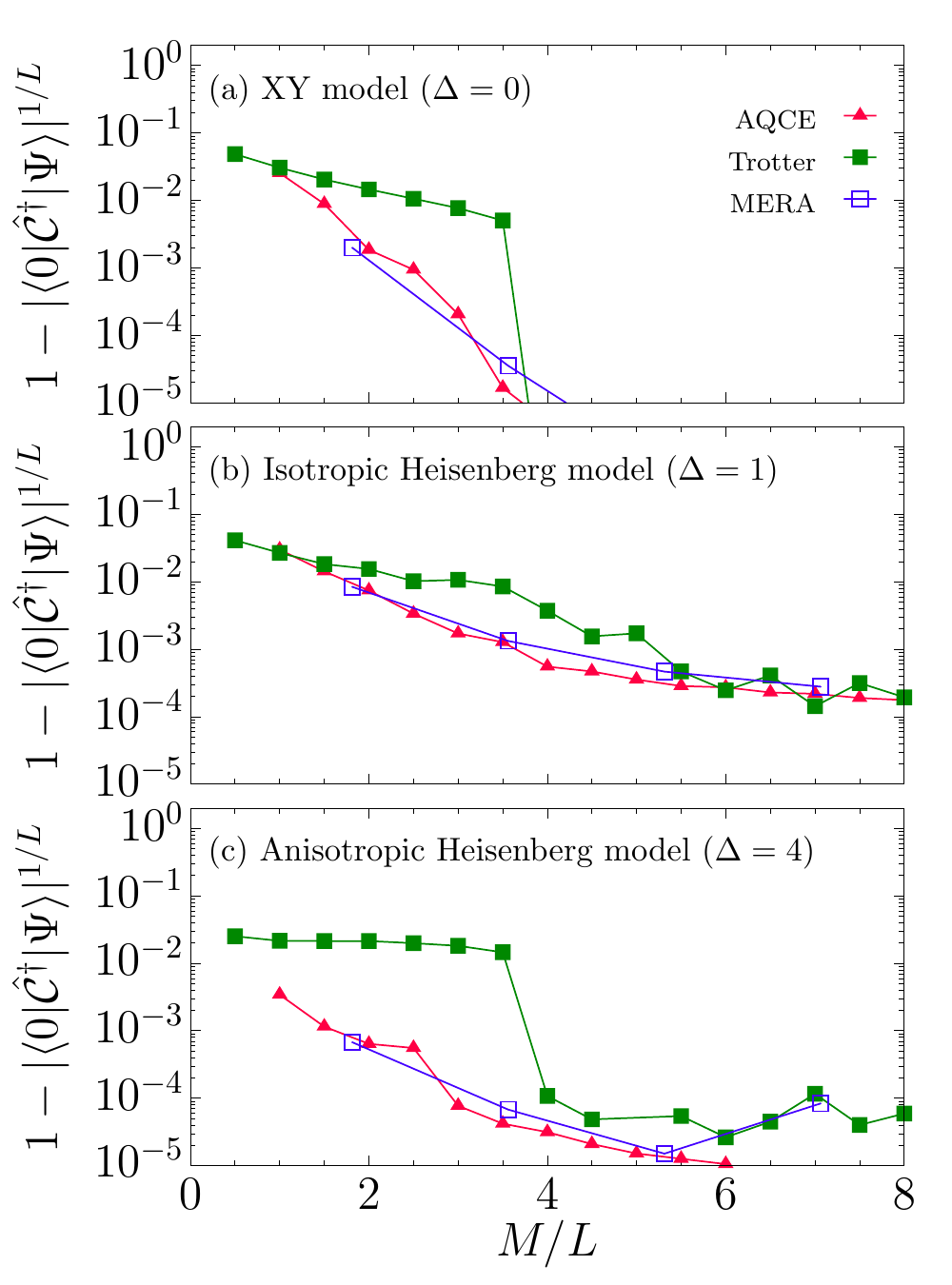}
  \caption{(a) Fidelity per site between the ground state $|\Psi\rangle$ of the one-dimensional $S=1/2$ XY model and 
  the quantum circuit states $\hat{\mathcal{C}} \vert 0 \rangle$ obtained by the AQCE algorithm and with the Trotter- 
  and MERA-like circuit structures for $L=16$. 
  (b) Same as (a) but for the ground state $|\Psi\rangle$ of the one-dimensional $S=1/2$ isotropic 
  antiferromagnetic Heisenberg model. 
  (c) Same as (a) but for the ground state $|\Psi\rangle$ of the one-dimensional $S=1/2$ anisotropic 
  antiferromagnetic Heisenberg model with $\Delta=4$. 
  The results obtained by the AQCE algorithm in (a) and (b) are the same as those shown in Fig.~\ref{fig:olap:xy} and 
  Fig.~\ref{fig:olap:heisen}, respectively. 
  Note that the ground states in (a) and (b) are at criticality, while the ground state in (c) is away from criticality.  
  }
  \label{fig:olap:compare}
\end{figure}

Figure~\ref{fig:olap:compare}(a) shows the fidelity between the ground state $|\Psi\rangle$ 
of the XY model and the optimized quantum circuit states $\hat{\mathcal{C}} \vert 0 \rangle$ obtained by the AQCE algorithm and 
with the Trotter- and MERA-like circuit structures for $L=16$. 
First, we find that the quantum circuit state with the Trotter-like circuit structure can represent numerically exactly the 
ground state $|\Psi\rangle$ at $M=4L$ (corresponding to $L^2/4$ for $L=16)$, 
which is consistent with the previous study using the DQAP ansatz~\cite{Shirakawa2021}. 
This is understood simply because the Trotter-like circuit and the DQAP ansatz have the same circuit structure, although these two 
approaches employ different optimization schemes to determine the optimal two-qubit unitary operators: In the DQAP ansatz, 
each two-qubit unitary operator is parametrized with a single variational parameter [i.e., $\hat{\cal{D}}$ with $\alpha_1=\alpha_2=\alpha_3$ in Eq.~(\ref{eq:exgate})]
and the variational parameters are optimized 
so as to minimize the expectation value of energy, while the optimal unitary operators in the Trotter-like circuit are 
determined essentially deterministically by maximizing the fidelity of the ground state. 
We also find in Fig.~\ref{fig:olap:compare}(a) that the quantum circuit state $\hat{\mathcal{C}}|0\rangle$ obtained by the AQCE algorithm 
is much better than that with the Trotter-like circuit structure when $M < 4 L$ and 
it is competitive in terms of the accuracy 
with that with the MERA-like circuit structure.

Figure~\ref{fig:olap:compare}(b) shows the fidelity between the ground state $|\Psi\rangle$ 
of the isotropic antiferromagnetic Heisenberg model and the optimized quantum circuit states $\hat{\mathcal{C}} \vert 0 \rangle$ 
obtained by the AQCE algorithm and 
with the Trotter- and MERA-like circuit structures for $L=16$.
Similar to the results for the XY model in Fig.~\ref{fig:olap:compare}(a), 
the quantum circuit state $\hat{\mathcal{C}}|0\rangle$ obtained by the AQCE algorithm exhibits the 
better accuracy than that with the Trotter-like circuit structure and it is compatible with 
that with the MERA-like circuit structure when $M \leq 5 L$. 
However, for $M > 5L$, all these three quantum circuit states show the similar accuracy that is  
improved approximately exponentially with increasing $M$.

The ground states in these two cases are both at criticality and the MERA is known to be best suited for describing 
such a quantum state~\cite{Evenbly2011}. Therefore, it is also interesting to study a case for which the ground state is 
away from criticality. 
Figure~\ref{fig:olap:compare}(c) shows the fidelity between the ground state $|\Psi\rangle$ of 
the anisotropic antiferromagnetic Heisenberg model with $\Delta=4$ and the optimized quantum circuit states 
$\hat{\mathcal{C}} \vert 0 \rangle$ obtained by the AQCE algorithm and with the Trotter- and MERA-like circuit structures for $L=16$.
In this case, the ground state is gapped and is less entangled as compared to those of the previous two models with 
$\Delta=0$ and $\Delta=1$.
Therefore, one expects that the number of two-qubit unitary operators required to achieve given accuracy for $\Delta=4$ 
is smaller than that for $\Delta =0$ and $\Delta =1$.
We indeed find in Fig.~\ref{fig:olap:compare}(c) that the fidelity is much closer to 1 when the number of two-qubit unitary operators is small
for the optimized quantum circuit states obtained by the AQCE algorithm and with the MERA-like circuit structure, 
but not for the optimized quantum circuit state with the Trotter-like circuit structure. 
In the case of the Trotter-like circuit, the fidelity first remains almost constant with increasing $M$ until $M=4L$ at which  
the fidelity suddenly jumps to a larger value and then again remains almost constant afterward.

It is interesting to observe in Figs.~\ref{fig:olap:compare}(b) and \ref{fig:olap:compare}(c) 
that the fidelity becomes approximately independent of the quantum circuit structures employed 
when $M/L$ is larger than $4$ or $5$.
A possible reason for this is due to the effect of the barren plateau phenomena.
It is known that the unitary 2-design can be realized in polynomial time for a quantum circuit 
where two-qubit unitary operators are randomly distributed~\cite{Harrow2009}.
As shown in Fig.~\ref{fig:qce:circuit}(b), the distribution of two-qubit unitary operators in the quantum circuit 
obtained by the AQCE algorithm for $L=8$ is quite random. 
Therefore, it is naturally expected that the quantum circuit obtained 
exhibits the unitary 2-design and thus might suffer from the barren plateaus phenomena.   
Since the fidelities for other quantum circuits also exhibit similar values, 
we expect that all of them might suffer from the barren plateau phenomena.
Indeed, we find that in all cases, the change of the quantum circuit during the optimization iteration
is very small when $M$ is large. This implies that a better quantum circuit can be generated more efficiently  
when the number of two-qubit unitary operators is small enough not to exhibit the unitary 2-design. 
However, we should note that the small improvement of fidelity with further increasing $M$ also simply 
implies a trapping of a local minimum of the cost function.

\subsection{Quantum circuit encoding of classical data} \label{sec:res:data}

In this section, we demonstrate that the AQCE algorithm is also useful to construct a optimal 
quantum circuit to represent classical data such as a classical image. 
It is well known that there are several ways to encode classical data to a quantum state 
(for example, see Ref.~\onlinecite{Schuld2019}). However, it is usually not obvious
how to optimally prepare such a quantum state encoding particular classical data
in a quantum circuit with a less number of quantum gates. 
We show that the AQCE algorithm can be a promising approach for this purpose. 

One way to express classical data in a quantum state 
is the amplitude encoding~\cite{Schuld2016}, where
the classical data ${\bm x} = \{ x_0, x_1, \Compactcdots, x_n, \Compactcdots, x_{N-1} \}$
is described by using a quantum state
\begin{equation}
%  \vert {\bm x} \rangle = \sum_{n=0}^{N-1} \bar{x}_n \vert n \rangle,
  \vert \Psi_{\rm c} \rangle = \sum_{n=0}^{N-1} \bar{x}_n \vert n \rangle.
  \label{eq:amp:encode}
\end{equation}
Here, $\vert n \rangle$ is the basis labelled by Eq.~(\ref{eq:computational:label})
with $L \geq \log_2 N$ and 
\begin{equation}
  \bar{x}_n = x_n / \sqrt{V_x}
\end{equation}
with $V_x$ being a volume of ${\bm x}$ given by
\begin{equation}
  V_x = \sum_{n=0}^{N-1} \vert x_n \vert^2. \label{eq:x:volume}
\end{equation}
Each element $x_n$ in the classical data ${\bm x}$ is usually real number, but the amplitude encoding can also be applied to 
the case of complex number. 
There exist several proposals to implement the amplitude encoding~\cite{Grover2000,Shende2006,Plesch2011,Yuval2019}.
However, these are in general not best fit for a near-term application. 
A variational quantum algorithm using a parametrized quantum circuit has also been proposed recently~\cite{Nakaji2021}.

In the previous sections, we have demonstrated the quantum circuit encoding of a quantum state focusing on the ground state 
of a typical many-body Hamiltonian encountered in condensed matter physics and quantum statistical physics, which  
is in some sense simple. 
Instead, a quantum state given in Eq.~(\ref{eq:amp:encode}) representing classical data 
is relatively complicated and moreover there is no prior knowledge of such a quantum state. 
Therefore, the quantum circuit encoding of such a quantum state in Eq.~(\ref{eq:amp:encode}) is generally a difficult task 
in any means. 
Here, we employ the AQCE algorithm to demonstrate the quantum circuit encoding of a quantum state representing a classical image.

\begin{figure*}
  \includegraphics[width=\hsize]{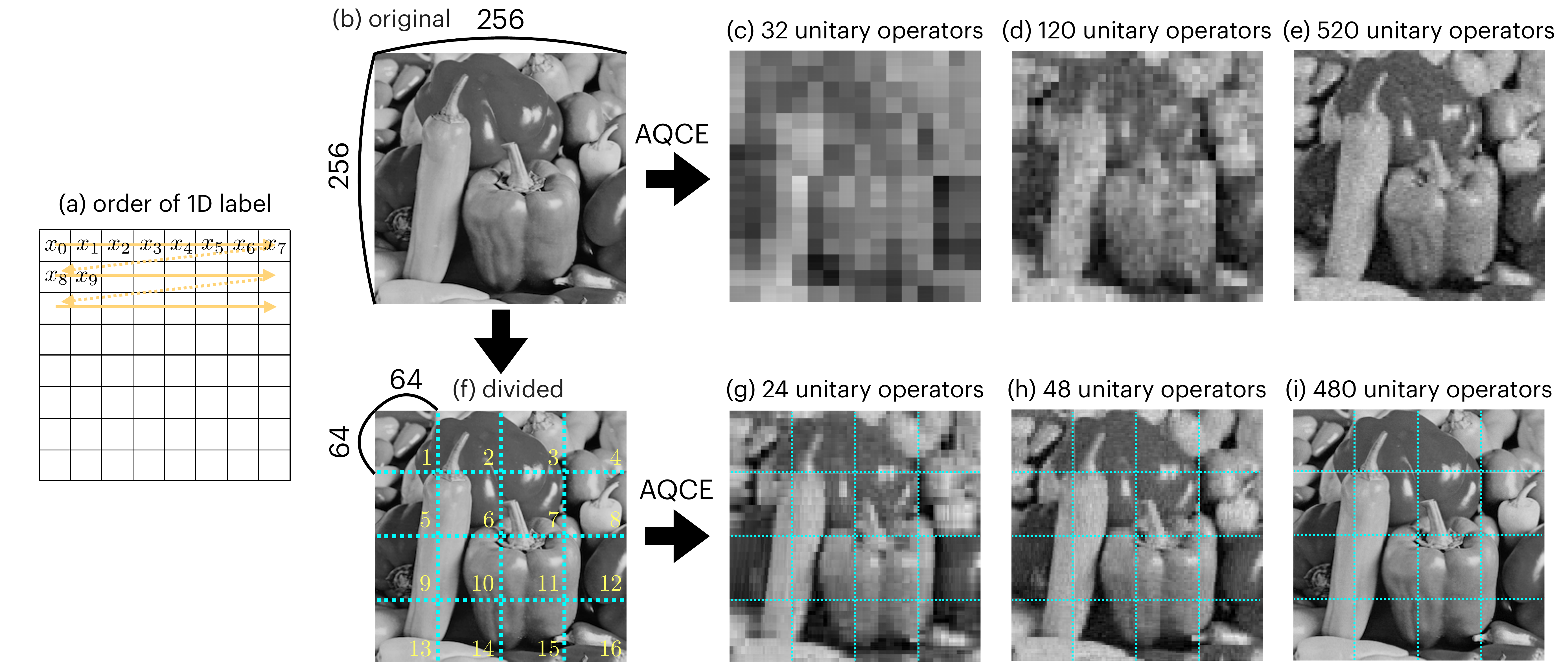}
  \caption{
    Quantum circuit encoding of a gray scale picture known as ``{\it Peppers}" \cite{ImageDatabase}.
    (a) Labeling of two-dimensional classical data (with $8\times 8$ pixels, as an example). 
    (b) Original picture with 256$\times$256 pixels.
    (c)-(e) Pictures reconstructed by decoding the quantum circuit states $\hat{\mathcal{C}}|0\rangle$ on $L=16$ qubits 
    with the different number $M$ of two-qubit unitary operators, $M=32$, 112, and 520.  
    (f) Original picture divided into 16 pieces ($m_s=1,2,\dots,16$ indicated by yellow in the picture) with $64 \times 64$ pixels each.
    (g)-(i) Pictures reconstructed by decoding each quantum circuit state $\hat{\mathcal{C}}^{(m_s)}|0\rangle$ on $L=12$ qubits 
    with the different number $M$ of two-qubit unitary operators, $M=24$, 42, and 450.
    }
  \label{fig:picture}
\end{figure*}

As an example of a classical image, we consider the gray scale picture shown in Fig.~\ref{fig:picture}(b),
which is also known as ``{\it Peppers}'' available in Ref.~\cite{ImageDatabase}.
The data size of this picture is $256 \times 256$ pixels 
and each pixel in the two-dimensional array is assigned to represent each part of the picture 
located at a position labeled  $(i_x,i_y)$ with $i_x = 0,1,2,\Compactcdots,255$ ($=2^8-1$) in the horizontal axis from left to right and 
$i_y = 0,1,2,\Compactcdots,255$ in the vertical axis from top to bottom,
as shown in Fig.~\ref{fig:picture}(a).
Therefore, the picture is fully given by a $2^{16}$ dimensional vector 
${\bm x}_{\rm Pep} = \{ x_0, x_1, x_2, \Compactcdots, x_s, \Compactcdots, x_{65535} \}$ of non-negative real numbers, 
where label $s= i_x + 256 i_y$.
This suggests that the data can be transformed into a quantum state $\vert \Psi_{\rm c} \rangle$ in the form given 
in Eq.~(\ref{eq:amp:encode}) with $L=16$ qubits.
Using numerical simulations, we perform the AQCE algorithm to encode the quantum state $\vert \Psi_{\rm c} \rangle$ 
into an optimal quantum circuit state $\hat{\mathcal{C}} \vert 0 \rangle$. 
For this end, we set the control parameters in the AQCE algorithm as $(M_0,  N, \delta M) =(16, 100, 8)$ with varying 
the total number $M$ of two-qubit unitary operators $\hat{\mathcal{U}}_m$ in the generated quantum circuit (defined as $M_{\rm max}$
in Sec.~\ref{sec:encode:aqce}).

\begin{figure}
  \includegraphics[width=\hsize]{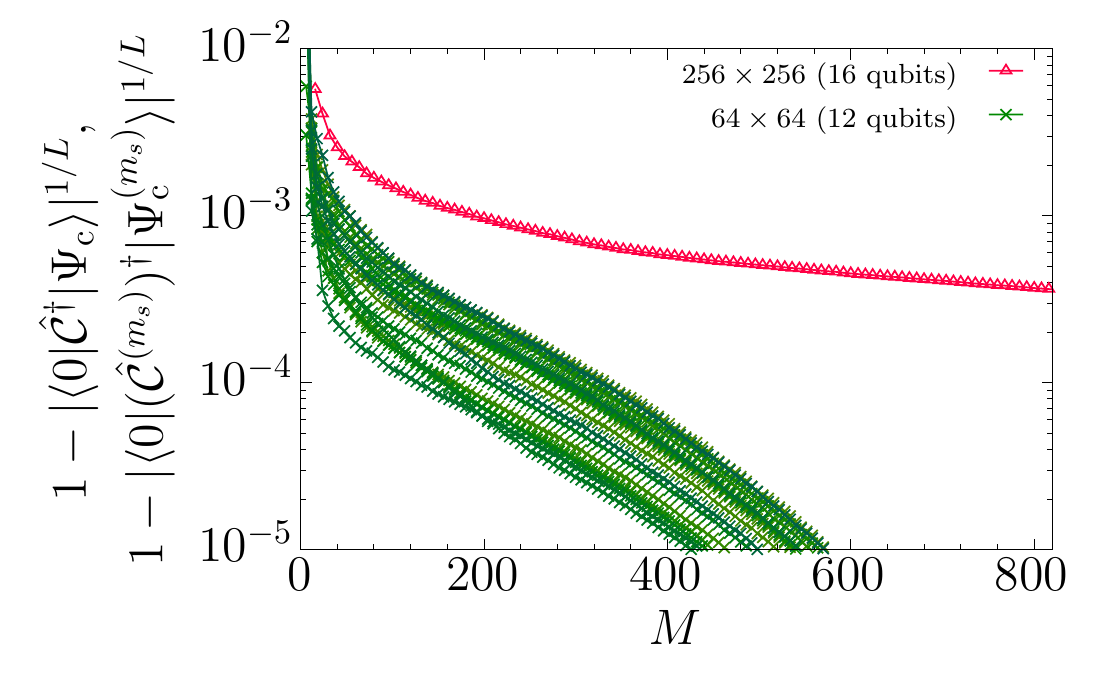}
  \caption{
  Fidelity per site between the quantum state $|\Psi_{\rm c}\rangle$ ($|\Psi_{\rm c}^{(m_s)}\rangle$) on $L=16$ ($L=12$) qubits 
  representing the original picture in Fig.~\ref{fig:picture}(b) 
  [the $m_s$th segment of the original picture in Fig.~\ref{fig:picture}(f)] and the quantum circuit state $\hat{\mathcal{C}}|0\rangle$ 
  ($\hat{\mathcal{C}}^{(m_s)}|0\rangle$) indicated by red triangles (green crosses). 
  Here, $m_s=1,2,\dots,16$ and thus 16 different results are shown for the case where the picture is divided into 16 pieces. 
  }
  \label{fig:olap:picture}
\end{figure}

Figures~\ref{fig:picture}(c)-\ref{fig:picture}(e) show the reconstructed pictures by decoding the quantum circuit states 
$\hat{\mathcal{C}} \vert 0 \rangle$ with the different number $M$ of two-qubit unitary operators. In reconstructing the classical data 
${\bm x}' = \{ x'_0, x'_1, \Compactcdots, x'_n, \Compactcdots, x'_{N-1} \}$ from the amplitude $\bar{x}_n'=\langle n|\hat{\mathcal{C}} \vert 0 \rangle$ 
of the quantum circuit state, 
we have to rescale back the amplitude with the volume $V_x$, i.e., $x'_n=\sqrt{V'_x}\bar{x}'_n$.
It turns out that when the number $M$ of two-qubit unitary operators is extremely small,
the reconstructed picture looks more like a mosaic, as shown in Fig.~\ref{fig:picture}(c) for $M=32$. However, 
as expected, the reconstructed pictures are improved with increasing $M$ [see Figs.~\ref{fig:picture}(d) and \ref{fig:picture}(e)]. 
To be more quantitative, we plot the fidelity between the quantum state $\vert \Psi_{\rm c} \rangle$ representing the original picture 
and the quantum circuit state $\hat{\mathcal{C}} \vert 0 \rangle$ in Fig.~\ref{fig:olap:picture}. 
The fidelity improves rather rapidly with increasing $M$ for $M$ up to 50, but the improvement becomes somewhat slower 
for $M>100$.

For the better performance, next we simply divide the original classical data 
${\bm x}_{\rm Pep} = \{ x_0, x_1, x_2, \Compactcdots, x_{65535} \}$ 
into 16 pieces, each representing a $64 \times 64$ pixels picture, as shown in Fig.~\ref{fig:picture}(f). 
This implies that each segment of the picture is given by a $2^{12}$ dimensional vector, i.e.,  
${\bm x}_{\rm Pep}^{(m_s)} = \{ x_0^{(m_s)}, x_1^{(m_s)}, x_2^{(m_s)}, \Compactcdots, x_{4095}^{(m_s)} \}$ with $m_s=1,2,\dots, 16$.  
Accordingly, a quantum state $\vert \tilde{\Psi}_{\rm c}\rangle$ for the whole picture is given by 
a direct product of quantum states $|\tilde{\Psi}_{\rm c}^{(m_s)}\rangle$ representing different segments of the original picture, i.e.,  
\begin{equation}
\vert \tilde{\Psi}_{\rm c}\rangle = \bigotimes_{m_s=1}^{16} |\Psi_{\rm c}^{(m_s)}\rangle, 
\label{eq:amp:sum}
\end{equation}
where
\begin{equation}
 \vert \Psi_{\rm c}^{(m_s)} \rangle = \sum_{n=0}^{2^{12}-1} \bar{x}_n^{(m_s)} \vert n^{(m_s)} \rangle
  \label{eq:amp:encode2}
\end{equation}
with $\bar{x}_n^{(m_s)} = x_n^{(m_s)} / \sqrt{V_x^{(m_s)}}$ and 
$V_x^{(m_s)} = \sum_{n=0}^{2^{12}-1} \vert x_n^{(m_s)} \vert^2$.  
Note that $\vert n^{(m_s)} \rangle$ in Eq.~(\ref{eq:amp:encode2}) is the basis labelled by Eq.~(\ref{eq:computational:label}) 
within the $m_s$th segment. Therefore, each $\vert \Psi_{\rm c}^{(m_s)} \rangle$ is properly normalized within the segment, 
i.e., $\langle \Psi_{\rm c}^{(m_s)}\vert \Psi_{\rm c}^{(m_s)} \rangle=1$. 
The quantum state $\vert \Psi_{\rm c}^{(m_s)} \rangle $ in Eq.~(\ref{eq:amp:encode2}) is now expressed with a smaller number of 
qubits $L=12$ and each $\vert \Psi_{\rm c}^{(m_s)} \rangle $ is encoded separately into a quantum circuit state
$\hat{\mathcal{C}}^{(m_s)}|0\rangle$ using the AQCE algorithm, which is expected to be easier than the case for $L=16$. 
We should however note that the total Hilbert space defining $\vert \tilde{\Psi}_{\rm c}\rangle$ in Eq.~(\ref{eq:amp:sum}) is now 
increased to $2^{12 \times 16} = 2^{192}$ from $2^{16}$ for $\vert \Psi_{\rm c}\rangle$ to represent the $2^{16}$ dimensional classical data,  
suggesting that the input classical data is mapped into a higher dimensional space via a feature map $\vert \tilde{\Psi}_{\rm c}\rangle$~\cite{Stoudenmire2016,Mitarai2018}. 
Although we do not perform any further explicit demonstration, this might find an interesting application of quantum machine learning 
based on a kernel method~\cite{Stoudenmire2016,Mitarai2018}.

We employ the AQCE algorithm to encode separately the quantum state $\vert \Psi_{\rm c}^{(m_s)} \rangle $ representing 
the $m_s$th segment of the picture
with the control parameters $(M_0,N,\delta M) = (12,100,6)$ and varying the total number $M$ of two-qubit unitary operators 
in the quantum circuit $\hat{\mathcal{C}}^{(m_s)}$. 
Figures~\ref{fig:picture}(g)-\ref{fig:picture}(i) show the reconstructed pictures by decoding the quantum circuit states 
$\hat{\mathcal{C}}^{(m_s)}|0\rangle$ with properly rescaling back the amplitude 
$\bar{x}_n^{(m_s)\prime}=\langle n^{(m_s)}|\hat{\mathcal{C}}^{(m_s)}|0\rangle$ by the $m_s$ dependent volume $\sqrt{V_x^{(m_s)}}$. 
We find that the original picture is reconstructed very efficiently with a much less number $M$ of two-qubit unitary operators, as 
compared with the results of encoding the whole picture without the segmentation shown 
in Figs.~\ref{fig:picture}(c)-\ref{fig:picture}(e). 
The fidelity between the quantum state $\vert \Psi_{\rm c}^{(m_s)} \rangle$ representing the $m_s$th segment of the picture 
and the quantum circuit state $\hat{\mathcal{C}}^{(m_s)} \vert 0 \rangle$ is also shown in Fig.~\ref{fig:olap:picture}.
We observe that the fidelity can be improved much more efficiently with increasing $M$ when the whole picture is divided into many 
pieces so as to decrease the dimension of the classical data that is to be encoded into a quantum circuit with a less number $L$ of 
qubits.

Recall now that a $64 \times 64$ pixels picture is given by a $2^{12} = 4096$ dimensional vector. 
While an SU(4) operator (i.e, a two-qubit unitary operator with ignoring a global phase) is parametrized 
by 15 independent real parameters, 
two consecutive single-qubit Euler rotations are redundant (see Fig.~\ref{fig:twogate}). 
Removing these redundancies, the number of the independent real parameters
for a quantum circuit with $M$ number of SU(4) operators is $9M + L \times 3$.  
Therefore, the number of the independent real parameters for the quantum circuit with $M = 480$ on $L=12$ qubits is 
almost equal to the dimension of the segmented picture. 
As shown in Fig.~\ref{fig:picture}(i), we indeed find that the reconstructed picture reproduces the original picture with 
a reasonable accuracy.

How is this quantum circuit encoding of classical data potentially useful in the context of quantum machine learning? 
In order to make good use of quantum computer for machine learning, classical data has to be implemented into a 
quantum device in the first place. As explained above, a quantum state representing classical data via, e.g.,
the amplitude encoding is generally too complicated to be prepared in a quantum device naively. 
The quantum circuit encoding proposed here can be employed for this purpose to approximately construct a quantum circuit 
representing a quantum state of classical data with controlled accuracy. 
This can be done on a classical computer and 
the obtained quantum circuit is implemented in a quantum device for further processing of machine learning. 
In the next section, we shall demonstrate experimentally some of this procedure.

\section{\label{sec:exp}Experimental demonstration using a quantum device}

Although the quantum-classical hybrid computation of the AQCE algorithm is in principle possible, 
we find that the implementation using currently available quantum devices is practically difficult.  
Therefore, here we instead experimentally demonstrate that
the AQCE algorithm indeed generates a quantum circuit that can be implemented on a real quantum device to produce 
a desired quantum state with reasonable accuracy. 
For this demonstration, we use a quantum device (ibmq\_lima) provided by IBM Quantum~\cite{IBM} and all experimental data 
were collected on 15 Octorber 2021.  

\subsection{Quantum states in the two-qubit space}\label{sec:twoqubits}

\begin{figure*}
  \includegraphics[width=0.9\hsize]{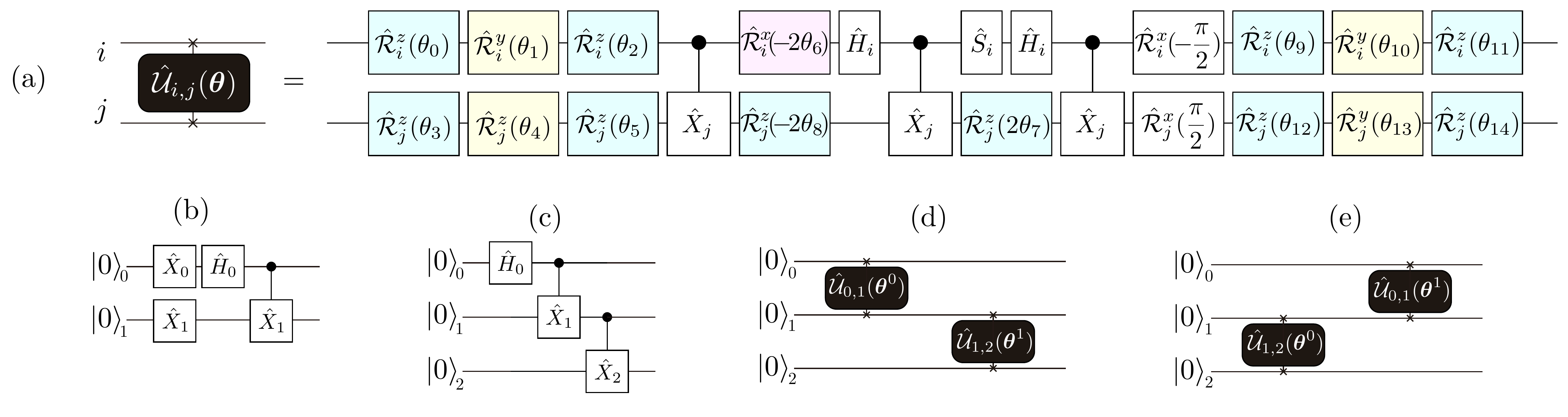}
  \caption{
  (a) Two-qubit unitary operator $\hat{\mathcal{U}}_{i,j}(\mbox{\boldmath{$\theta$}})$ acting on qubit $i$ and $j$ is 
  implemented on a quantum device by a standard set of quantum gates with 15 parameters 
  $\mbox{\boldmath{$\theta$}} = \{ \theta_0, \theta_1, \Compactcdots, \theta_{14} \}$ 
  for the rotation angles of single-qubit gates. 
  (b) Quantum circuit $\hat{\mathcal{U}}_{\text{2QS}}$ that generates the singlet state $\vert \Psi_{\text{2QS}} \rangle $ 
  in the two-qubit space.
  (c) Quantum circuit $\hat{\mathcal{U}}_{\rm GHZ}$ that generates the GHZ state $\vert \Psi_{\rm GHZ} \rangle$ in three-qubit space.
  (d), (e) Quantum circuit structures obtained by the AQCE algorithm for a quantum state in the three-qubit space, 
  containing two two-qubit unitary operators acting on qubits that are physically connected in the quantum device employed here. 
  }
  \label{fig:device:circuit}
\end{figure*}

We first consider one of the simplest quantum states, i.e., the singlet state in the two-qubit space (one of the Bell states) 
given by
\begin{equation}
  \vert \Psi_{\text{2QS}} \rangle = \frac{1}{\sqrt{2}} ( \vert 0 1 \rangle - \vert 1 0 \rangle ),
  \label{eq:ex:2qs}
\end{equation}
where $\vert 0 1 \rangle = |0\rangle_0\otimes|1\rangle_1$ and $\vert 1 0 \rangle = |1\rangle_0\otimes|0\rangle_1$, 
following the notation introduced at the beginning of Sec.~\ref{sec:encode:obj}. 
We apply the AQCE algorithm on a classical computer to encode the quantum state $\vert \Psi_{\text{2QS}}\rangle$ and 
obtain within the machine precision that  
\begin{equation}
  \vert \Psi_{\text{2QS}} \rangle = 
  \hat{\mathcal{U}}_{0,1}(\mbox{\boldmath{$\theta$}}) \vert 0 \rangle,  
\end{equation}
where the quantum circuit $\hat{\mathcal{U}}_{i,j}(\mbox{\boldmath{$\theta$}})$ operating on qubits $i$ and $j$ 
is given as 
\begin{equation}
  \begin{split}
  \hat{\mathcal{U}}_{i,j}(\mbox{\boldmath{$\theta$}}) & = 
  \hat{\mathcal{R}}^z_j (\theta_{14})
  \hat{\mathcal{R}}^z_j (\theta_{13})
  \hat{\mathcal{R}}^z_j (\theta_{12})
  \hat{\mathcal{R}}^z_i (\theta_{11})
  \hat{\mathcal{R}}^z_i (\theta_{10})
  \hat{\mathcal{R}}^z_i (\theta_9)
  \\ & \times 
  \hat{\mathcal{R}}^x_j (\pi/2)
  \hat{\mathcal{R}}^x_i (-\pi/2)
  \hat{C}_i(\hat{X}_j) 
  \hat{H}_i \hat{S}_i
  \\ & \times 
  \hat{\mathcal{R}}^z_j (2\theta_7)
  \hat{C}_i(\hat{X}_j)
  \hat{\mathcal{R}}^z_j (-2\theta_8)
  \hat{H}_i 
  \hat{\mathcal{R}}^x_i (-2\theta_6)
  \hat{C}_i(\hat{X}_j)
  \\
  &
  \times
  \hat{\mathcal{R}}^z_j (\theta_5) 
  \hat{\mathcal{R}}^y_j (\theta_4) 
  \hat{\mathcal{R}}^z_j (\theta_3) 
  \hat{\mathcal{R}}^z_i (\theta_2) 
  \hat{\mathcal{R}}^y_i (\theta_1) 
  \hat{\mathcal{R}}^z_i (\theta_0)
  \end{split}
  \label{eq:ex:c:qce}
\end{equation}
and the resulting set of parameters $\mbox{\boldmath{$\theta$}} = \{ \theta_0, \theta_1, \Compactcdots, \theta_{14} \}$ is displayed 
in Table.~\ref{tab:2q:params}.
The explicit form of the quantum circuit $\hat{\mathcal{U}}_{i,j}(\mbox{\boldmath{$\theta$}})$ and the associated quantum gates 
are shown in Fig.~\ref{fig:device:circuit}(a).
Note that the singlet state in Eq.~(\ref{eq:ex:2qs})
can also be prepared simply by
\begin{equation}
  \vert \Psi_{\text{2QS}} \rangle = \hat{\mathcal{U}}_{\text{2QS}} \vert 0 \rangle 
\end{equation}
with the quantum circuit
\begin{equation}
  \hat{\mathcal{U}}_{\text{2QS}} = \hat{C}_0(\hat{X}_1) \hat{H}_0 \hat{X}_1 \hat{X}_0,
  \label{eq:ex:c:sing:man}
\end{equation}
as shown in Fig.~\ref{fig:device:circuit}(b).

\begin{table}
  \caption{
  Sets of parameters $\mbox{\boldmath{$\theta$}} = \{ \theta_0, \theta_1, \Compactcdots, \theta_{14} \}$
  for the quantum circuits $\hat{\mathcal{U}}_{0,1}(\mbox{\boldmath{$\theta$}})$ in Eq.~(\ref{eq:ex:c:qce}) 
  [also see Fig.~\ref{fig:device:circuit}(a)] generated by the AQCE algorithm, 
  encoding the singlet state $\vert \Psi_{\rm 2QS} \rangle$ 
  and the random state $\vert \Psi_{\rm 2QR} \rangle$ in the two-qubit space.}
  \label{tab:2q:params}
  \begin{tabular}{lrr}
    \hline
    \hline
    {} & singlet state & random state \\
    \hline
    $\theta_0$ & 1.6823068 & 2.0216448 \\
    $\theta_1$ & 3.1415927 & 1.3683389 \\
    $\theta_2$ & 0 & $-$2.2863607 \\
    $\theta_3$ & $-$0.9758576 & $-$2.8429004 \\
    $\theta_4$ & 0 & 1.9027058 \\
    $\theta_5$ & $-$1.6678105 & $-$1.8420845 \\
    $\theta_6$ & 0.3926991 & 0.7086172 \\
    $\theta_7$ & 3.5342917 & 1.1534484 \\
    $\theta_8$ & 3.1355175 & 1.6383263 \\
    $\theta_9$ & $-$2.6094912 & $-$2.6132016 \\
    $\theta_{10}$ & $-$3.1415927 & $-$2.0676228 \\
    $\theta_{11}$ & 3.1204519 & 2.1424122 \\
    $\theta_{12}$ & $-$1.6869951 & $-$1.2293439 \\
    $\theta_{13}$ & $-$3.1415926 & $-$1.8418481 \\
    $\theta_{14}$ & 2.4721516 & $-$2.6729236 \\
    \hline
  \end{tabular}
\end{table}

\begin{figure}
  \begin{center}
    \includegraphics[width=\hsize]{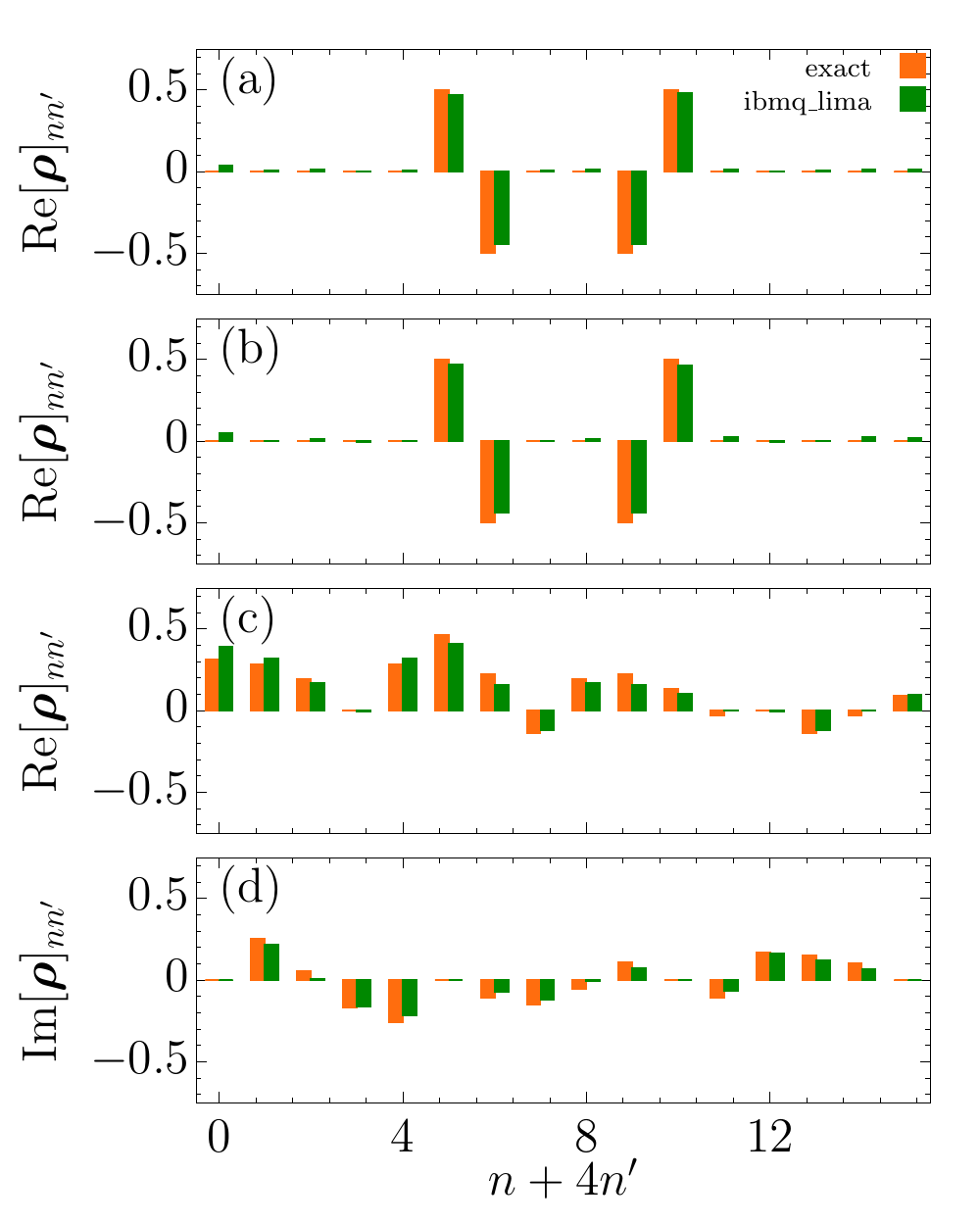}
    \caption{(a) 
    Density matrix $[\bm{\rho}]_{nn'}=\langle n|\hat{\rho}|n'\rangle$
    evaluated on the quantum device (ibmq\_lima) with the 
    quantum circuit $\hat{\mathcal{U}}_{\rm 2QS}$ describing the singlet state $\vert \Psi_{\text{2QS}}\rangle$ 
    in Eq.~(\ref{eq:ex:2qs}). 
    (b) Same as (a) but with the quantum circuit $\hat{\mathcal{U}}_{0,1}(\mbox{\boldmath{$\theta$}}) $ in Eq.~(\ref{eq:ex:c:qce}) 
    generated by the AQCE algorithm encoding the singlet state $\vert \Psi_{\text{2QS}}\rangle$. 
    The parameter set $\mbox{\boldmath{$\theta$}}$ is provided in the column ``singlet state" of Table~\ref{tab:2q:params}. 
    Note that only the real part of density matrix is shown in (a) and (b) because the imaginary part is zero 
    for the singlet state $\vert \Psi_{\text{2QS}}\rangle$. 
    (c) Real and (d) imaginary parts of density matrix $[\bm{\rho}]_{nn'}=\langle n|\hat{\rho}|n'\rangle$ evaluated on the quantum 
    device (ibmq\_lima) with the quantum circuit $\hat{\mathcal{U}}_{0,1}(\mbox{\boldmath{$\theta$}}) $ in Eq.~(\ref{eq:ex:c:qce}) 
    generated by the AQCE algorithm encoding the random state $\vert \Psi_{\text{2QR}}\rangle$ in Eq.~(\ref{eq:ex:2rnd}). 
    The parameter set $\mbox{\boldmath{$\theta$}}$ is provided in the column ``random state" of Table~\ref{tab:2q:params}.
    For comparison, the exact results are also shown by orange bars.
    The density matrix is evaluated on the quantum device by the quantum state tomography, measuring 16 different sets of Pauli 
    strings with length two, and the results shown here are obtained from the averaged values over 4096 measurements 
    of each Pauli string. 
    $|n\rangle$ and $|n'\rangle$ with $n,n'=0,1,2,3$ are the basis states of $L=2$ qubits labeled as in Eq.~(\ref{eq:computational:label}).
    }
    \label{fig:ex:2q}
  \end{center}
\end{figure}

By using the quantum device,
we evaluate in Figs.~\ref{fig:ex:2q}(a) and \ref{fig:ex:2q}(b) the density matrix, 
$[\bm{\rho}]_{nn'}=\langle n|\hat{\rho}|n'\rangle$, 
of the singlet state generated by the quantum 
circuits $\hat{\mathcal{U}}_{\text{2QS}}$  in Eq.~(\ref{eq:ex:c:sing:man}) 
and $\hat{\mathcal{U}}_{0,1}(\mbox{\boldmath{$\theta$}})$ in Eq.~(\ref{eq:ex:c:qce}), respectively. 
Here, $|n\rangle$ and $|n'\rangle$ with $n,n'=0,1,2,3$ are the basis states of $L=2$ qubits labeled as in Eq.~(\ref{eq:computational:label}).  
To evaluate the density matrix,
we perform the quantum state tomography by measuring 16 different sets of Pauli strings (including the identity operator) 
with length two [see Eqs.~(\ref{eq:dopt}) and (\ref{eq:dopt2})]. Each Pauli string is measured on the quantum device 
4096 times and the density matrix $[\bm{\rho}]_{nn'}$ shown in Figs.~\ref{fig:ex:2q}(a) and \ref{fig:ex:2q}(b) 
is evaluated from the averaged values over these measurements. 
These results are also compared with the exact values. 
We find that the density matrices evaluated on the quantum device with the two different quantum circuits, one  
obtained by the AQCE algorithm, are rather similar and can both reproduce the exact result with reasonable accuracy. 

For qualitative comparison, we now introduce the following quantity: 
\begin{equation}
  Q = ({\rm Tr}[ \hat{\rho}_{A} \hat{\rho}_{B}])^{1/2}, \label{eq:qfid}
\end{equation}
where $\hat{\rho}_A$ and $\hat{\rho}_B$ are the density operators of two quantum states $A$ and $B$. 
Here we employ this quantity as fidelity of two quantum states because some of the eigenvalues of the 
density matrix evaluated from our experimental measurements are negative.   
This can be justified when two quantum states $A$ and $B$ are pure states, i.e., 
$\hat{\rho}_A = \vert \Psi_A \rangle \langle \Psi_A \vert$ and $\vert \Psi_B \rangle \langle \Psi_B \vert$, 
because in this case $Q = \vert \langle \Psi_A \vert \Psi_B \rangle \vert$.   
Using the density matrix evaluated experimentally in Fig.~\ref{fig:ex:2q}(b), 
we find that the fidelity $Q$ for the exact singlet state $\vert \Psi_{\text{2QS}} \rangle$ and the singlet state generated
by the quantum circuit $\hat{\mathcal{U}}_{0,1}(\mbox{\boldmath{$\theta$}})$ is as large as 0.9512.
This is indeed comparable to the fidelity $Q=0.9607$ for the exact singlet state $\vert \Psi_{\text{2QS}} \rangle$ 
and the singlet state generated by the quantum circuit $\hat{\mathcal{U}}_{\rm 2QS}$.

Next, we consider the case where a quantum state is more complex in the sense that the associated density matrix has 
many nonzero elements. 
To this end, we examine a random state described by the following state in the two-qubit space:
\begin{equation}
  \begin{split}
  \vert \Psi_{\text{2QR}} \rangle = & ( 0.36179353 + {\rm i} 0.42519915 ) \vert 0 0 \rangle \\
  & + ( 0.14876111 + {\rm i} 0.33156910 ) \vert 1 0 \rangle \\
  & + (-0.02356009 + {\rm i} 0.68066637 ) \vert 0 1 \rangle \\
  & + ( 0.23101109 - {\rm i} 0.19752287 ) \vert 1 1 \rangle,
  \end{split}
  \label{eq:ex:2rnd}
\end{equation} 
where $\vert 0 0 \rangle = |0\rangle_0\otimes|0\rangle_1$, etc. 
The coefficients are randomly determined as follows: We first use a random generator in the c++ standard library
for the normal distribution with the mean 0 and the standard deviation 1 to determine the real and imaginary parts of each 
coefficient and then normalize the resulting state~\cite{Mezzadri2006}.
We apply the AQCE algorithm on a classical computer to encode the quantum state $\vert \Psi_{\text{2QR}} \rangle$ 
and obtain the quantum circuit
$\hat{\mathcal{U}}_{0,1}(\mbox{\boldmath{$\theta$}})$ with the 
parameter set $\mbox{\boldmath{$\theta$}} = \{ \theta_1, \theta_2, \Compactcdots, \theta_{14} \}$
given in Table~\ref{tab:2q:params}, which can represent $\vert \Psi_{\text{2QR}} \rangle$ exactly within the machine precision.  
The density matrix $[\bm{\rho}]_{nn'}$ 
of the random state generated by the quantum 
circuits $\hat{\mathcal{U}}_{0,1}(\mbox{\boldmath{$\theta$}})$ is evaluated on the quantum device in 
Figs.~\ref{fig:ex:2q}(c) and \ref{fig:ex:2q}(d) by using the quantum state tomography described above.  
Similar to the case of the singlet state, we find that the results evaluated on the quantum device are rather well compared 
with the exact values. 
Indeed, we find that the fidelity $Q$ for the exact random state $\vert \Psi_{\text{2QR}} \rangle$ and 
the random state generated by the quantum circuit $\hat{\mathcal{U}}_{0,1}(\mbox{\boldmath{$\theta$}})$ is as large as 0.9592.

\subsection{Quantum states in the three-qubit space}

The demonstrations shown above are focused on quantum states in the two-qubit space. 
It is also highly interesting to continue a similar demonstration for a quantum state in a larger Hilbert space. 
Let us now consider the GHZ state in the three-qubit space given by
\begin{equation}
  \vert \Psi_{\rm GHZ} \rangle = \frac{1}{\sqrt{2}} ( \vert 0 0 0 \rangle + \vert 1 1 1 \rangle ), 
  \label{eq:ghz}
\end{equation}
where $\vert 0 0 0 \rangle = |0\rangle_0\otimes|0\rangle_1\otimes|0\rangle_2$ 
and $\vert 1 1 1 \rangle = |1\rangle_0\otimes|1\rangle_1\otimes|1\rangle_2$, 
following the notation introduced at the beginning of Sec.~\ref{sec:encode:obj}. 
It is known that the GHZ state can be prepared simply by 
\begin{equation}
  \vert \Psi_{\rm GHZ} \rangle = \hat{\mathcal{U}}_{\rm GHZ} \vert 0 \rangle 
\end{equation}
with the quantum circuit 
\begin{equation}
  \hat{\mathcal{U}}_{\rm GHZ} =
  \hat{C}_1(\hat{X}_2)
  \hat{C}_0(\hat{X}_1)
  \hat{H}_0
  \label{eq:ex:c:ghzs:man}
\end{equation}
acting on qubits 0, 1, and 2, as shown in Fig.~\ref{fig:device:circuit}(c). 

We also perform the AQCE algorithm on a classical computer to encode the GHZ state into a quantum circuit. 
Considering a set of bonds $\mathbb{B}$ in the AQCE algorithm, it is wise to include only pairs of qubits that are physically 
connected in the quantum device so as to decrease the number of extra quantum gates. 
In the quantum device employed in this demonstration, there are only two pairs of qubits: 
$\mathbb{B} = \{ \{ 0,1 \}, \{ 1,2 \} \}$.
However, it is not obvious in advance how many two-qubit unitary operators $\hat{\mathcal{U}}_{i,j}(\mbox{\boldmath{$\theta$}})$
are necessary to encode the GHZ state. 
By performing the AQCE algorithm on a classical computer, we find within the machine precision that
\begin{equation}
\vert \Psi_{\text{GHZ}} \rangle = \hat{\mathcal{U}}_{0,1,2}(\mbox{\boldmath{$\theta$}})|0\rangle
  \label{eq:ex:ghz}
  \end{equation}
with the quantum circuit $\hat{\mathcal{U}}_{0,1,2}(\mbox{\boldmath{$\theta$}})$ being composed of two two-qubit unitary operators, 
\begin{equation}
  \hat{\mathcal{U}}_{0,1,2}(\mbox{\boldmath{$\theta$}}) =
  \hat{\mathcal{U}}_{1,2}(\mbox{\boldmath{$\theta$}}^1) 
  \hat{\mathcal{U}}_{0,1}(\mbox{\boldmath{$\theta$}}^0),
  \label{eq:ex:c:qce:3}
\end{equation}
where $\mbox{\boldmath{$\theta$}} = \{ \mbox{\boldmath{$\theta$}}^{0}, \mbox{\boldmath{$\theta$}}^{1} \}$ and 
the resulting sets of parameters 
$\mbox{\boldmath{$\theta$}}^m = \{ \theta_0^m, \theta_1^m, \Compactcdots, \theta_{14}^m \}$ ($m=0$ and $1$) 
for $\hat{\mathcal{U}}_{0,1}(\mbox{\boldmath{$\theta$}}^0)$ and 
$\hat{\mathcal{U}}_{1,2}(\mbox{\boldmath{$\theta$}}^1)$ are given in Table~\ref{tab:ghz:params}. 
The schematic structure of the quantum circuit is shown in Fig.~\ref{fig:device:circuit}(d). 
Notice in Table~\ref{tab:ghz:params} that $\theta_0^1=\theta_1^1=\theta^1_2=0$ because these parameters correspond to the 
first Euler rotation (acting on qubit 1) of the second two-qubit unitary 
operator $ \hat{\mathcal{U}}_{1,2}(\mbox{\boldmath{$\theta$}}^1)$, 
which can be absorbed into the last Euler rotation of the first two-qubit unitary operator
$ \hat{\mathcal{U}}_{0,1}(\mbox{\boldmath{$\theta$}}^1) $. 
We should also note that since the GHZ state is translational symmetric, the quantum circuit with the structure shown 
in Fig.~\ref{fig:device:circuit}(e) is topologically equivalent. 
The AQCE algorithm select one of them and, in this particular demonstration, 
the quantum circuit with the structure shown in Fig.~\ref{fig:device:circuit}(d) is selected.

\begin{table}
  \caption{Sets of parameters $\mbox{\boldmath{$\theta$}}^m = \{ \theta_0^m, \theta_1^m, \Compactcdots, \theta_{14}^m \}$ 
  ($m=0$ and $1$) 
  for the quantum circuit $\hat{\mathcal{U}}_{0,1,2}(\mbox{\boldmath{$\theta$}})$ in Eq.~(\ref{eq:ex:c:qce:3}) 
  [also see Fig.~\ref{fig:device:circuit}(d)] generated by the AQCE algorithm, 
  encoding the GHZ state $\vert \Psi_{\rm GHZ}\rangle$ 
  in the three-qubit space.    }\label{tab:ghz:params}
  \begin{tabular}{lrr}
    \hline
    \hline
    {} & $m=0$ & $m=1$ \\
    \hline
$\theta_{0}^{m}$  &    0.70081942 &    0 \\ 
$\theta_{1}^{m}$  &    1.59343588 &    0 \\ 
$\theta_{2}^{m}$  & $-$2.99819974 &    0 \\ 
$\theta_{3}^{m}$  &    3.01209222 &    0.03670498 \\
$\theta_{4}^{m}$  &    1.45398911 &    1.57079633 \\ 
$\theta_{5}^{m}$  & $-$2.86368415 &    0 \\ 
$\theta_{6}^{m}$  &    0.25868788 &    0.25573854 \\ 
$\theta_{7}^{m}$  &    0.14505637 &    0 \\ 
$\theta_{8}^{m}$  & $-$0.63764672 & $-$2.35619449 \\ 
$\theta_{9}^{m}$  & $-$1.71718177 &    0 \\ 
$\theta_{10}^{m}$ & $-$2.80049545 & $-$3.14159265 \\ 
$\theta_{11}^{m}$ & $-$2.60125707 &    0.04066063 \\ 
$\theta_{12}^{m}$ & $-$2.13304918 & $-$1.57079633 \\ 
$\theta_{13}^{m}$ &    1.68590984 & $-$1.57079633 \\ 
$\theta_{14}^{m}$ & $-$2.04727870 & $-$1.11421776 \\ 
    \hline
  \end{tabular}
\end{table}

\begin{figure}
  \includegraphics[width=\hsize]{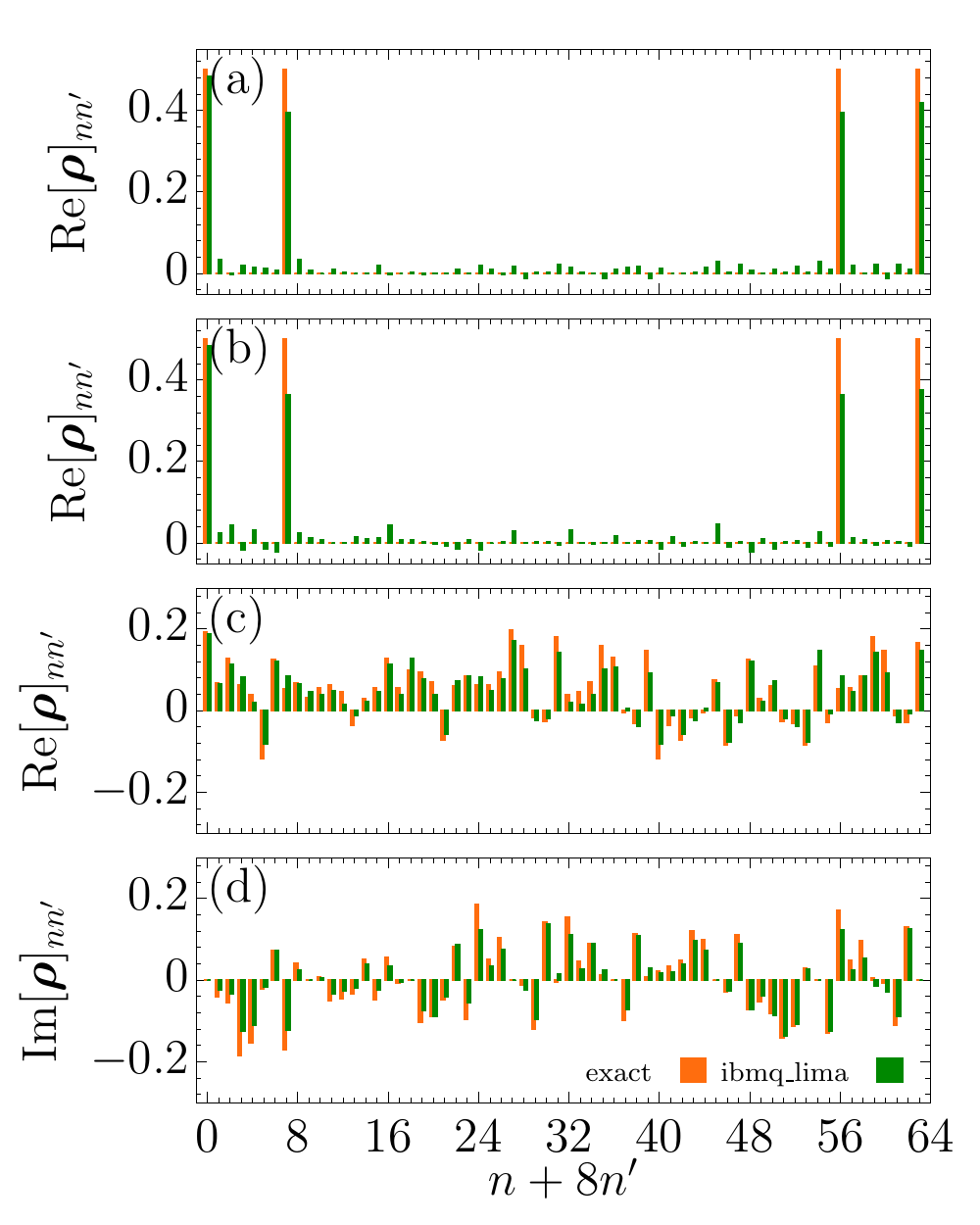}
  \caption{
    (a) Density matrix $[\bm{\rho}]_{nn'}=\langle n|\hat{\rho}|n'\rangle$
    evaluated on the quantum device (ibmq\_lima) with the 
    quantum circuit $\hat{\mathcal{U}}_{\rm GHZ}$ describing the GHZ state $\vert \Psi_{\text{GHZ}}\rangle$ 
    in Eq.~(\ref{eq:ghz}). 
    (b) Same as (a) but with the quantum circuit $\hat{\mathcal{U}}_{0,1,2}(\mbox{\boldmath{$\theta$}}) $ in Eq.~(\ref{eq:ex:c:qce:3}) 
    generated by the AQCE algorithm encoding the GHZ state $\vert \Psi_{\text{GHZ}}\rangle$. 
    The parameter set $\mbox{\boldmath{$\theta$}}$ is provided in Table~\ref{tab:ghz:params}. 
    Note that only the real part of density matrix is shown in (a) and (b) 
    because the imaginary part is zero for the GHZ state $\vert \Psi_{\rm GHZ} \rangle$. 
    (c) Real and (d) imaginary parts of density matrix $[\bm{\rho}]_{nn'}=\langle n|\hat{\rho}|n'\rangle$ evaluated on the quantum 
    device (ibmq\_lima) with the quantum circuit $\hat{\mathcal{U}}^\prime_{0,1,2}(\mbox{\boldmath{$\theta$}}) $ 
    in Eq.~(\ref{eq:ex:c:qce:4}) 
    generated by the AQCE algorithm encoding the random state $\vert \Psi_{\text{3QR}}\rangle$ in Eq.~(\ref{eq:3rnd}). 
    The parameter set $\mbox{\boldmath{$\theta$}}$ is provided in Table~\ref{tab:3rnd}.
    For comparison, the exact results are also shown by orange bars.
    The density matrix is evaluated on the quantum device by the quantum state tomography, measuring 64 different sets of Pauli 
    strings with length three, and the results shown here are obtained from the averaged values over 4096 measurements 
    of each Pauli string. 
    $|n\rangle$ and $|n'\rangle$ with $n,n'=0,1,2,\dots,7$ are the basis states of $L=3$ qubits labeled as in 
     Eq.~(\ref{eq:computational:label}).
  }
  \label{fig:ex:3q}
\end{figure}

Next, using the quantum device, we evaluate in Figs.~\ref{fig:ex:3q}(a) and \ref{fig:ex:3q}(b) the density matrix, 
$[\bm{\rho}]_{nn'}=\langle n|\hat{\rho}|n'\rangle$, 
of the GHZ state generated by the quantum 
circuits $\hat{\mathcal{U}}_{\text{GHZ}}$  in Eq.~(\ref{eq:ex:c:ghzs:man}) 
and $\hat{\mathcal{U}}_{0,1,2}(\mbox{\boldmath{$\theta$}})$ in Eq.~(\ref{eq:ex:c:qce:3}), respectively. 
Here, $|n\rangle$ and $|n'\rangle$ with $n,n'=0,1,2,\dots,7$ are the basis states of $L=3$ qubits labeled as in Eq.~(\ref{eq:computational:label}). 
Similar to the cases of $L=2$ qubits discussed in Sec.~\ref{sec:twoqubits}, we 
evaluate the density matrix by performing the quantum state tomography, where 64 different sets of Pauli strings 
(including the identity operator) with length three [see Eqs.~(\ref{eq:dopt}) and (\ref{eq:dopt2}) for $L=2$ qubits] are measured. 
The density matrix $[\bm{\rho}]_{nn'}$ shown in Figs.~\ref{fig:ex:3q}(a) and \ref{fig:ex:3q}(b) 
is evaluated from the averaged values of Pauli strings measured 4096 times each. 
Although the number of quantum gates in the quantum circuit $\hat{\mathcal{U}}_{0,1,2}(\mbox{\boldmath{$\theta$}})$ is 
much larger than that in the quantum circuit $\hat{\mathcal{U}}_{\text{GHZ}}$, we find that 
the density matrices evaluated on the quantum device with these two different quantum circuits 
are rather similar and are in reasonable agreement with the exact values. 
More quantitatively, using the density matrix evaluated experimentally in Fig.~\ref{fig:ex:3q}(b),
we find that the fidelity $Q$ for the exact GHZ state $\vert \Psi_{\rm GHZ} \rangle$ and
the GHZ state generated by the quantum circuit $\hat{\mathcal{U}}_{0,1,2}(\mbox{\boldmath{$\theta$}})$
is as large as 0.8906. This is comparable to the fidelity $Q = 0.9189$ for the exact GHZ state $\vert \Psi_{\rm GHZ} \rangle$ 
and the GHZ state generated by the quantum circuit $\hat{\mathcal{U}}_{\rm GHZ}$.

Finally, we examine a random state in the three-qubit space: 
\begin{equation}
  \begin{split}
    \vert \Psi_{\text{3QR}} \rangle = &
        ( -0.41507377 + {\rm i} 0.14526187 ) \vert 000 \rangle \\
    & + (  0.03169105 + {\rm i} 0.35848024 ) \vert 1 0 0 \rangle \\
    & + ( -0.23166622 + {\rm i} 0.21332733 ) \vert 0 1 0 \rangle \\
    & + ( -0.32248929 - {\rm i} 0.06104028 ) \vert 1 1 0 \rangle \\
    & + ( -0.11551530 + {\rm i} 0.13972069 ) \vert 0 0 1 \rangle \\
    & + (  0.26960898 - {\rm i} 0.03973709 ) \vert 1 0 1 \rangle \\
    & + (  0.00215509 + {\rm i} 0.44364270 ) \vert 0 1 1 \rangle \\
    & + (  0.01417350 + {\rm i} 0.40747913 ) \vert 1 1 1 \rangle,
  \end{split}
  \label{eq:3rnd}
\end{equation}
where $\vert 0 0 0\rangle = \vert 0\rangle_0\otimes \vert 0\rangle_1 \otimes \vert 0\rangle_2$, etc. and 
the coefficients are randomly determined as in the case of the random state $\vert \Psi_{\rm 2QR} \rangle$ in Eq.~(\ref{eq:ex:2rnd}).
We perform the AQCE algorithm on a classical computer and obtain within the machine precision 
that 
\begin{equation}
  \vert \Psi_{\text{3QR}} \rangle =
    \hat{\mathcal{U}}^{\prime}_{0,1,2}(\mbox{\boldmath{$\theta$}}) \vert 0 \rangle  
\end{equation} 
with the quantum circuit 
\begin{equation}
  \hat{\mathcal{U}}^{\prime}_{0,1,2}(\mbox{\boldmath{$\theta$}}) =
  \hat{\mathcal{U}}_{0,1}(\mbox{\boldmath{$\theta$}}^1)
  \hat{\mathcal{U}}_{1,2}(\mbox{\boldmath{$\theta$}}^0), 
  \label{eq:ex:c:qce:4}
\end{equation}
where $\mbox{\boldmath{$\theta$}} = \{ \mbox{\boldmath{$\theta$}}^{0}, \mbox{\boldmath{$\theta$}}^{1} \}$ and 
the resulting sets of parameters 
$\mbox{\boldmath{$\theta$}}^m = \{ \theta_0^m, \theta_1^m, \Compactcdots, \theta_{14}^m \}$ ($m=0$ and $1$) 
for $\hat{\mathcal{U}}_{1,2}(\mbox{\boldmath{$\theta$}}^0)$ and 
$\hat{\mathcal{U}}_{0,1}(\mbox{\boldmath{$\theta$}}^1)$ are given in Table~\ref{tab:3rnd}. 
The schematic structure of the quantum circuit $ \hat{\mathcal{U}}^{\prime}_{0,1,2}(\mbox{\boldmath{$\theta$}})$ is 
shown in Fig.~\ref{fig:device:circuit}(e). 
Here, we should note that, depending of the initialization process, the AQCE algorithm also finds a quantum circuit 
forming the structure shown in Fig.~\ref{fig:device:circuit}(d) with a different set of parameters $\mbox{\boldmath{$\theta$}}$, 
which can encode the random state $\vert \Psi_{\text{3QR}} \rangle$ exactly within the machine precision. 
This implies that two two-qubit unitary operators are enough to encode any quantum state in the three-qubit space.

\begin{table}
  \caption{
    Sets of parameters $\mbox{\boldmath{$\theta$}}^m = \{ \theta_0^m, \theta_1^m, \Compactcdots, \theta_{14}^m \}$ 
  ($m=0$ and $1$) 
  for the quantum circuit $\hat{\mathcal{U}}^\prime_{0,1,2}(\mbox{\boldmath{$\theta$}})$ in Eq.~(\ref{eq:ex:c:qce:4}) 
  [also see Fig.~\ref{fig:device:circuit}(e)] generated by the AQCE algorithm, 
  encoding the random state $\vert \Psi_{\rm 3QR}\rangle$ 
  in the three-qubit space. 
  }
  \label{tab:3rnd}
  \begin{tabular}{lrr}
    \hline
    \hline
    {} & $m=0$ & $m=1$ \\
    \hline
$\theta_{0}^{m}$  &    1.39099869 &    0.12699636 \\ 
$\theta_{1}^{m}$  &    1.22253363 &    1.49657252 \\ 
$\theta_{2}^{m}$  & $-$1.22510250 & $-$0.96112628 \\ 
$\theta_{3}^{m}$  & $-$1.04474694 &    0 \\ 
$\theta_{4}^{m}$  &    1.85347535 &    0 \\ 
$\theta_{5}^{m}$  & $-$2.24417198 &    0 \\ 
$\theta_{6}^{m}$  & $-$1.06037512 & $-$0.39222573 \\ 
$\theta_{7}^{m}$  & $-$0.87968547 &    0.60984155 \\ 
$\theta_{8}^{m}$  & $-$0.05457889 & $-$0.07696758 \\ 
$\theta_{9}^{m}$  &    0.03359139 &    0.48406694 \\ 
$\theta_{10}^{m}$ &    2.27862931 & $-$0.36703453 \\ 
$\theta_{11}^{m}$ &    0.49867804 &    0.19553219 \\ 
$\theta_{12}^{m}$ &    2.89140237 & $-$1.17312888 \\ 
$\theta_{13}^{m}$ & $-$0.80188802 & $-$2.29176295 \\ 
$\theta_{14}^{m}$ &    1.52534544 & $-$3.06220240 \\ 
    \hline
  \end{tabular}
\end{table}

We perform the quantum state tomography on the quantum device to evaluate 
the density matrix $[\bm{\rho}]_{nn'}$ 
of the random state generated by the quantum 
circuit $\hat{\mathcal{U}}^{\prime}_{0,1,2}(\mbox{\boldmath{$\theta$}})$. 
The 64 different Pauli strings with length three are measured 4096 times each and the density matrix shown in 
Figs.~\ref{fig:ex:3q}(c) and \ref{fig:ex:3q}(d) is obtained from the averaged values of these measurements. 
We find that the fidelity $Q$ for the exact random state $\vert \Psi_{\text{3QR}} \rangle$
and the random state generated by the quantum circuit $\hat{\mathcal{U}}^{\prime}_{0,1,2}(\mbox{\boldmath{$\theta$}})$
is as large as 0.9051, suggesting good accordance with the exact result.

\section{Summary} \label{sec:summary}

We have proposed the quantum circuit encoding algorithm to encode a given quantum state $|\Psi\rangle$ 
onto a quantum circuit $\hat{\mathcal{C}}$ composed of $K$-qubit unitary operators $\hat{\mathcal{U}}_m$ 
by maximizing the absolute value of the fidelity $F=\langle 0| \hat{\mathcal{C}}^\dag |\Psi\rangle$. 
The fidelity $|F|$ can be maximized deterministically by sequentially optimizing each unitary operator $\hat{\mathcal{U}}_m$ 
one by one via the SVD of the fidelity tensor matrix ${\bm F}_m$, 
a similar scheme used for the optimization in the tensor network method. 
The most demanding part of the algorithm is to construct the fidelity tensor matrix ${\bm F}_m$ and we have shown 
how a quantum computer can be utilized for this task. 
The AQCE algorithm proposed here determines not only the form of the individual unitary operators but also the optimal location of qubits 
in the circuit on which each unitary operator acts. Therefore, it allows us to  
generate an optimal quantum circuit of a given quantum state automatically. 
The elementary quantum gates are algebraically assigned when the encoded quantum circuit is 
composed of two-qubit unitary operators. We emphasize that the AQCE algorithm proposed here does not rely on any parametrized 
quantum circuit as in variational quantum algorithms such as the variational quantum eigensolver
and thus the associated parameter optimization is not required.

Using numerical simulations, we have demonstrated the AQCE 
algorithm to encode a ground state of a quantum many-body system, 
such as the spin-1/2 isotropic antiferromagnetic Heisenberg model and the spin-1/2 XY model in one spatial dimension, 
onto a quantum circuit composed of two-qubit unitary operators. 
We have also compared the results with the quantum circuit encoding of the same quantum state onto 
a quantum circuit in a given circuit structure 
such as the Trotter-like and MERA-like circuit structures and found that the quantum circuit generated by the AQCE algorithm is 
better than the Trotter-like circuit and is equally competitive with the MERA-like circuit.

We have also demonstrated that the AQCE algorithm can be applied to encode a quantum state 
representing classical data such as a classical image. As a concrete example, we considered a gray scale picture of 
$256\times256$ pixels, which can be expressed as a quantum state $|\Psi_{\rm c}\rangle$ on $16$ qubits by 
using the amplitude encoding, and thus can be encoded onto a quantum circuit $\hat{\mathcal{C}}$ by employing 
the AQCEalgorithm. 
Although the picture reconstructed by decoding the quantum circuit state $\hat{\mathcal{C}}|0\rangle$ improves its quality 
systematically with increasing the number of two-qubit unitary operators in the quantum circuit $\hat{\mathcal{C}}$, 
the improvement is relatively slow if the size of the picture is large. 
Therefore, we have also made a different attempt by dividing the original picture into 16 pieces, which thus allows us to 
represent each segment of the picture of $64\times64$ pixels with a quantum state $|\Psi_{\rm c}^{(m_s)}\rangle$ on $12$ qubits 
for $m_s=1,2,\dots,16$. 
This implies that the original classical data is represented by a direct product of 16 quantum states 
$|\Psi_{\rm c}^{(m_s)}\rangle$, which is thus defined in a higher dimensional space than the input classical data.
We have encoded each quantum state $|\Psi_{\rm c}^{(m_s)}\rangle$ separately onto a different quantum circuit 
$\hat{\mathcal{C}}^{(m_s)}$ and found that the quality of the reconstructed picture by decoding all these quantum circuit states 
$\hat{\mathcal{C}}^{(m_s)}|0\rangle$ is much improved. 
This is encouraging for a near-term application because, depending on available quantum devices, 
one can adjust the number of qubits by dividing classical data into multiple pieces. 
In the context of quantum machine learning, 
the AQCE algorithm would be potentially useful for finding an optimal quantum circuit, which can be 
done classically, in order to prepare a 
quantum state representing classical data that is to be processed on a quantum computer for machine learning.

Moreover, we have used the quantum device provided by IBM Quantum to demonstrate experimentally that quantum circuits 
generated by the AQCE algorithm can be implemented on a real quantum device to produce a desired quantum state 
with reasonable accuracy. For this purpose, we have considered the well-known quantum states, such as the singlet state and 
the GHZ state, as well as random states in the two- and three-qubit spaces, and shown that the density matrix evaluated
on the quantum device for the quantum circuits obtained by the AQCE algorithm is indeed compatible with the 
exact values.

As clearly demonstrated for several examples, 
the AQCE algorithm can encode a given quantum state onto a quantum circuit with controlled 
accuracy by varying the number $M$ of unitary operators $\hat{\mathcal{U}}_m$ in the quantum circuit. However, we have observed 
that the improvement of accuracy with increasing $M$ becomes sometimes slower when the number of qubits is large. 
There are two possible ways to further improve the AQCE algorithm. One is to improve the procedure of increasing the number of 
unitary operators by $\delta M$ in the enlargement step of the algorithm. 
The procedure adopted as a prototype algorithm in this paper is to simply insert $\delta M$ new 
unitary operators at the end of the quantum circuit (see Fig.~\ref{fig:aqce}). We have found that this simple strategy 
is not the most efficient. 
Instead, one may as well insert these new unitary operators in any location among already existing unitary operators. 
However, this is certainly more costly if a brute-force search is used. 

Another way to improve the AQCE algorithm is related to how to generate and update unitary operators in the quantum circuit. 
In all the demonstrations, 
a quantum state is encoded directly onto a quantum circuit composed of unitary operators acting only on two qubits. 
However, as described 
in Secs.~\ref{sec:encode}, the space on which unitary operators act is not necessary the two-qubit space but the AQCE algorithm 
can encode a quantum state more generally onto a quantum circuit composed of $K$-qubit unitary operators with $K>2$. 
One possible strategy is to encode a quantum state first onto a quantum circuit composed of unitary operators acting on a large 
qubit space, and these unitary operators are then encoded into unitary operators acting on a smaller qubit space. 
We have found that this procedure can improve the accuracy significantly when the number of qubits is large and more details will 
be reported elsewhere.

\section*{Acknowledgement}
The authors are grateful to K. Seki and Y. Otsuka for fruitful discussion.
The calculation has been performed on the RIKEN supercomputer system (HOKUSAI GreatWave) and 
the supercomputer Fugaku installed in RIKEN R-CCS. 
This work was supported by Grant-in-Aid for Scientific Research (A) (No.~JP21H03455),
for Young Scientists (B) (No.~JP17K14359), and
for Transformative Research Areas (A) (No.~21H05191) from MEXT, Japan, 
and by JST PRESTO (No. JPMJPR1911), Japan. 
This work was also supported in part by MEXT Q-LEAP (No.~JPMXS0120319794), Japan, and 
by the COE research grant in computational science from
Hyogo Prefecture and Kobe City through Foundation for Computational Science.

\appendix

\section{Decomposition of a general two-qubit unitary operator} \label{app:assign}

In this appendix, following Ref.~\onlinecite{Kraus2001}, we briefly summarize the derivation of Eq.~(\ref{eq:twogate:physical}). 
Let us first introduce the magic basis
$\{ \vert \phi_0 \rangle, \vert \phi_1 \rangle, \vert \phi_2 \rangle, \vert \phi_3 \rangle \}$
defined on the two-qubit system $\mathbb{I} = \{ i, j \}$: 
\begin{align}
  & \vert \phi_0 \rangle = \frac{1}{\sqrt{2}} ( \vert 00 \rangle + \vert 11 \rangle ), \\
  & \vert \phi_1 \rangle = - \frac{\rm i}{\sqrt{2}} ( \vert 00 \rangle - \vert 11 \rangle ), \\
  & \vert \phi_2 \rangle = \frac{1}{\sqrt{2}} ( \vert 01 \rangle - \vert 10 \rangle ), \\
  & \vert \phi_3 \rangle = - \frac{\rm i}{\sqrt{2}} ( \vert 01 \rangle + \vert 10 \rangle ), 
\end{align}
where $\vert 0 0 \rangle = |0\rangle_i\otimes|0\rangle_j$, $\vert 1 1 \rangle = |1\rangle_i\otimes|1\rangle_j$, 
$\vert 0 1 \rangle = |0\rangle_i\otimes|1\rangle_j$, and $\vert 1 0 \rangle = |1\rangle_i\otimes|0\rangle_j$. 
The unitary transformation $\hat{\mathcal{M}}$ from the magic basis
$\{ \vert \phi_0 \rangle, \vert \phi_1 \rangle, \vert \phi_2 \rangle, \vert \phi_3 \rangle \}$
to the computational basis
$\{ \vert 0 \rangle, \vert 1 \rangle, \vert 2 \rangle, \vert 3 \rangle \} = \{ \vert 00 \rangle, \vert 1 0 \rangle, \vert 0 1 \rangle, \vert 11 \rangle \}$
is given by
\begin{equation}
  \hat{\mathcal{M}} = \sum_{n=0}^3 \sum_{k=0}^3 \vert n \rangle [ {\bm M} ]_{nk} \langle \phi_k \vert,
\end{equation}
where
\begin{equation}
  {\bm M} =
  \frac{1}{\sqrt{2}}
  \left(
  \begin{array}{cccc}
    1 & -{\rm i} & 0 & 0 \\
    0 & 0 & -1 & -{\rm i} \\
    0 & 0 & 1 & -{\rm i} \\
    1 & {\rm i} & 0 & 0 \\
  \end{array}
  \right).
\end{equation}
Any state $\vert \psi \rangle$ on the two-qubit system $\mathbb{I}$ can be represented by 
$\vert \psi \rangle = \sum_{k=0}^3 \mu_k \vert \phi_k \rangle$, where $\mu_k$ is generally complex.

Now let $\{ \vert \psi_k \rangle \}_{k=0}^3$ be a basis where $\langle \psi_k \vert \psi_{k^{\prime}} \rangle = \delta_{kk^{\prime}}$ and
$\vert \psi_k \rangle$ are all maximally entangled. We then call such a basis $\{ \vert \psi_k \rangle \}_{k=0}^3$ maximaly entangled basis.
One can find several interesting properties of this representation.
Here we only list the minimum knowledge to construct the quantum circuit representation of any unitary operator acting on two qubits.
\begin{itemize}
\item[(i)] If $\vert \psi \rangle$ is real in the magic basis ($\mu_k \in \mathbb{R}$), $\vert \psi \rangle$ is maximally entangled.
  If $\vert \psi \rangle$ is maximally entangled, one can choose $\vert \psi \rangle$ real in the magic basis except for the global phase factor.
\item[(ii)] $\sum_{k=0}^3 \mu_k^2 = 0$ for $\vert \psi \rangle = \sum_{k=0}^3 \mu_k \vert \phi_k \rangle$ if and only if 
$\vert \psi \rangle$ is a product state.
\item[(iii)] Let $\{ \vert \psi_k \rangle \}_{k=0}^3$ be a maximally entangled basis.
  Then, one can obtain local unitary operators $\hat{\mathcal{R}}_i$ and $\hat{\mathcal{R}}_j$ and a phase $\xi_k$ such that
  \begin{equation}
    \hat{\mathcal{R}}_i \hat{\mathcal{R}}_j {\rm e}^{{\rm i}\xi_k} \vert \psi_k \rangle = \vert \phi_k \rangle.
    \label{eq:magic:th3}
  \end{equation}
\item[(iv)] For any unitary operator $\hat{\mathcal{U}}$ acting on two qubits, there exist a phase $\varepsilon_k$ and two maximally entangled bases
  $\{ \vert \psi_k \rangle \}_{k=0}^3$ and $\{ \vert \psi_k^{\prime} \rangle \}_{k=0}^3$ such that
  \begin{equation}
    \hat{\mathcal{U}} \vert \psi_k \rangle = {\rm e}^{{\rm i} \varepsilon_k} \vert \psi_k^{\prime} \rangle.
    \label{eq:magic:th4}
  \end{equation}
\item[(v)] The magic states $\vert \phi_k \rangle$ are the eigenstates of the operator
  \begin{equation}
    \hat{\mathcal{D}} = {\rm e}^{-{\rm i}( \alpha_x \hat{X}_i \hat{X}_j + \alpha_y \hat{Y}_i \hat{Y}_j + \alpha_z \hat{Z}_i \hat{Z}_j )},
  \end{equation}
  namely, $\hat{\mathcal{D}} \vert \phi_k \rangle = {\rm e}^{-{\rm i}\lambda_k} \vert \phi_k \rangle$ with  
  $\lambda_k$ being given as
  \begin{align}
    & \lambda_0 = \alpha_x - \alpha_y + \alpha_z + 2 \pi n_0, \label{eq:twogate:lambda0}\\
    & \lambda_1 = - \alpha_x + \alpha_y + \alpha_z + 2 \pi n_1, \label{eq:twogate:lambda1} \\
    & \lambda_2 = - \alpha_x - \alpha_y - \alpha_z + 2 \pi n_2, \label{eq:twogate:lambda2} \\
    & \lambda_3 = \alpha_x + \alpha_y - \alpha_z + 2 \pi n_3, \label{eq:twogate:lambda3}
  \end{align}
\end{itemize}
where $n_k$ ($k=0,1,2,3$) are integers.
The proof of the above properties (i)-(v) is outlined in Ref.~\onlinecite{Kraus2001}.
Here, we briefly comment on 
how to find $\hat{\mathcal{R}}_i$, $\hat{\mathcal{R}}_j$, and ${\rm e}^{{\rm i}\xi_k}$ in (iii) and 
$\{ \vert \psi_k \rangle \}_{k=0}^3$, $\{ \vert \psi_k^{\prime} \rangle \}_{k=0}^3$, and ${\rm e}^{{\rm i} \varepsilon_k}$ in (iv).

In property (iii), 
one can always choose $\vert \psi_k \rangle={\rm e}^{i\eta_k} \vert \bar{\psi}_k \rangle$, where $\vert \bar{\psi}_k \rangle$ is a real 
in the magic basis [property (i)]. 
If we define $\vert \mu \rangle = (\vert \bar{\psi}_0 \rangle + {\rm i} \vert \bar{\psi}_1 \rangle)/\sqrt{2}$
and $\vert \nu \rangle = ( \vert \bar{\psi}_0 \rangle - {\rm i} \vert \bar{\psi}_1 \rangle)/\sqrt{2}$,
then $\vert \mu \rangle$ and $\vert \nu \rangle$ are product states such that 
$\vert \mu \rangle = \vert a \rangle_i \vert b \rangle_j \equiv \vert a b \rangle$
and $\vert \nu \rangle = \vert \bar{a} \rangle_i \vert \bar{b} \rangle_j \equiv \vert \bar{a} \bar{b} \rangle$~\cite{note_ab_bases}.
Since $\vert \bar{\psi}_0 \rangle$ and $\vert \bar{\psi}_1 \rangle$ are maximally entangled states and $\langle \mu \vert \nu \rangle = 0$, 
$\vert a \rangle_i$ and $\vert \bar{a} \rangle_i$ ($\vert b \rangle_j$ and $\vert \bar{b} \rangle_{j}$)
are orthogonal to each other. Similarly, the remaining states $\vert \bar{\psi}_2 \rangle$ and $\vert \bar{\psi}_3 \rangle$ are represented
by using the linear combination of $\vert a \rangle_i \vert \bar{b} \rangle_j \equiv \vert a \bar{b} \rangle$ and
$\vert \bar{a} \rangle_i \vert b \rangle_j \equiv \vert \bar{a} b \rangle$.
Without loss of generality, one can find that $\vert \bar{\psi}_2 \rangle = ( {\rm e}^{{\rm i}\delta} \vert a \bar{b} \rangle - {\rm e}^{-{\rm i}\delta} \vert \bar{a} b \rangle)/\sqrt{2}$ 
and thereby $\vert \bar{\psi}_3 \rangle = -{\rm i} ( {\rm e}^{{\rm i}\delta} \vert a \bar{b} \rangle + {\rm e}^{-{\rm i}\delta} \vert \bar{a} b \rangle )/\sqrt{2}$.
In this case, if we define
\begin{align}
  & \hat{\mathcal{R}}_i = ( \vert 0 \rangle_i ) ({}_i \langle a \vert) + (\vert 1 \rangle_i)({}_i \langle \bar{a} \vert) {\rm e}^{{\rm i}\delta}, \\
  & \hat{\mathcal{R}}_j = ( \vert 0 \rangle_j ) ({}_j \langle b \vert) + (\vert 1 \rangle_j)({}_j \langle \bar{b} \vert) {\rm e}^{-{\rm i}\delta},
\end{align}
we can obtain the relation in Eq.~(\ref{eq:magic:th3}) by choosing appropriately the phase factor $\xi_k$.
In this way, we can obtain the local unitary operators $\hat{\mathcal{R}}_i$ and $\hat{\mathcal{R}}_j$, and the phase factors $\xi_k$ 
in (iii).

In property (iv), for a given unitary operator $\hat{\mathcal{U}}$, let $\vert \psi_k \rangle$ be the eigenstates of the operator 
$\hat{\mathcal{W}} = \hat{\mathcal{U}}^t \hat{\mathcal{U}}$ with the corresponding eigenvalues ${\rm e}^{2{\rm i}\varepsilon_k}$, 
where $\hat{\mathcal{U}}^t$ is the transpose of $\hat{\mathcal{U}}$.
From the fact that $\hat{\mathcal{W}}^{\dagger} \hat{\mathcal{W}} = \hat{1}$ and $\hat{\mathcal{W}}^t = \hat{\mathcal{W}}$, 
the eigenstates $\vert \psi_k \rangle$ can be chosen as real in the magic basis and hence $\{ \vert \psi_k \rangle \}_{k=0}^3$ is a 
maximally entangled basis.
From the eigenvalue equation $( \hat{\mathcal{W}} - {\rm e}^{2{\rm i}\varepsilon_k} ) \vert \psi_k \rangle = 0$, one can readily show 
that $\vert \psi^{\prime}_k \rangle \equiv {\rm e}^{-{\rm i}\varepsilon_k} \hat{\mathcal{U}} \vert \psi_k \rangle$ is real in the magic basis,
suggesting that $\{ \vert \psi_{k'} \rangle \}_{k=0}^3$ is also a maximally entangled basis. 
Therefore, we obtain the maximally entangled bases $\{ \vert \psi_k \rangle \}_{k=0}^3$ and $\{ \vert \psi_k^{\prime} \rangle \}_{k=0}^3$, 
and the phase factor $\varepsilon_k$ in (iv)

Now, let $\vert \psi_k \rangle$ and $\vert \psi_k^{\prime} \rangle$ be the states that satisfy property (iv) 
for a unitary operator $\hat{\mathcal{U}}$.
From property (iii), one can find the set of the local operators $\hat{\mathcal{R}}_i$ and $\hat{\mathcal{R}}_j$ and the phase $\xi_k$ 
for $\vert \psi_k \rangle$ such that
\begin{equation}
  \vert \psi_k \rangle = {\rm e}^{-{\rm i}\xi_k} \hat{\mathcal{R}}_i^{\dagger} \hat{\mathcal{R}}_j^{\dagger} \vert \phi_k \rangle.
  \label{eq:magic:local:1}
\end{equation}
Similarly, we can also find $\hat{\mathcal{R}}_i^{\prime}$, $\hat{\mathcal{R}}_j^{\prime}$, and $\xi_k^{\prime}$ for $\vert \psi_k^{\prime} \rangle$
such that 
\begin{equation}
  \vert \psi_k^{\prime} \rangle = {\rm e}^{-{\rm i}\xi^{\prime}_k} \hat{\mathcal{R}}_i^{\prime} \hat{\mathcal{R}}_j^{\prime} \vert \phi_k \rangle.
  \label{eq:magic:local:2}
\end{equation}
Inserting Eqs.~(\ref{eq:magic:local:1}) and (\ref{eq:magic:local:2}) into (\ref{eq:magic:th4}),
we obtain that 
\begin{equation}
  \hat{\mathcal{U}} {\rm e}^{-{\rm i}\xi_k} \hat{\mathcal{R}}_i^{\dagger} \hat{\mathcal{R}}_j^{\dagger} \vert \phi_k \rangle 
  = {\rm e}^{{\rm i}\varepsilon_k} {\rm e}^{-{\rm i}\xi^{\prime}_k} \hat{\mathcal{R}}_i^{\prime} \hat{\mathcal{R}}_j^{\prime}
  \vert \phi_k \rangle.
\end{equation}
Multiplying $( \hat{\mathcal{R}}_i^{\prime} \hat{\mathcal{R}}_j^{\prime} )^\dag$ from the left,
we obtain that 
\begin{equation}
  ( \hat{\mathcal{R}}_i^{\prime} )^{\dagger}
  ( \hat{\mathcal{R}}_j^{\prime} )^{\dagger}
  \hat{\mathcal{U}}
  \hat{\mathcal{R}}_i^{\dagger}
  \hat{\mathcal{R}}_j^{\dagger} \vert \phi_k \rangle
  =
  {\rm e}^{-{\rm i}(\xi^{\prime}_k-\xi_k-\varepsilon_k)} \vert \phi_k \rangle.
  \label{eq:magic:rrurr}
\end{equation}
Equation~(\ref{eq:magic:rrurr}) suggests that
the operator $( \hat{\mathcal{R}}_i^{\prime} )^{\dagger} ( \hat{\mathcal{R}}_j^{\prime} )^{\dagger} \hat{\mathcal{U}} \hat{\mathcal{R}}_i^{\dagger} \hat{\mathcal{R}}_j^{\dagger}$
becomes identical to ${\rm e}^{-{\rm i}\alpha_0}\hat{\mathcal{D}}$ if we choose
\begin{equation}
  \alpha_0+\lambda_k = \xi^{\prime}_k - \xi_k - \varepsilon_k + 2 \pi n_k
  \label{eq:twogate:solution}
\end{equation}
for $k=0,1,2,3$. Inserting the Eqs.~(\ref{eq:twogate:lambda0})-(\ref{eq:twogate:lambda3}) into Eqs.(\ref{eq:twogate:solution}),
we can determine the phase factors $\alpha_{\mu}$ ($\mu=0,x,y,z$)
and therefore obtain the following form of the unitary operator $\hat{\mathcal{U}}$: 
\begin{equation}
  \hat{\mathcal{U}} = {\rm e}^{-{\rm i}\alpha_0} \hat{\mathcal{R}}_i^{\prime} \hat{\mathcal{R}}_j^{\prime} \hat{\mathcal{D}} \hat{\mathcal{R}}_i \hat{\mathcal{R}}_j. 
  \label{eq:twogate:physical2}
\end{equation}
Since $\hat{\mathcal{R}}_i^{\prime}$, $\hat{\mathcal{R}}_j^{\prime}$, $\hat{\mathcal{R}}_i$, and $\hat{\mathcal{R}}_j$ are
the single-qubit unitary operators, there alway exist the representations by using the Euler rotations (see Appendix~\ref{app:phase}). 
The two-qubit unitary operator $\hat{\mathcal{D}}$ is represented by the elementary single- 
and two-qubit gates~\cite{Vidal2004,Mark2008}, as given in Eq.~(\ref{eq:exgate:decomposed}) and also in Fig.~\ref{fig:twogate}(b).

Finally, we note the practical procedure to determine the quantum gates explicitly for a given unitary operator $\hat{\mathcal{U}}$.  
In order to assign quantum gates properly for a unitary operator $\hat{\mathcal{U}}$, 
one has to select an appropriate set of integers $n_k$ for $k=0,1,2,3$  
in Eq.~(\ref{eq:twogate:solution}). In addition, there might also be additional phases when the single-qubit unitary operators are
represented by the standard single-qubit rotation gates, as described in Appendix~\ref{app:phase}. 
Since the number of possible combinations is limited, one can always find the appropriate set of integers $n_k$ by checking 
all combinations, in practice.

\section{Phase factors in a single-qubit unitary operator} \label{app:phase}

As shown in Eqs.~(\ref{eq:twogate:physical}) and (\ref{eq:twogate:physical2}), a two-qubit unitary operator $\hat{\mathcal{U}}$ is 
generally decomposed into $\hat{\mathcal{U}} = {\rm e}^{-{\rm i}\alpha_0} \hat{\mathcal{R}}_i^{\prime} \hat{\mathcal{R}}_j^{\prime} \hat{\mathcal{D}}  \hat{\mathcal{R}}_i \hat{\mathcal{R}}_j$, where 
$\hat{\mathcal{R}}_i^{\prime}$, 
$\hat{\mathcal{R}}_j^{\prime}$,
$\hat{\mathcal{R}}_i$,
and
$\hat{\mathcal{R}}_j$
are single-qubit unitary operators. It is well known that  
these single-qubit operators can be assigned by using the Euler rotation operator 
given as
\begin{equation}
  \hat{\mathcal{R}}(\theta_1,\theta_2,\theta_3) =
      {\rm e}^{-{\rm i}\theta_3 \hat{Z}/2}
      {\rm e}^{-{\rm i}\theta_2 \hat{Y}/2}
      {\rm e}^{-{\rm i}\theta_1 \hat{Z}/2}, 
      \label{eq:R_Euler}
\end{equation}
where $\hat{Y}$ and $\hat{Z}$ are the Pauli operators acting on the target qubit $i$ or $j$.
However, this is true only when we introduce the overall phase factor 
$\theta_0$, as explicitly shown below.

Let us now denote 
$\hat{\mathcal{V}} \in \{ \hat{\mathcal{R}}_i^{\prime}, \hat{\mathcal{R}}_j^{\prime}, \hat{\mathcal{R}}_i, \hat{\mathcal{R}}_j \}$
and assume that $\hat{\mathcal{V}}$ is given in the computational basis as 
\begin{equation}
  \hat{\mathcal{V}} = \sum_{\sigma=0,1} \sum_{\sigma^{\prime}=0,1} \vert \sigma \rangle [ {\bm V} ]_{\sigma\sigma^{\prime}} \langle  \sigma^{\prime} \vert,
\end{equation}
where $\bm V$ is the matrix representation of $\hat{\mathcal{V}}$ 
with the element $[ {\bm V} ]_{\sigma\sigma^{\prime}}=v_{\sigma\sigma'}$. 
The single-qubit operator $\hat{\mathcal{V}}$ can be assigned by using the Euler rotation operator as
\begin{equation}
  \hat{\mathcal{V}} = {\rm e}^{-{\rm i}\theta_0/2} \hat{\mathcal{R}}(\theta_1,\theta_2,\theta_3).
      \label{eq:R_Euler2}
\end{equation}
On the other hand, the matrix representation ${\bm R}$ of $\hat{\mathcal{R}}(\theta_1,\theta_2,\theta_3)$
in the computational basis is given by 
\begin{equation}
  {\bm R} = \left(
  \begin{array}{cc}
    {\rm e}^{-{\rm i}(\theta_3+\theta_1)/2} \cos (\theta_2/2) & -{\rm e}^{-{\rm i}(\theta_3-\theta_1)/2} \sin (\theta_2/2) \\
    {\rm e}^{{\rm i}(\theta_3-\theta_1)/2} \sin (\theta_2/2) & {\rm e}^{{\rm i}(\theta_3+\theta_1)/2} \cos (\theta_2/2) \\
  \end{array}
  \right).
\end{equation}
Therefore, one can deterime $\theta_0$, $\theta_1$, $\theta_2$, and $\theta_3$
by solving the following simultaneous nonlinear equations: 
\begin{align}
  v_{00} = & {\rm e}^{-{\rm i}(\theta_0+\theta_3+\theta_1)/2} \cos (\theta_2/2), \\
  v_{10} = & {\rm e}^{-{\rm i}(\theta_0-\theta_3+\theta_1)/2} \sin (\theta_2/2), \\
  v_{01} = & -{\rm e}^{-{\rm i}(\theta_0+\theta_3-\theta_1)/2} \sin (\theta_2/2), \\
  v_{11} = & {\rm e}^{-{\rm i}(\theta_0-\theta_3-\theta_1)/2} \cos (\theta_2/2).
\end{align}

We can readily find that the solution of these equations is given as 
\begin{align}
  \theta_0 = & {\rm i} \ln \left( v_{00} v_{11} - v_{10} v_{01} \right) + \pi m_0 / 2, \\
  \theta_1 = & {\rm i} \ln \left( - \frac{v_{00}v_{10}}{v_{11}v_{01}} \right) + \pi m_1 / 2, \\
  \theta_2 = & \arccos ( \frac{1}{2} \vert v_{00} v_{11} + v_{10} v_{01} \vert ) + \pi m_2 / 2, \\
  \theta_3 = & {\rm i} \ln \left( - \frac{v_{00}v_{01}}{v_{11}v_{10}} \right) + \pi m_3 / 2
\end{align}
for $ v_{01}  v_{10} \neq 0$ and $ v_{00}  v_{11} \neq 0$. 
Here, $m_i$ ($i=0,1,2,3$) is an integer number that is determined so as to reproduce the sign
of the original matrix elements $v_{\sigma\sigma'}$.

For $ v_{01}  =  v_{10}  = 0$ but $ v_{00} v_{11}  \neq 0$, 
we can set $\theta_2 = 0$ and thus the Euler rotation is simply the rotation around the $z$-axis. 
Therefore, only $\theta_1 + \theta_3$ is relevant and we set $\theta_3 = 0$ without losing generality. 
The solution is thus given as  
\begin{align}
  \theta_0 = & {\rm i} \ln \left( v_{00} v_{11} \right) + \pi m_0, \\
  \theta_1 = & 2 {\rm i} \ln \left( \frac{v_{00}}{v_{11}} \right) + \pi m_1, \\
  \theta_2 = & 0, \\
  \theta_3 = & 0.
\end{align}

For $ v_{00}  =  v_{11}  = 0$ but $ v_{01} v_{10}  \neq 0$, 
we can set $\theta_2 = \pi$ without losing generality.
Similar to the previous case, we can also set $\theta_3=0$ 
because ${\rm e}^{-{\rm i}\pi\hat{Y}/2}=-{\rm i}\hat{Y}$.
The solution is thus given as  
\begin{align}
  \theta_0 = & {\rm i} \ln \left( - v_{10} v_{01} \right) + \pi m_0, \\
  \theta_1 = &  {\rm i} \ln \left(  - \frac{v_{10}}{v_{01}}  \right) +  \pi m_1  , \\
  \theta_2 = & \pi, \\
  \theta_3 = & 0.
\end{align}

\bibliography{qce}

\begin{thebibliography}{67}
\bibitem [{pha()}]{phase}%
  \BibitemOpen
  \href@noop {} {}\bibinfo {note} {Apart from a global phase factor that is
  irrelevant to the discussion here.}\BibitemShut {Stop}%
\bibitem [{not()}]{note_ab_bases}%
  \BibitemOpen
  \href@noop {} {}\bibinfo {note} {In practice, we can obtain, for example,
  $\vert a \rangle_i$ and $\vert b \rangle_j$ by performing the SVD of the
  matrix $\mu_{nn^{\prime}}$, where $\mu_{nn^{\prime}}$ is defined as $\vert
  \mu \rangle = ( \vert \bar{\Psi}_0 \rangle + {\rm i} \vert \bar{\Psi}_1
  \rangle ) / \sqrt{2} = \sum_{n=0}^1 \sum_{n^{\prime}=0}^1 \mu_{nn^{\prime}}
  \vert n \rangle_i \vert n^{\prime} \rangle_j$.}\BibitemShut {Stop}%
\end{thebibliography}

\end{document}